\documentclass[11pt,a4paper]{article}
\pdfoutput=1
\usepackage{amsfonts}
\usepackage{amsmath}
\usepackage{amssymb}
\usepackage{amsthm}
\usepackage{graphicx}
\usepackage{booktabs}
\usepackage{theorem}
\usepackage{array}
\usepackage[numbers, sort]{natbib}
\usepackage{hyperref}
\usepackage{color}
\usepackage{color}
\usepackage{accents}
\usepackage[footnotesize]{caption}

\allowdisplaybreaks[1]

\makeatletter

\newcommand{\Rmnum}[1]{\expandafter\@slowromancap\romannumeral #1@}
\makeatother

\begin{document}

\begin{titlepage}

%\vspace*{-15mm}
\begin{flushright}
%MPP-2010-52\\
\end{flushright}
%\vspace*{0.7cm}

\begin{center}
{ \bf\Large Renormalization Group Running of Lepton Mixing
Parameters in See-Saw Models with $S_4$ Flavor Symmetry }
\\[8mm]
Gui-Jun Ding$^{a,b}$ % \footnote{E-mail:\texttt{dinggj@ustc.edu.cn}}
, Dong-Mei Pan$^{a}$

\end{center}
\vspace*{0.50cm} \centerline{\it$^{a}$Department of Modern Physics,}
\centerline{\it University of Science and Technology of China,
Hefei, Anhui 230026, China} \vspace*{0.3cm} \centerline{\it
$^{b}$Department of Physics,} \centerline{\it University of
Wisconsin-Madison,1150 University Avenue, Madison, WI 53706, USA }
\vspace*{1.20cm}
\begin{abstract}

\noindent

We study the renormalization group running of the tri-bimaximal
mixing predicted by the two typical $S_4$ flavor models at leading
order. Although the textures of the mass matrices are completely
different, the evolution of neutrino mass and mixing parameters is
found to display approximately the same pattern. For both normal
hierarchy and inverted hierarchy spectrum, the quantum corrections
to both atmospheric and reactor neutrino mixing angles are so small
that they can be neglected. The evolution of the solar mixing angle
$\theta_{12}$ depends on $\tan\beta$ and neutrino mass spectrum, the
deviation from its tri-bimaximal value could be large. Taking into
account the renormalization group running effect, the neutrino
spectrum is constrained by  experimental data on $\theta_{12}$ in addition to
the self-consistency conditions of the models, and
the inverted hierarchy spectrum is disfavored for large $\tan\beta$.
The evolution of light-neutrino masses is approximately described by
a common scaling factor.

\end{abstract}

\end{titlepage}

\setcounter{footnote}{0}

\section{\label{sec:int}Introduction}

The neutrino physics has made great progress in the past decades.
The mass square differences $\Delta m^2_{\rm sol}$, $\Delta
m^2_{\rm atm}$ and the mixing angles have been measured with good
accuracy \cite{Strumia:2006db,Schwetz:2008er,Fogli:Indication}.
A global fit to the current neutrino oscillation data demonstrates
that the observed lepton mixing matrix is remarkably compatible with
the tri-bimaximal (TB) mixing pattern \cite{TBmix}, which suggests
the following values of the mixing angles:
\begin{equation}
\label{1}\sin^2\theta^{TB}_{12}=\frac{1}{3},~~~\sin^2\theta^{TB}_{23}=\frac{1}{2},~~\sin\theta^{TB}_{13}=0
\end{equation}
The question of how to achieve TB mixing has been the subject of
intense theoretical speculation. Recently it has been found that the
flavor symmetry based on the discrete group is particularly suitable
to reproduce this specific mixing pattern in leading order (LO).
Various discrete flavor symmetry models have been built, please see
the Refs.\cite{Altarelli:2010gt,Ishimori:2010au} for a review. A
common feature of these model is to produce TB mixing at leading
order, and the leading order predictions are always subjected to
corrections due to higher dimensional operators in both the driving
superpotential and the Yukawa superpotentials. These models provide
an elegant description of neutrino mixing at very high energy scale,
whereas the neutrino experiments are performed at low energy scale.
In order to compare the model predictions with experimental data,
one has to perform a renormalization group (RG) running from the
high energy scale where the theory is defined to the electroweak
scale $M_Z$. Moreover, we note that RG effects have interesting
implications for model building, the lepton mixing angles can be
magnified \cite{Balaji:2000au}, even the bimaximal mixing at high
energy can be compatible with low energy experiment
\cite{Antusch:2002hy}. Therefore, in a consistent flavor model
building, we have to guarantee that the successful leading order
predictions are not destroyed by the RG running corrections. The aim
of this work is to analyze the RG corrections on the TB mixing
pattern in two typical $S_4$ flavor models
\cite{Bazzocchi:2009pv,Ding:2009iy} in addition to the next to
leading order corrections arising from high dimensional operators
and to confront them with experimental values. We shall see that the
running of the neutrino parameters is strongly constrained by the
flavor symmetry as well, and this result holds very generally for the discrete
flavor symmetry models.

The $S_4$ flavor symmetry is very interesting. From the group theory
point of view \cite{Lam:2008rs}, it is the minimal group which can
produce the TB mixing in a natural way, namely without ad hoc
assumptions. It is remarkable that we have more alternatives to
realize the exact TB mixing than those in the $A_4$ flavor model
\cite{Bazzocchi:2009pv,Ding:2009iy,Ma:2005pd,Meloni:2009cz,Altarelli:2009gn,Grimus:2009pg}.
In particular, the $\mathbf{2}$ dimensional irreducible
representation of $S_4$ group can be utilized to describe the quark
sector. Moreover, the group $S_4$ as a flavor symmetry, as has been
shown for example in Refs.
\cite{Dutta:2009bj,Hagedorn:2010th,Ishimori:2008fi,Toorop:2010yh,Ding:2010pc},
can also give a successful description of the quark and lepton
masses and mixing angles within the framework of grand unified
theory (GUT). We note that $S_4$ as a flavor symmetry has been
investigated long ago \cite{Pakvasa:1978tx,Hagedorn:2006ug}, but
with different aims and different results.

%We note that the RG corrections to TB mixing (but not in the context of flavor symmetry) have considered in
The paper is organized as follows. In section \ref{sec:RGE_review},
we briefly review the RG equations for the type I see-saw mechanism.
Then we give a concise introduction to the Bazzocchi-Merlo-Morisi
(BMM) $S_4$ model \cite{Bazzocchi:2009pv} and the $S_4$ model of
Ding \cite{Ding:2009iy} in section \ref{sec:models}, where the main
features of these models are shown. In section
\ref{sec:RGE_running}, Our results of RG effects on the neutrino
mixing parameters for these two interesting models are presented.
Finally we draw our conclusions in section \ref{sec:conclusion}.

\section{\label{sec:RGE_review}Running of neutrino parameters in type I see-saw scenario}

The running of neutrino masses and lepton mixing angles is very
important and has been studied extensively in the literature
\cite{rge1,rge2,rge3,rge4,rge5,Chakrabortty:2008zh,Antusch:2003kp,Antusch:2005gp,Xing:2007fb,Bergstrom:2010qb,Blennow:2011mp}
in the past years. In particular, Antusch et al. have developed the
Mathematica package REAP in Ref. \cite{Antusch:2005gp}, which can
solve renormalization group equations (RGE) and provide numerical
values for the neutrino mass and mixing parameters. In this section,
we present the RGEs for neutrino parameters in the minimal
supersymmetric standard model (MSSM) extended by three singlet
(right-handed) heavy neutrinos. The superpotential is given by
\begin{equation}
\label{2}W=D^cY_dQH_d+U^cY_uQH_u+E^cY_eLH_d+N^cY_{\nu}LH_u+\frac{1}{2}N^cMN^c
\end{equation}
where $Q$ and $L$ are the left-handed quark and lepton doublets
chiral superfields, respectively, $U^c$, $D^c$, $N^c$ and $E^c$ are
right-handed up-type quark, down-type quark, heavy neutrino and
charged lepton singlet superfields, respectively, $H_u$ and $H_d$ are
the well-known two Higgs doublets in MSSM. The Yukawa matrices
$Y_u$, $Y_d$, $Y_{\nu}$ and $Y_e$ are general complex $3\times3$
matrices and the $3\times3$ heavy neutrino mass matrix $M$ is
symmetric. Integrating out all the heavy singlet neutrinos, one gets
the usual dimension-5 effective neutrino mass operator
%\begin{equation}
%\label{3}{\cal L}_{\kappa}=\frac{1}{4}\kappa_{fg}({\overline{\ell^{fC}_L}}\cdot h_u)(\ell^{g}_L\cdot h_u)+h.c.
%\end{equation}
\begin{equation}
\label{3}{\cal L}_{\kappa}=-\frac{1}{4}\kappa_{fg}(L^{f}\cdot H_u)(L^{g}\cdot H_u)
\end{equation}
%the superscript $C$ denotes charge conjugation of the fermion field,
where $f$ and $g$ are family indices, and the dot indicates the
$SU(2)_L$ invariant contractions. After electroweak symmetry
breaking, this operator leads to the light-neutrino masses
\begin{equation}
\label{4}m_{\nu}(\mu)=-\frac{1}{4}\kappa(\mu)v^2\sin^2\beta
\end{equation}
where $\mu$ is the renormalization scale, $v=246$ GeV and
$\tan\beta=v_u/v_d$ is the ratio of vacuum expectation values (VEV)
of the Higgs doublets. Above the heaviest neutrino mass scale, the
light-neutrino mass matrix reads
\begin{equation}
\label{5}m_{\nu}(\mu)=-\frac{1}{2}Y^T_{\nu}(\mu)M^{-1}(\mu)Y_{\nu}(\mu)v^2\sin^2\beta
\end{equation}
When we evolved the energy from high energy scale down to the low
experimental observation scale, the heavy singlet neutrinos involved
in the see-saw mechanism have to be integrated out one by one, thus
one has to consider a series of effective theories
\cite{Antusch:2005gp}. In general, the light-neutrino mass matrix
can be written  as\footnote{We use the GUT charge normalization for
the gauge coupling $g_1$.}
\begin{equation}
\label{6}m_{\nu}=-\frac{1}{4}\left(\accentset{(n)}{\kappa}+2\accentset{(n)}{Y}^T_{\nu}\accentset{(n)}{M}^{-1}\accentset{(n)}{Y}_{\nu}\right)v^2\sin^2\beta
\end{equation}
where the superscript $(n)$ denotes a quantity below the $n$th mass
threshold. In the MSSM, the two parts $\accentset{(n)}{\kappa}$ and
$2\accentset{(n)}{Y}^T_{\nu}\accentset{(n)}{M}^{-1}\accentset{(n)}{Y}_{\nu}$
evolve in the same way
\begin{eqnarray}
\label{8}16\pi^2\frac{d \accentset{(n)}{X}}{d t}=\left(Y^{\dagger}_eY_e+\accentset{(n)}{Y}^{\dagger}_{\nu}\accentset{(n)}{Y}_{\nu}\right)^{T}\accentset{(n)}{X}+\accentset{(n)}{X}\left(Y^{\dagger}_eY_e+\accentset{(n)}{Y}^{\dagger}_{\nu}\accentset{(n)}{Y}_{\nu}\right)+\Big[2\,{\rm Tr}\big(\accentset{(n)}{Y}^{\dagger}_{\nu}\accentset{(n)}{Y}_{\nu}+3Y^{\dagger}_uY_u\big)-\frac{6}{5}g^2_1-6g^2_2\Big]\accentset{(n)}{X}
\end{eqnarray}
where $t=\ln(\mu/\mu_0)$, and $\accentset{(n)}{X}$ stands for
$\accentset{(n)}{\kappa}$ or
$2\accentset{(n)}{Y}^T_{\nu}\accentset{(n)}{M}^{-1}\accentset{(n)}{Y}_{\nu}$.
The RG equations for the Yukawa couplings $Y_u$, $Y_d$, $Y_{\nu}$,
$Y_e$ and the right-handed neutrino mass matrix $M$ are given by
\begin{eqnarray}
\nonumber&&16\pi^2\frac{d\accentset{(n)}{Y}_{\nu}}{dt}=\accentset{(n)}{Y}_{\nu}\left[3\accentset{(n)}{Y}^{\dagger}_{\nu}\accentset{(n)}{Y}_{\nu}+Y^{\dagger}_eY_e+{\rm Tr}(\accentset{(n)}{Y}^{\dagger}_{\nu}\accentset{(n)}{Y}_{\nu})+3{\rm Tr}(Y^{\dagger}_uY_u)-\frac{3}{5}g^2_1-3g^2_2\right]\\
\nonumber&&16\pi^2\frac{dY_e}{dt}=Y_e\left[3Y^{\dagger}_eY_e+\accentset{(n)}{Y}^{\dagger}_{\nu}\accentset{(n)}{Y}_{\nu}+3{\rm Tr}(Y^{\dagger}_dY_d)+{\rm Tr}(Y^{\dagger}_eY_e)-\frac{9}{5}g^2_1-3g^2_2\right]\\
\nonumber&&16\pi^2\frac{dY_u}{dt}=Y_u\left[Y^{\dagger}_dY_d+3Y^{\dagger}_uY_u+{\rm Tr}(\accentset{(n)}{Y}^{\dagger}_{\nu}\accentset{(n)}{Y}_{\nu})+3{\rm Tr}(Y^{\dagger}_uY_u)-\frac{13}{15}g^2_1-3g^2_2-\frac{16}{3}g^2_3\right]\\
\nonumber&&16\pi^2\frac{dY_d}{dt}=Y_d\left[3Y^{\dagger}_dY_d+Y^{\dagger}_uY_u+3{\rm Tr}(Y^{\dagger}_dY_d)+{\rm Tr}(Y^{\dagger}_eY_e)-\frac{7}{15}g^2_1-3g^2_2-\frac{16}{3}g^2_3\right]\\
\label{9}&&16\pi^2\frac{d\accentset{(n)}{M}}{dt}=2(\accentset{(n)}{Y}_{\nu}\accentset{(n)}{Y}^{\dagger}_{\nu})\accentset{(n)}{M}+2\accentset{(n)}{M}(\accentset{(n)}{Y}_{\nu}\accentset{(n)}{Y}^{\dagger}_{\nu})^{T}
\end{eqnarray}
In the full theory above the highest see-saw scale, the superscript
$(n)$ has to be omitted, and the RG equations for MSSM without
singlet neutrinos can be recovered by setting the neutrino Yukawa
couplings and the mass matrix of the singlets to be zero. Below the
SUSY breaking scale, which is taken to be 1000 GeV in this work, we
go to the standard model region. Since all the heavy right-handed
neutrinos have already been integrated out at this scale, the
neutrino masses are described by the effective dimension-5 operator,
and the neutrino mass matrix $m_{\nu}$ evolves as
\begin{equation}
\label{10}16\pi^2\frac{dm_{\nu}}{dt}=-\frac{3}{2}(Y^{\dagger}_eY_e)^{T}m_{\nu}-\frac{3}{2}m_{\nu}(Y^{\dagger}_eY_e)+\left[2{\rm Tr}(3Y^{\dagger}_uY_u+3Y^{\dagger}_dY_d+Y^{\dagger}_eY_e)-3g^2_2+\lambda\right]m_{\nu}
\end{equation}
where $\lambda$ is the Higgs self-interaction coupling\footnote{We
use the convention that the Higgs self-interaction term in the
Lagrangian is $-\frac{\lambda}{4}(H^{\dagger}H)^2$.}. In order to
calculate the RG evolution of the effective neutrino mass matrix, we
have to solve the RG equations for all the parameters of the theory
simultaneously\footnote{The running of the gauge couplings has to be
taken into account as well, the corresponding $\beta$ functions are
well-known .}. At the mass threshold, we should integrate out the
corresponding heavy neutrino and perform the tree-level matching
condition for the effective coupling constant between the effective
theories
\begin{equation}
\label{7}\accentset{(n)}{\kappa}_{gf}\Big|_{M_n}=\;\;\accentset{(n+1)}{\kappa}_{gf}\Big|_{M_n}+2(\;\;\accentset{(n+1)}{Y}_{\nu}^{\,\,T}\,)_{gn}M^{-1}_n(\;\;\accentset{(n+1)}{Y}_{\nu}\;)_{nf}\Big|_{M_n}~~~({\rm no\,sum\,over}\, n)
\end{equation}
%matching procedure shown in Eq.(\ref{7}).

\section{\label{sec:models}Variants of the Two $S_4$ Models\label{sec:BMM}}

In this section, we recapitulate the main features of the $S_4$
flavor model of BMM \cite{Bazzocchi:2009pv} and Ding
\cite{Ding:2009iy}. Both models generate neutrino masses via type I
see-saw mechanism, and the neutrino TB mixing is produced at LO. For
an introduction to the group theory of $S_4$ we refer to
Refs.\cite{Ding:2009iy,Ding:2010pc}, the same conventions for the
$S_4$ representation matrix and Clebsch-Gordan coefficient are used
in this work.

\subsection{\label{sec:BMM}BMM $S_4$ model}
In this model the flavor symmetry $S_4$ is accompanied by the cyclic
group $Z_5$ and the Froggatt-Nielsen symmetry $U(1)_{FN}$. The $S_4$
flavor symmetry is spontaneously broken to the subgroup $Z_2\times
Z_2$ in the neutrino sector and to nothing in the charged lepton one
at leading order. This misalignment between the flavor symmetry
breaking in the neutrino and charged lepton sectors is exactly the
origin of the TB mixing. Furthermore, the auxiliary symmetry $Z_5$
eliminating some dangerous terms, with the interplay of the
continuous $U(1)_{FN}$, is responsible for the hierarchy among the
charged lepton masses. The leptonic fields and the flavon fields of
the model and their transformation properties under the flavor
symmetry are shown in Table \ref{table:BMM_transformation}.

\begin{table}[hptb]
\begin{center}
\begin{tabular}{|c|c|c|c|c|c|c||c||c|c||c|c||c|}
  \hline\hline
%  &&&&&&&&&&&& \\[-0,3cm]
  & $\ell$ & $e^c$ & $\mu^c$ & $\tau^c$ & $\nu^c$ & $H_{u,d}$ & $\theta$ & $\psi$ & $\eta$ & $\Delta$ & $\varphi$ & $\xi'$ \\
%  &&&&&&&&&&&& \\[-0,3cm]
  \hline
%  &&&&&&&&&&&& \\[-0,3cm]
  $S_4$ & $3_1$ & $1_2$ & $1_2$ & $1_1$ & $3_1$ & $1_1$ & $1_1$ & $3_1$ & 2 & $3_1$ & $2$ & $1_2$  \\
%  &&&&&&&&&&&& \\[-0,3cm]
  $Z_5$ & $\omega^4$ & $1$ & $\omega^2$ & $\omega^4$ & $\omega$ & 1 & 1 & $\omega^2$ & $\omega^2$ & $\omega^3$ & $\omega^3$ & 1  \\
%  &&&&&&&&&&&& \\[-0,3cm]
  $U(1)_{FN}$ & 0 & 1 & 0 & 0 & 0 & 0 & -1 & 0 & 0 & 0 & 0 & 0  \\
  \hline\hline
  \end{tabular}
\caption{\label{table:BMM_transformation}Transformation properties
of the leptonic fields and flavons in the BMM model
\cite{Bazzocchi:2009pv}. Note that $\omega$ is the fifth root of
unity, i.e. $\omega=e^{i2\pi/5}$.}
\end{center}
\end{table}

By introducing a $U(1)_R$ symmetry, the authors in Ref.
\cite{Bazzocchi:2009pv} have shown that the flavon fields develop
the following vacuum alignment at LO
\begin{eqnarray}
\nonumber\langle\psi\rangle&=&\left(
             \begin{array}{c}
               0 \\
               1 \\
               0 \\
             \end{array}
           \right)v_{\psi},~~~\langle\eta\rangle=\left(
             \begin{array}{c}
               0 \\
               1 \\
             \end{array}
           \right)v_\eta\\
\nonumber&&        \\
\nonumber\langle\Delta\rangle&=&\left(
             \begin{array}{c}
               1 \\
               1 \\
               1 \\
             \end{array}
           \right)v_\Delta,~~~
\langle\varphi\rangle=\left(
             \begin{array}{c}
               1 \\
               1 \\
             \end{array}
           \right)v_{\varphi}\\
\nonumber&&  \\
\label{11}\langle\xi'\rangle&=&v_{\xi'},~~~
\langle\theta\rangle=v_\theta
\end{eqnarray}
%with
%\begin{equation}
%v_{\Delta}^2=-\frac{g_3}{3g_2}v_{\varphi}^2\quad\quad v_\psi=-\dfrac{f_2}{2f_1}v_\eta\quad\quad v_{\xi'}=\frac{h_1}{M_{\xi'}}v_\eta v_{\varphi}\quad\quad|\theta|^2=M^2_{FI}/g_{FN}
%\end{equation}
%and $v_{\varphi}$ and $v_{\eta}$ undetermined. To produce the mass hierarchy among the charged fermions, all the VEVs (scaled by the cutoff $\Lambda$)
%should be of the same order of magnitude, and it belong to the range $0.01<VEV/\Lambda<0.05$.
The superpotential of the model in the lepton sector is
\begin{eqnarray}
\nonumber&&w_{\ell}=\sum^{4}_{i=1}\frac{\theta}{\Lambda}\frac{y_{e,i}}{\Lambda^3}\,e^c(\ell X_i)_{1_2}H_d+\frac{y_{\mu}}{\Lambda^2}\mu^c(\ell\psi\eta)_{1_2}H_d+\frac{y_{\tau}}{\Lambda}\tau^c(\ell\psi)_{1_1}H_d+...\\
\label{12}&&w_{\nu}=y(\nu^c\ell)_{1_1}H_u+x_d(\nu^c\nu^c\varphi)_{1_1}+x_t(\nu^c\nu^c\Delta)_{1_1}+...
\end{eqnarray}
where the subscript $1_1$ and $1_2$ denote the contraction in $1_1$
and $1_2$, respectively, and dots stand for higher dimensional
operators, which are suppressed by additional powers of the cutoff
$\Lambda$. The composite $X$ is given by
\begin{equation}
\label{13}X=\{\psi\psi\eta,\,\psi\eta\eta,\,\Delta\Delta\xi',\,\Delta\varphi\xi'\}
\end{equation}
Taking into account the vacuum alignment in Eq.(\ref{11}), the mass matrix for the charged lepton reads
\begin{equation}
\label{14}m_{\ell}=\frac{v_du}{\sqrt{2}}\left(\begin{array}{ccc}y^{(1)}_eu^2t&y^{(2)}_eu^2t&y^{(3)}_eu^2t\\
0&y_{\mu}u&0\\
0&0&y_{\tau}
\end{array}\right)
\end{equation}
where $y^{(i)}_e$ is the linear combination of the $y_{e,i}$
contributions. The parameter $u$ parameterizes the ratio
$v_{\psi}/\Lambda$, $v_{\eta}/\Lambda$,
$v_{\Delta}/\Lambda$,$v_{\varphi}/\Lambda$ and $v_{\xi'}/\Lambda$,
which should be of the same order of magnitude to produce the mass
hierarchy among the charged fermions. The parameter $t$ denotes the
ratio $v_{\theta}/\Lambda$. It has been shown that the parameters
$u$ and $t$ belong to the range $0.01<u,t<0.05$
\cite{Bazzocchi:2009pv}. The first term in $w_{\nu}$ is the neutrino
Dirac-Yukawa coupling, and the last two terms determine the mass
matrix of the heavy right-handed neutrinos. Straightforwardly we
have
\begin{equation}
\label{15}m^D_{\nu}=\frac{1}{\sqrt{2}}\left(\begin{array}{ccc}1&0&0\\
0&0&1\\
0&1&0\\
\end{array}\right)yv_{u},~~~~~M_{N}=\left(\begin{array}{ccc}2c&b-c&b-c\\
b-c&b+2c&-c\\
b-c&-c&b+2c
\end{array}\right)
\end{equation}
where $b=2x_dv_{\varphi}$ and $x_t=2x_tv_{\Delta}$. Integrating out
the heavy neutrino $\nu^c$, the light-neutrino effective mass matrix
is given by the see-saw formula
\begin{eqnarray}
\nonumber&&m_{\nu}=-(m^{D}_{\nu})^{T}M^{-1}_Nm^{D}_{\nu}=\frac{y^2v^2_u}{4}\left(\begin{array}{ccc}
\frac{b+c}{b(b-3c)}&\frac{-b+c}{b(b-3c)}&\frac{-b+c}{b(b-3c)}\\
\frac{-b+c}{b(b-3c)}&\frac{-b^2+4bc+3c^2}{b(b^2-9c^2)}&\frac{b^2-2bc+3c^2}{b(b^2-9c^2)}\\
\frac{-b+c}{b(b-3c)}&\frac{b^2-2bc+3c^2}{b(b^2-9c^2)}&\frac{-b^2+4bc+3c^2}{b(b^2-9c^2)}
\end{array}\right)
\end{eqnarray}
The light-neutrino mass matrix $m_{\nu}$ can be diagonalized by
\begin{equation}
\label{16}U^{T}_{\nu}m_{\nu}U_{\nu}={\rm diag}(m_{\nu_1},
m_{\nu_2},m_{\nu_3})
\end{equation}
where $m_{\nu1,\nu2,\nu3}$ are the light-neutrino masses
\begin{eqnarray}
\nonumber&&m_{\nu_1}=\frac{y^2v^2_u}{2}\frac{1}{|-b+3c|}\\
\nonumber&&m_{\nu_2}=\frac{y^2v^2_u}{2}\frac{1}{2|b|}\\
\label{17}&&m_{\nu_3}=\frac{y^2v^2_u}{2}\frac{1}{|b+3c|}
\end{eqnarray}
The unitary matrix $U_{\nu}$ can be written as
\begin{equation}
\label{18}U_{\nu}=iU_{TB}U_P
\end{equation}
where $U_{TB}$ is the TB mixing matrix
\begin{equation}
\label{19}U_{TB}=\left(\begin{array}{ccc}\sqrt{2/3}&1/\sqrt{3}&0\\
-1/\sqrt{6}&1/\sqrt{3}&1/\sqrt{2}\\
-1/\sqrt{6}&1/\sqrt{3}&-1/\sqrt{2}
\end{array}\right)
\end{equation}
and $U_P$=diag($e^{i\alpha_1/2}$, $e^{i\alpha_2/2}$,
$e^{i\alpha_3/2}$) is a matrix of phase with $\alpha_1={\rm
arg}(-b+3c)$, $\alpha_2={\rm arg}(b)$ and $\alpha_3={\rm
arg}(b+3c)$. Therefore the lepton mixing matrix is the TB mixing
apart from the negligible corrections of order $u^2t^2$ from the
charged lepton sector. We note that the right-handed neutrino mass
matrix $M_N$ is diagonalized by TB mixing as well, the mass
eigenvalues are given by $M_1=|-b+3c|$, $M_2=2|b|$ and $M_3=|b+3c|$.
Comparing with the light-neutrino masses in Eq.(\ref{15}), we have
the interesting relation
\begin{equation}
\label{add_EPJC_1}m_{\nu_i}=\frac{y^2v^2_{u}}{2M_i}
\end{equation}
The Yukawa coupling $y$ is of ${\cal O}(1)$, and we use $|\Delta
m^2_{atm}|^{1/2}$ as the typical light-neutrino mass scale, then we
obtain
\begin{equation}
\label{add_EPJC_2}M_i\sim 10^{14\div15}{\rm GeV}
\end{equation}
The coefficients $x_t$ and $x_d$ are expected to be of ${\cal
O}(1)$, as a consequence the VEVs $v_{\varphi}$ and $v_{\Delta}$
should be of the same order as the right-handed neutrino mass $M_i$.
It is obvious that the model is rather constrained, there are only
three independent parameters, which can be chosen to be
$|b|=y^2v^2_u/(4m_{\nu2})$, $Z$ and $\Omega$. The latter two are
defined according to
\begin{equation}
\label{20}\frac{c}{b}=Ze^{i\Omega}
\end{equation}
We can easily express $Z$ and $\Omega$ in terms of the neutrino masses as
\begin{eqnarray}
\nonumber&&Z=\frac{1}{3}\sqrt{\frac{2m^2_{\nu2}}{m^2_{\nu1}}+\frac{2m^2_{\nu2}}{m^2_{\nu3}}-1}\\
\label{21}&&\cos\Omega=\frac{\frac{m^2_{\nu2}}{m^2_{\nu3}}-\frac{m^2_{\nu2}}{m^2_{\nu1}}}{\sqrt{\frac{2m^2_{\nu2}}{m^2_{\nu1}}+\frac{2m^2_{\nu2}}{m^2_{\nu3}}-1}}
\end{eqnarray}
The above relations hold for both normal hierarchy (NH) and inverted
hierarchy (IH) spectrum. Taking into account the experimentally
measured mass difference $\Delta m^2_{\rm
sol}=m^2_{\nu2}-m^2_{\nu1}$ and $\Delta m^2_{\rm
atm}=|m^2_{\nu3}-m^2_{\nu1}(m^2_{\nu2})|$, we have only one free
parameter left, which is conveniently chosen to be the lightest
neutrino mass. Imposing the constraint $|\cos\Omega|\leq1$, we
obtain the following limits for the lightest neutrino mass
\footnote{The same parameter space is obtained by the neutrino mass
sum rule method \cite{Barry:2010yk}.}
\begin{eqnarray}
\nonumber&&m_{\nu1}\geq0.011 {\,\rm eV},~~~~~~~~~{\rm NH}\\
\label{23}&&m_{\nu3}\geq0.029{\,\rm eV},~~~~~~~~~{\rm IH}
\end{eqnarray}
where the central values of $\Delta m^2_{\rm sol}$ and $\Delta
m^2_{\rm atm}$ are used. We would like to stress that the mass
squared differences are running quantities, therefore the bounds in
Eq.(\ref{23}) would change somewhat at low energy after considering
the RG effects.

The model is so predictive that we can express the Majorana
phases in terms of the lightest neutrino mass as well. In the
standard parametrization \cite{pdg}, the lepton PMNS mixing matrix
is defined by
\begin{eqnarray}
\nonumber&&U_{PMNS}={\rm diag}(e^{i\delta_{e}},e^{i\delta_{\mu}},e^{i\delta_{\tau}})\left(\begin{array}{ccc}c_{12}c_{13}&s_{12}c_{13}&s_{13}e^{-i\delta}\\
-s_{12}c_{23}-c_{12}s_{23}s_{13}e^{i\delta}&c_{12}c_{23}-s_{12}s_{23}s_{13}e^{i\delta}&s_{23}c_{13}\\
s_{12}s_{23}-c_{12}c_{23}s_{13}e^{i\delta}&-c_{12}s_{23}-s_{12}c_{23}s_{13}e^{i\delta}&c_{23}c_{13}\end{array}\right)\\
\label{add1}&&~~~~~\times{\rm diag}(e^{-i\varphi_{1}/2},e^{-i\varphi_{2}/2},1)
\end{eqnarray}
where $c_{ij}=\cos\theta_{ij}$, $s_{ij}=\sin\theta_{ij}$ with
$\theta_{ij}\in[0,\pi/2]$, the unphysical phases $\delta_e$,
$\delta_{\mu}$ and $\delta_{\tau}$ can be absorbed into charged
lepton fields, $\delta$ is the Dirac CP violating phase,
$\alpha_{21}$ and $\alpha_{31}$ are the two Majorana CP violating
phases, all the three CP violating phases $\delta$, $\varphi_{1}$
and $\varphi_{2}$ are allowed to vary in the range of $0\sim2\pi$.
Recalling that the leptonic mixing matrix is given by Eq.(\ref{18}),
we can identify the two CP violating phases as
\begin{equation}
\label{add2}\varphi_1=\alpha_3-\alpha_1,~~~~~~~\varphi_2=\alpha_3-\alpha_2
\end{equation}
As a result, we have
\begin{eqnarray}
\nonumber&&\cos\varphi_1=\frac{-1+9Z^2}{\sqrt{1+81Z^4-18Z^2\cos2\Omega}}\,,~~~~~\sin\varphi_1=\frac{-6Z\sin\Omega}{\sqrt{1+81Z^4-18Z^2\cos2\Omega}}\\
\label{add3}&&\cos\varphi_2=\frac{1+3Z\cos\Omega}{\sqrt{1+9Z^2+6Z\cos\Omega}}\,,~~~~~~~~\sin\varphi_2=\frac{3Z\sin\Omega}{\sqrt{1+9Z^2+6Z\cos\Omega}}
\end{eqnarray}
Since we can only determine $\cos\Omega$ from the neutrino mass
spectrum, the Majorana phases $\varphi_1$, $\varphi_2$ can take two
sets of values corresponding to $\sin\Omega>0$ and $\sin\Omega<0$
respectively. We note that the Dirac CP phase is undetermined
because the reactor angle is vanishing in TB mixing. The above
successful leading order results are corrected by the NLO
contributions, which consists of the higher dimensional operators in
both the driving superpotential and Yukawa superpotentials. It has
been shown that all the three leptonic mixing angles receive
corrections of order $u$ \cite{Bazzocchi:2009pv}.

\subsection{The $S_4$ model of Ding}
The total flavor symmetry of this model is $S_4\times Z_3\times Z_4$
\cite{Ding:2009iy}. It is remarkable that the realistic pattern of
fermion masses and flavor mixing in both the lepton and quark sector
have been reproduced in this model, and the mass hierarchies are
determined by the spontaneous breaking of the flavor symmetry
without invoking a Froggatt-Nielsen $U(1)$ symmetry. The leptonic
fields and the flavons of the model and their classifications under
the flavor symmetry are shown in Table \ref{tab:S4trans}, where the
quark fields have been omitted.

\begin{table}[hptb]
\begin{center}
\begin{tabular}{|c|c|c|c|c|c|c|c|c|c|c|c|c|}\hline\hline
& $\ell$  & $e^{c}$ & $\mu^{c}$ &  $\tau^{c}$ &$\nu^c$ &
$H_{u,d}$&$\varphi$ & $\chi$ & $\zeta$ & $\eta$ & $\phi$ & $\Delta$
\\\hline

$\rm{S_4}$& $3_1$& $1_1$ & $1_2$&$1_1$ &  $3_1$  &  $1_1$&$3_1$ &
$3_2$& $1_2$ & 2   & $3_1$   & $1_2$  \\\hline

$\rm{Z_{3}}$& $\omega$ & $\omega^2$& $\omega^2$&  $\omega^2$  & $1$
&1 & 1 &1 &1 &  $\omega^2$  & $\omega^2$   & $\omega^2$
\\\hline

$\rm{Z_{4}}$& 1 &i & -1&  -i  & 1  & 1& i &i & 1&  1  & 1  & -1
\\\hline\hline
\end{tabular}
\end{center}
\caption{\label{tab:S4trans}The transformation rules of the leptonic
fields and the flavons under the symmetry groups $S_4$, $Z_3$ and
$Z_4$ in the $S_4$ model of Ref. \cite{Ding:2009iy}, where $\omega$ is the third root of unity, i.e.
$\omega=e^{i\frac{2\pi}{3}}=(-1+i\sqrt{3})/2$.}
\end{table}

In this model the $S_4$ symmetry is broken down to Klein four and
$Z_3$ subgroups in the neutrino and charged lepton sector,
respectively, at LO, this specific breaking scheme require flavon
fields develop the following vacuum configuration
\begin{eqnarray}
\nonumber&&\langle\varphi\rangle=(0,V_{\varphi},0),~~~\langle\chi\rangle=(0,V_{\chi},0)~~~\langle\zeta\rangle=V_{\zeta}\\
\label{24}&&\langle\eta\rangle=(V_{\eta},V_{\eta}),~~~\langle\phi\rangle=(V_{\phi},V_{\phi},V_{\phi}),~~~\langle\Delta\rangle=V_{\Delta}
\end{eqnarray}
We have demonstrated that this particular vacuum alignment is a
natural solution to the scalar potential, all the VEVs (scaled by
the cutoff $\Lambda$) $V_{\varphi}/\Lambda$, $V_{\chi}/\Lambda$,
$V_{\zeta}/\Lambda$, $V_{\eta}/\Lambda$, $V_{\phi}/\Lambda$ and
$V_{\Delta}/\Lambda$ are of the same order of magnitude about ${\cal
O}(\lambda^2_c)$, and this vacuum configuration is stable under the
higher order corrections, please see Ref. \cite{Ding:2009iy} for
detail. Then the most general superpotential in the lepton sector,
which is compatible with the representation assignment of Table
\ref{tab:S4trans}, is given by
\begin{eqnarray}
\nonumber&&w_{\ell}=\frac{y_{e_1}}{\Lambda^3}\;e^{c}(\ell\varphi)_{1_1}(\varphi\varphi)_{1_1}h_d+\frac{y_{e_2}}{\Lambda^3}\;e^{c}((\ell\varphi)_2(\varphi\varphi)_2)_{1_1}h_d+\frac{y_{e_3}}{\Lambda^3}\;e^{c}((\ell\varphi)_{3_1}(\varphi\varphi)_{3_1})_{1_1}h_d\\
\nonumber&&~~+\frac{y_{e_4}}{\Lambda^3}\;e^{c}((\ell\chi)_2(\chi\chi)_2)_{1_1}h_d+\frac{y_{e_5}}{\Lambda^3}\;e^{c}((\ell\chi)_{3_1}(\chi\chi)_{3_1})_{1_1}h_d+\frac{y_{e_6}}{\Lambda^3}\;e^{c}(\ell\varphi)_{1_1}(\chi\chi)_{1_1}h_d\\
\nonumber&&~~+\frac{y_{e_7}}{\Lambda^3}\;e^{c}((\ell\varphi)_2(\chi\chi)_2)_{1_1}h_d+\frac{y_{e_8}}{\Lambda^3}\;e^{c}((\ell\varphi)_{3_1}(\chi\chi)_{3_1})_{1_1}h_d+\frac{y_{e_9}}{\Lambda^3}\;e^{c}((\ell\chi)_2(\varphi\varphi)_2)_{1_1}h_d\\
\nonumber&&~~+\frac{y_{e_{10}}}{\Lambda^3}\;e^{c}((\ell\chi)_{3_1}(\varphi\varphi)_{3_1})_{1_1}h_d+\frac{y_{\mu}}{2\Lambda^2}\mu^{c}(\ell(\varphi\chi)_{3_2})_{1_2}h_d+\frac{y_{\tau}}{\Lambda}\tau^{c}(\ell\varphi)_{1_1}h_d+...\\
\label{25}&&w_{\nu}=\frac{y_{\nu_1}}{\Lambda}((\nu^{c}\ell)_2\eta)_{1_1}h_u+\frac{y_{\nu_2}}{\Lambda}((\nu^{c}\ell)_{3_1}\phi)_{1_1}h_u+\frac{1}{2}M(\nu^c\nu^c)_{1_1}+...
\end{eqnarray}
where $(...)_{1_1,1_2, 2, 3_1,3_2}$ stands for the $1_1$, $1_2$,
$2$, $3_1$ and $3_2$ products, respectively. We note that and one can
always set $M$ to be real and positive by global phase
transformation of the lepton fields, and a priori $M$ should be of
the same order as the cutoff scale $\Lambda$. Taking into account
the vacuum alignment in Eq.(\ref{24}), we find that the charged
lepton mass matrix is diagonal at LO,
\begin{equation}
\label{26}m_{\ell}=\frac{v_d}{\sqrt{2}}\left(\begin{array}{ccc}
y_e\frac{v^3_{\varphi}}{\Lambda^3}&0&0\\
0&y_{\mu}\frac{v_{\varphi}v_{\chi}}{\Lambda^2}&0\\
0&0&y_{\tau}\frac{v_{\varphi}}{\Lambda}
\end{array}\right)
\end{equation}
where $y_e$ is the result of all the different contributions of
$y_{e_i}$. The neutrino Dirac and Majorana mass matrices can be
straightforwardly read out as
\begin{equation}
\label{27}m^{D}_{\nu}=\frac{v_{u}}{\sqrt{2}}\left(\begin{array}{ccc}2b&a-b&a-b\\
a-b&a+2b&-b\\
a-b&-b&a+2b\end{array}\right),~~~~~~~M_N=\left(\begin{array}{ccc}
M&0&0\\
0&0&M\\
0&M&0\end{array}\right)
\end{equation}
where $a=y_{\nu_1}\frac{v_{\eta}}{\Lambda}$ and
$b=y_{\nu_2}\frac{v_{\phi}}{\Lambda}$. As a result, the light-neutrino mass matrix is given by
\begin{equation}
\label{28}m_{\nu}=-(m^{D}_{\nu})^{T}M^{-1}_{N}m^{D}_{\nu}=-\frac{v^2_u}{2M}\left(\begin{array}{ccc}2a^2-4ab+6b^2&a^2+2ab-3b^2&a^2+2ab-3b^2\\
a^2+2ab-3b^2&a^2-4ab-3b^2&2a^2+2ab+6b^2\\
a^2+2ab-3b^2&2a^2+2ab+6b^2&a^2-4ab-3b^2
\end{array}\right)
\end{equation}
We can see that the mass matrix $m_{\nu}$ is exactly diagonalized by the TB mixing matrix
\begin{equation}
\label{29}U^{T}_{\nu}m_{\nu}U_{\nu}={\rm diag}(m_{\nu1},m_{\nu2},m_{\nu3})
\end{equation}
The unitary matrix $U_{\nu}$ is
\begin{equation}
\label{30}U_{\nu}=U_{TB}\,{\rm
diag}(e^{-i\alpha_1/2},e^{-i\alpha_2/2},e^{-i\alpha_3/2})
\end{equation}
The phases $\alpha_{1}$, $\alpha_2$ and $\alpha_3$ are closely related to the Majorana phase
\begin{equation}
\label{31}\alpha_1={\rm arg}(-(a-3b)^2/M),~~~~\alpha_2={\rm arg}(-4a^2/M),~~~~\alpha_3={\rm arg}((a+3b)^2/M)
\end{equation}
and the neutrino masses are given by
\begin{equation}
\label{32}m_{\nu1}=|(a-3b)^2|v^2_u/(2M),~~~m_{\nu2}=2|a|^2v^2_u/M,~~~m_{\nu3}=|(a+3b)^2|v^2_u/(2M)
\end{equation}
It is interesting to estimate the order of magnitude for the right-handed neutrino mass $M$. Since the parameters $a$ and $b$ are
expected to be of order $\lambda^2_c$, with this and using
$\sqrt{|\Delta m^2_{atm}|}\simeq0.05$ eV as the light-neutrino mass
scale in the see-saw formula, we obtain
\begin{equation}
\label{add_EPJC_3}M\sim 10^{12/13}{\rm GeV}
\end{equation}
Similar to the analysis of section \ref{sec:BMM}, we define
\begin{equation}
\label{33}\frac{b}{a}=R\,e^{i\Phi}
\end{equation}
Straightforwardly we can express $R$ and $\cos\Phi$ as functions of the neutrino masses
\begin{eqnarray}
\nonumber&&R=\frac{1}{3}\sqrt{\frac{2m_{\nu1}}{m_{\nu2}}+\frac{2m_{\nu3}}{m_{\nu2}}-1}\\
\label{34}&&\cos\Phi=\frac{\frac{m_{\nu3}}{m_{\nu2}}-\frac{m_{\nu1}}{m_{\nu2}}}{\sqrt{\frac{2m_{\nu1}}{m_{\nu2}}+\frac{2m_{\nu3}}{m_{\nu2}}-1}}
\end{eqnarray}
In exactly the same way section \ref{sec:BMM}, the Majorana phases in standard parameterization are determined as
\begin{equation}
\label{add4} \varphi_1=\alpha_1-\alpha_3,~~~~~\varphi_2=\alpha_2-\alpha_3
\end{equation}
with
\begin{eqnarray}
\nonumber&& \cos\varphi_1=\frac{-(1-9R^2)^2+36R^2\sin^2\Phi}{(1+9R^2)^2-36R^2\cos^2\Phi},~~~~~\sin\varphi_1=\frac{12R(1-9R^2)\sin\Phi}{(1+9R^2)^2-36R^2\cos^2\Phi}\\
\label{add5}&&\cos\varphi_2=\frac{-1-9R^2\cos2\Phi-6R\cos\Phi}{1+9R^2+6R\cos\Phi},~~~~\sin\varphi_2=\frac{6R(1+3R\cos\Phi)\sin\Phi}{1+9R^2+6R\cos\Phi}
\end{eqnarray}
Consequently, all the low energy parameters in the neutrino sector
can be expressed in terms of the lightest neutrino mass. Imposing
the condition $|\cos\Phi|\leq1$, we get the following constraint on
the lightest neutrino mass:
\begin{eqnarray}
\nonumber&&m_{\nu1}\geq0.011\;{\rm eV},~~~~~{\rm NH}\\
\label{35}&&m_{\nu3}>0.0\;{\rm eV},~~~~~~~~~~{\rm IH}
\end{eqnarray}
The NLO corrections have been analyzed in detail in Ref.
\cite{Ding:2009iy}. It is shown that both the neutrino masses and
mixing angles receive corrections of order
$\varepsilon\sim\lambda^2_c$ with respect to leading order result,
where $\varepsilon$ parameterizes the ratio $VEV/\Lambda$ and
$\lambda_c$ is the Cabibbo angle.

\section{\label{sec:RGE_running}RG running effects in $S_4$ flavor models}
As has been shown, the TB mixing is achieved in both BMM model and
the $S_4$ model of Ding at LO. In this section, we turn to a
quantitative discussion of RG effects, and compare them with the NLO
corrections and the experimental data. For definiteness we shall
assume a supersymmetry breaking scale of 1 TeV, below which the SM
is valid. We note that the mass hierarchy between top and bottom is
produced via the spontaneous breaking of flavor symmetry in both
models, and $\tan\beta$ should be small. As a result, we shall take
the parameter $\tan\beta$ to be 10 apart except where explicitly indicated
otherwise. To study the running of the neutrino mixing parameters
from the GUT scale to the electroweak scale, the Mathematica package
REAP is used \cite{Antusch:2005gp}. This package numerically solves
the RG equations of the quantities relevant for neutrino mass and
mixing, and it has been widely used for different purposes
\cite{Boudjemaa:2008jf}. The package can be downloaded from
http://users.physik.tu-muenchen.de/rge/REAP/index.html, and
Mathematica version 5 or higher is required. We note that the
approximate analytical solutions based on leading log approximation
to the RG equations have been derived in
Refs.\cite{Antusch:2005gp,Antusch:2003kp}, which allows one to
understand the generic behavior of the renormalization effects.
However, due to enhancement/suppression factors and possible
cancelations, the exact numerical solutions may differ considerably
from those estimates. Therefore throughout this paper we adopt a
numerical approach, exploiting the convenient REAP package.

As has been demonstrated above, we generally need to introduce
flavon fields to break the flavor symmetry in order to generate
fermion masses and flavor mixing. In the unbroken phase of flavor
symmetry, the flavons are active fields, therefore the corresponding
RG equations should be modified in principle. However, the
superpotentials of the models in Eq.(\ref{12}) and Eq.(\ref{25})
contain all the possible LO terms allowed by the symmetries, the
invariance under the flavor symmetry $S_4$ is maintained until we
move down to the scale of the VEV of the flavon fields, which is of
the order of GUT scale. We conclude that the flavor structures of
the models are preserved above the scale of the VEV of the flavon
fields, the contributions of the flavon fields in the RG running can
be absorbed by the redefinition of the model parameters
\cite{Lin:2009sq}. In the following, we will discuss the RG
evolution of neutrino masses and mixing parameters in both $S_4$
flavor models, stating from the initial conditions of neutrino Dirac
and Majorana mass matrices described in section \ref{sec:models} at
the GUT scale. In particularly, the parameter spaces are scanned.

\subsection{\label{sec:RG_BMM}RG effects in the BMM models}

\begin{figure}[hptb]
\begin{center}
\begin{tabular}{rr}
\includegraphics[scale=.33]{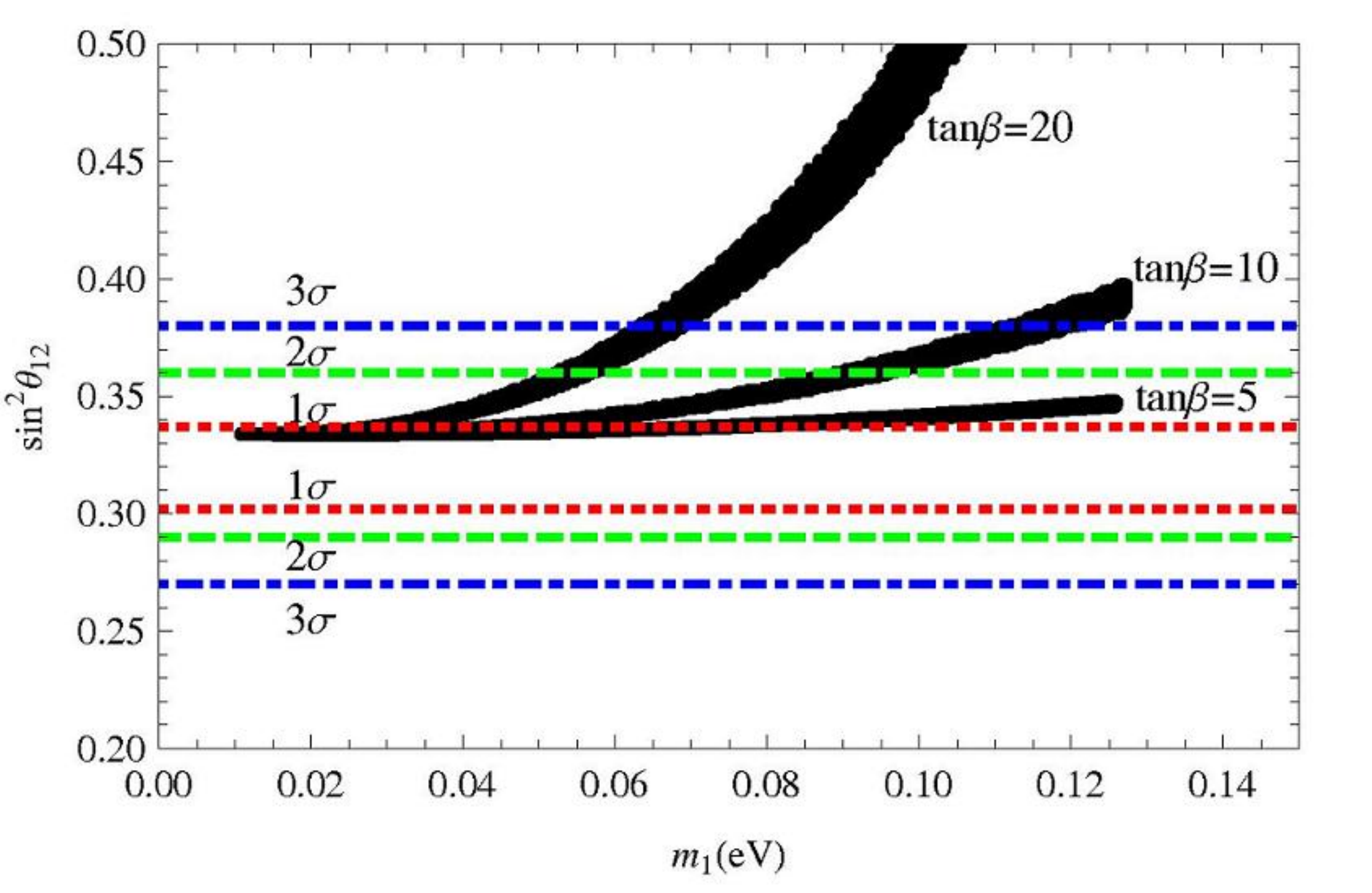}&\hspace{0.5cm}\includegraphics[scale=.33]{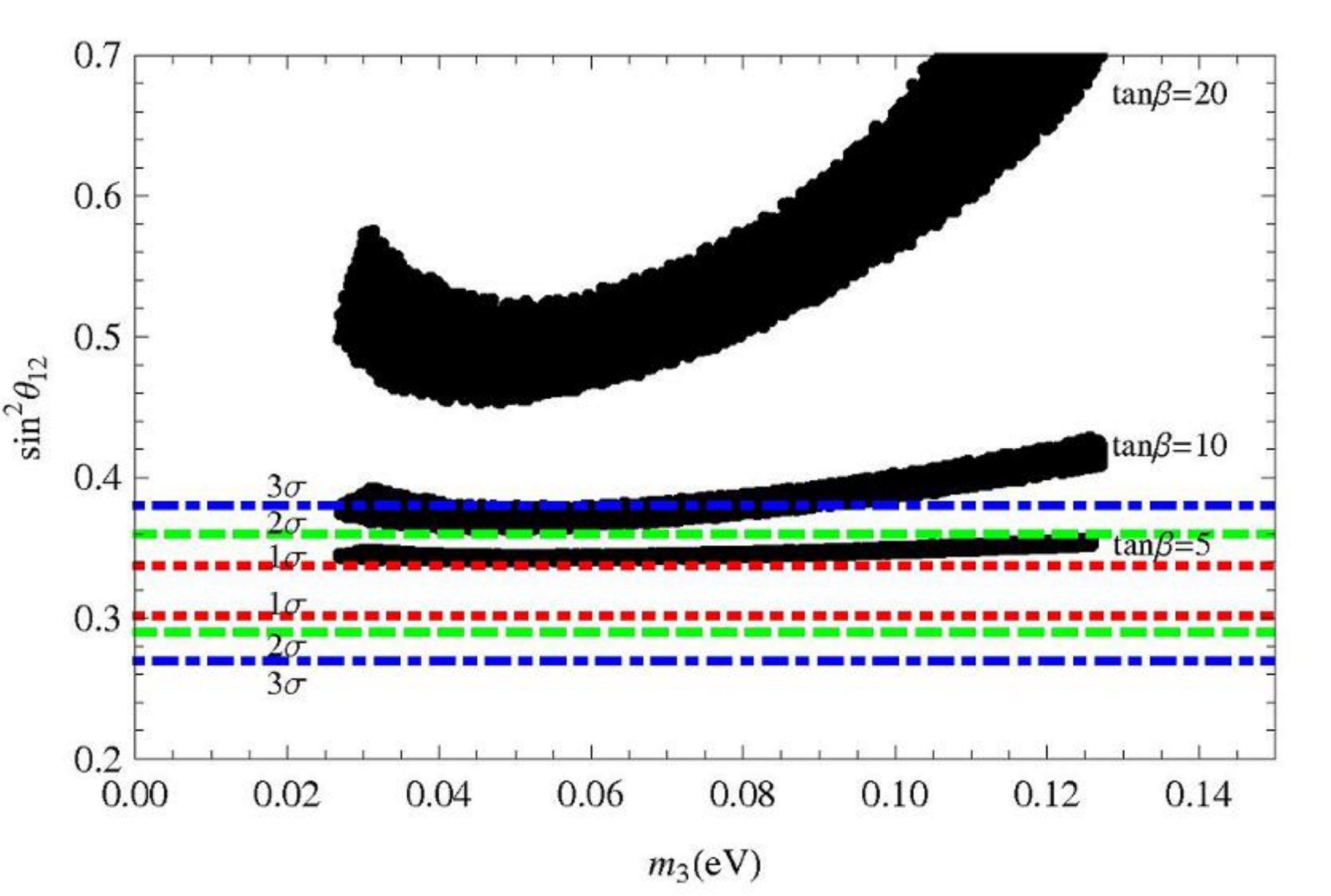}\\
\includegraphics[scale=.33]{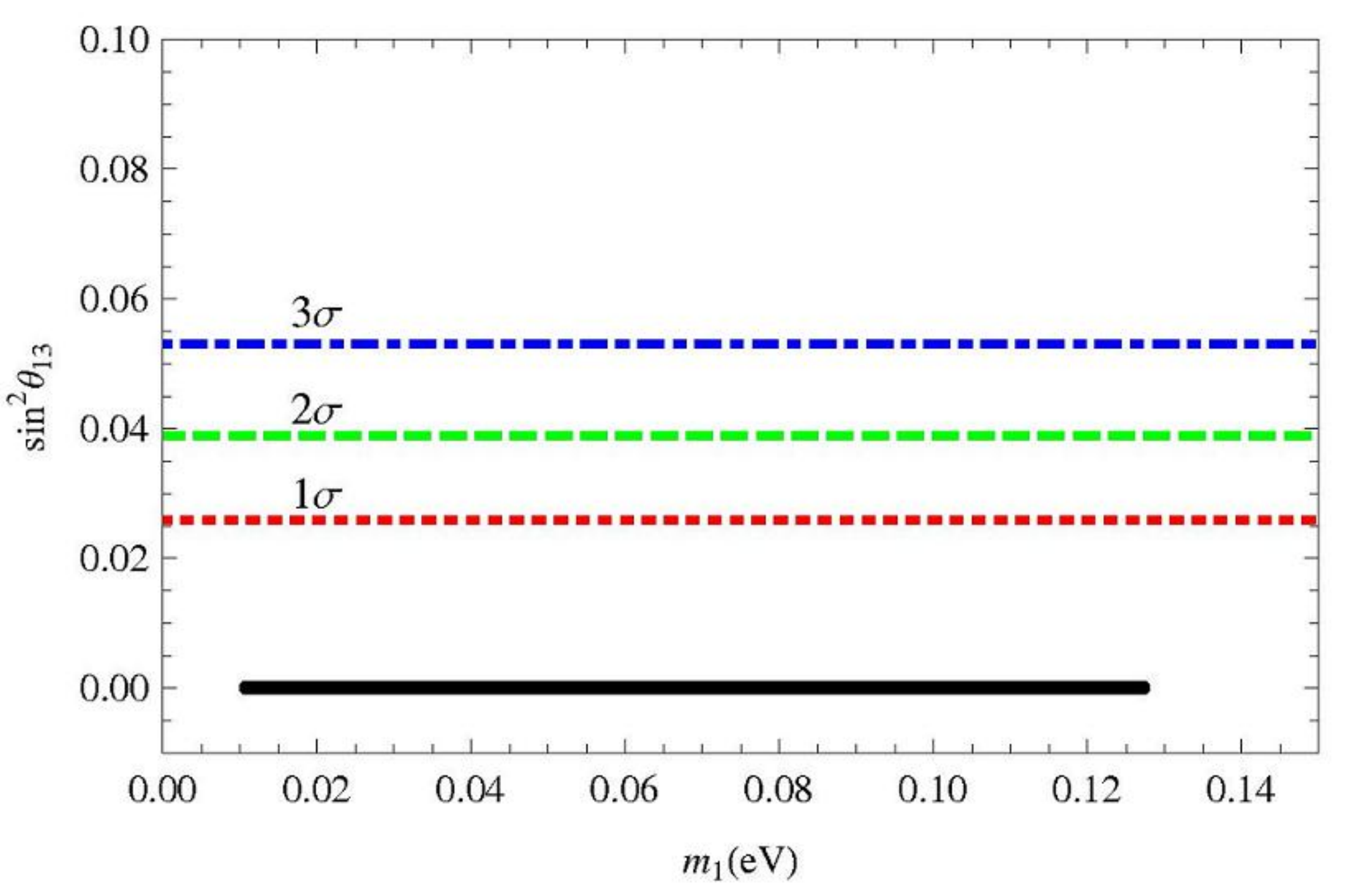}&\hspace{0.5cm}\includegraphics[scale=.33]{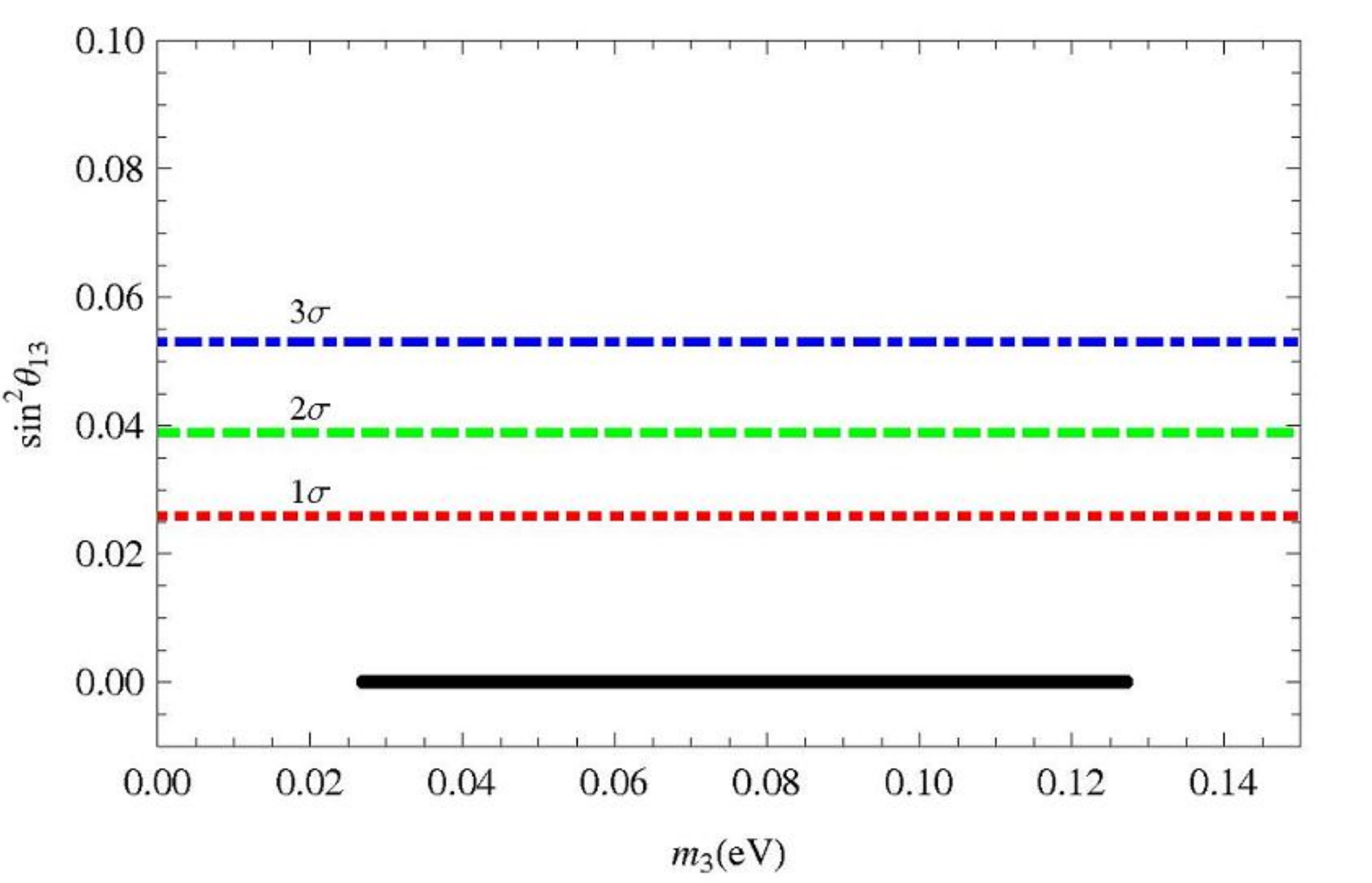}\\
\includegraphics[scale=.33]{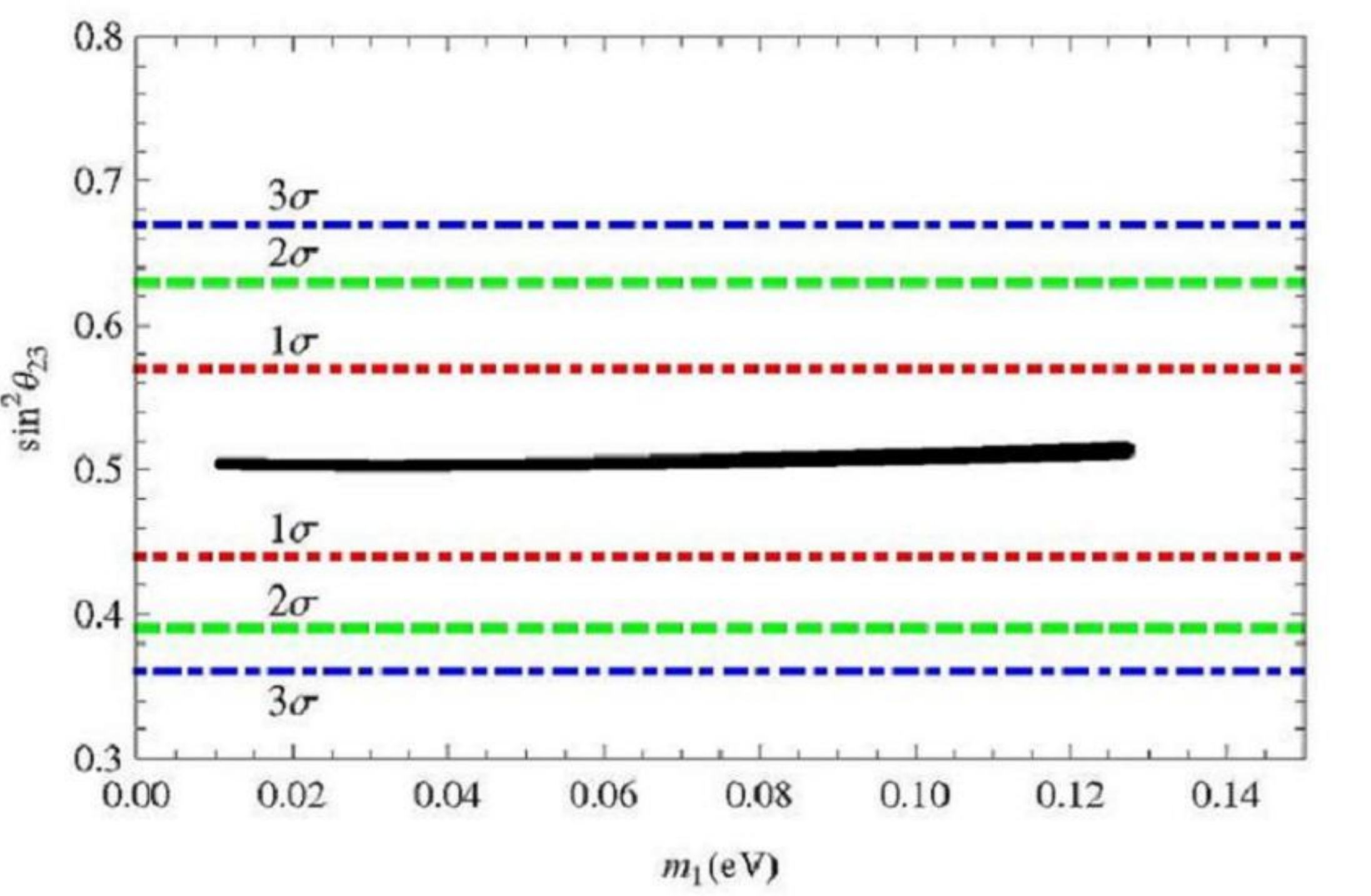}&\hspace{0.5cm}\includegraphics[scale=.33]{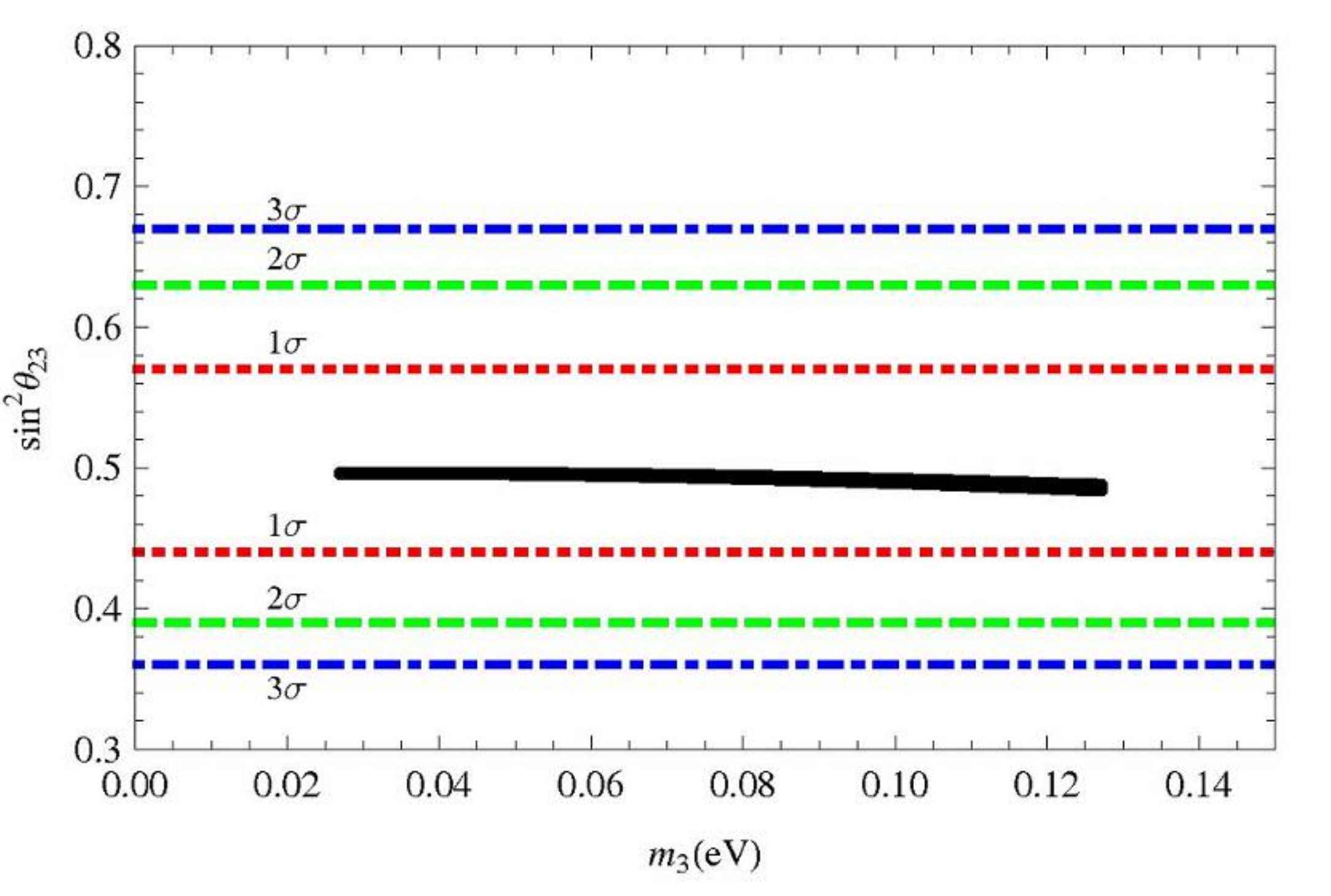}
\end{tabular}
\caption{\label{fig:RGE_BMM_mass1}RGE corrections to the neutrino
mixing angles in the BMM $S_4$ model with $\tan\beta=10$. The left
column of the plots are the results for NH spectrum, and the right
column for the IH case. In the case of $\sin^2\theta_{12}$,
$\tan\beta=5$ and $\tan\beta=20$ are considered.}
\end{center}
\end{figure}

In this section we report results of the calculations of the RG
evolution of the neutrino mixing parameters in the BMM model.
Without loss of generality, we choose the Yukawa coupling $y=1$ for
our numerical analysis. The GUT scale neutrino mass squared
differences $m^2_{\nu2}-m^2_{\nu1}$ and $|m^2_{\nu3}-m^2_{\nu1}|$
are treated as random numbers in the range of $3.5\times10^{-5}{~\rm
eV^2}\sim2.5\times10^{-4}{~\rm eV^2}$ and $1.0\times10^{-3}{~\rm
eV^2}\sim8.3\times10^{-3}{~\rm eV^2}$ respectively, \footnote{We
shall show later that the mass squared difference at the GUT scale
is about a factor of $1.2\sim3$ larger than its low energy value in
the whole spectrum.}, and the lightest neutrino mass is varied from
the lowest bound determined by Eq.(\ref{21}) or Eq.(\ref{34}) to 0.2
eV which is the future sensitivity of KATRIN experiment
\cite{katrin}. The RG corrected neutrino mixing angles as functions
of the lightest neutrino mass are shown in Fig.
\ref{fig:RGE_BMM_mass1} for both NH and IH spectrum \footnote{The
results are independent of the sign of $\sin\Omega$, the reason is
explained later.}. These plots display only the points corresponding
to choices of the parameters reproducing $\Delta m^2_{\rm atm}$,
$\Delta m^2_{\rm sol}$ and the mixing angles within the $3\sigma$
interval.

We see that the lightest neutrino mass is still bounded from below,
and the concrete values of the lower bounds are about 0.0107 eV and
0.027 eV, respectively, for the NH and IH spectrum, and these values
are found to be almost independent of $\tan\beta$. It is remarkable
that all the mixing parameters and $J_{CP}$ are predicted to lie in
a relative narrow range. For both NH and IH spectrum, it is obvious
that the RG changes of the atmospheric and reactor angles are very
small, the corresponding allowed regions lie within the current
$1\sigma$ bounds. In particular, the RG corrections to
$\sin^2\theta_{23}$ and $\sin^2\theta_{13}$ are of the same order or
even smaller than the NLO contributions. On the other hand, the
running of the solar neutrino mixing angle displays a different
pattern. The RG change of $\sin^2\theta_{12}$ is much larger than
those of $\sin^2\theta_{23}$ and $\sin^2\theta_{13}$, which is a
general property of the RG evolution
\cite{Antusch:2005gp,Antusch:2003kp}, consequently the deviation
from its TB value can be large. In the case of NH spectrum and large
$\tan\beta$, $\sin^2\theta_{12}$ is within the $3\sigma$ limit only
for smaller values of neutrino mass. Taking into account the lower
bound on the lightest neutrino mass, $m_{\nu1}$ is constrained to
lie in certain region, which decreases with $\tan\beta$. This point
can be clearly seen from Fig.\ref{fig:RGE_BMM_mass1}. For IH
spectrum, the RG effect of $\theta_{12}$ is even larger due to the
nearly degeneracy of $m_{\nu1}$ and $m_{\nu2}$. For example, for
$\tan\beta=10$, $\sin^2\theta_{12}$ is very close or above the
$3\sigma$ upper bound in the allowed region of $m_{\nu3}$, and the
values of $\sin^2\theta_{12}$ goes completely beyond the $3\sigma$
limit for larger $\tan\beta$. As a result, the IH spectrum is
strongly disfavored for $\tan\beta>10$ in the BMM model. We note
that possible large deviation of solar neutrino mixing angle from
the TB value, and small change of atmospheric and reactor angles
under RG running are predicted as well in the Altarelli-Feruglio
$A_4$ model \cite{Lin:2009sq}. In Ref.\cite{Lin:2009sq}, the authors
perform a general analysis of running effects on lepton mixing
parameters in flavor models with type I see-saw, they show that, for
the mass-independent mixing pattern, the running contribution from
the neutrino Yukawa coupling $Y_{\nu}$ can be absorbed by a small
shift on neutrino mass eigenvalues leaving mixing angles unchanged,
consequently the RG change of mixing angle is due to the
contribution coming from the charged lepton sector. This is exactly
the reason why similar results to the $A_4$ model are obtained here.

\begin{figure}[hptb]
\begin{center}
\begin{tabular}{rr}
\includegraphics[scale=.33]{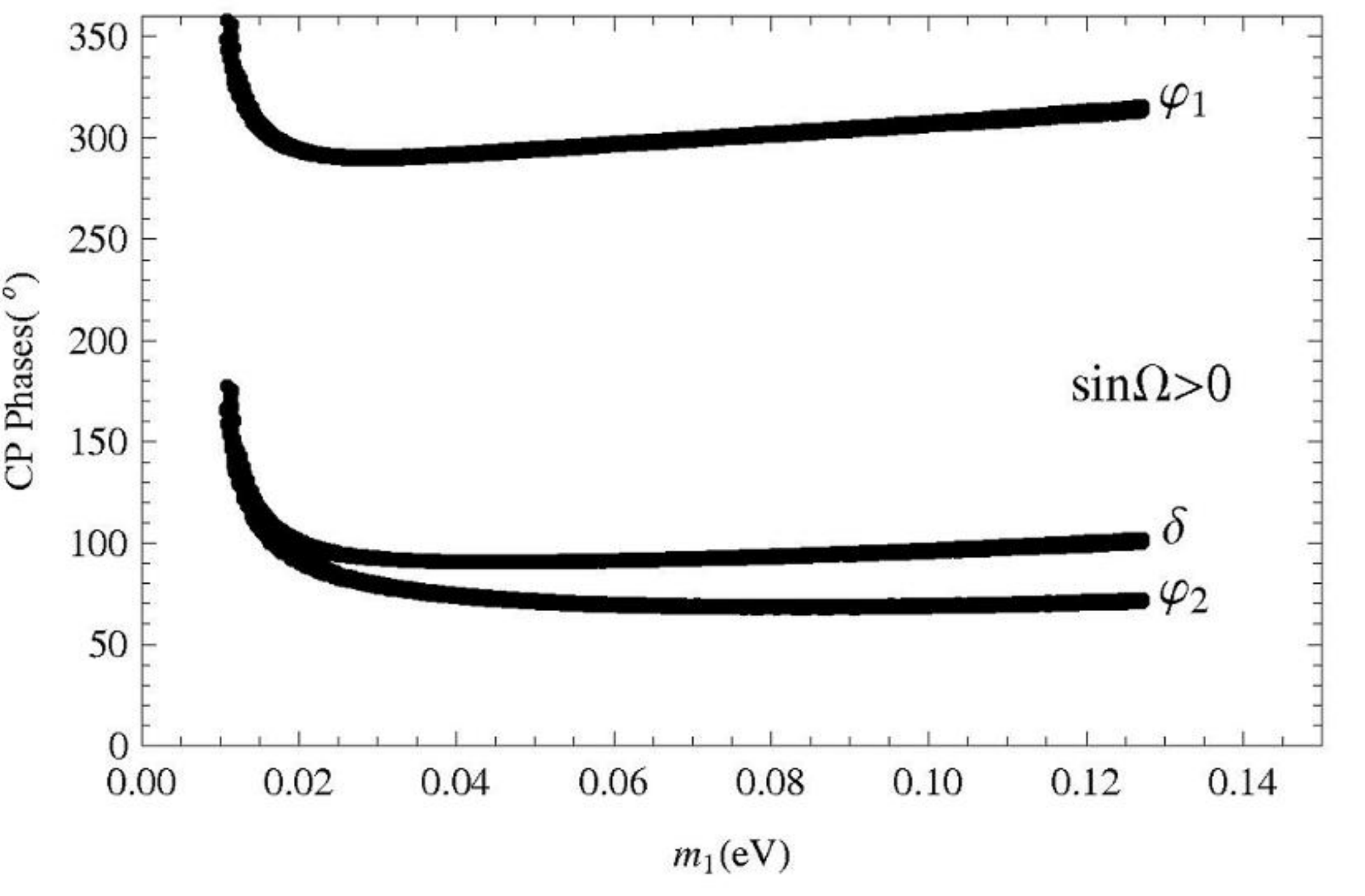}&\hspace{0.5cm}\includegraphics[scale=.33]{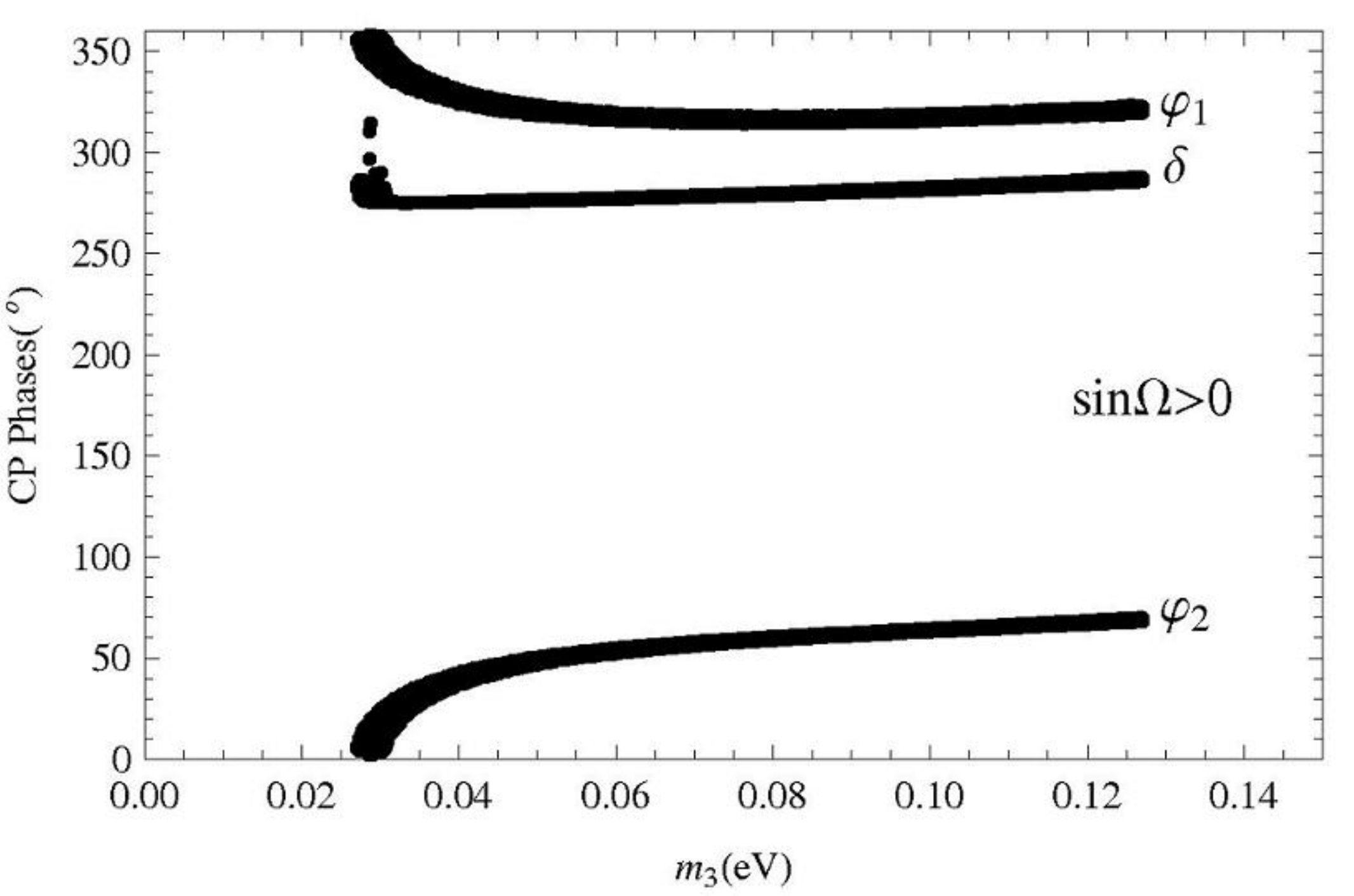}\\
\includegraphics[scale=.33]{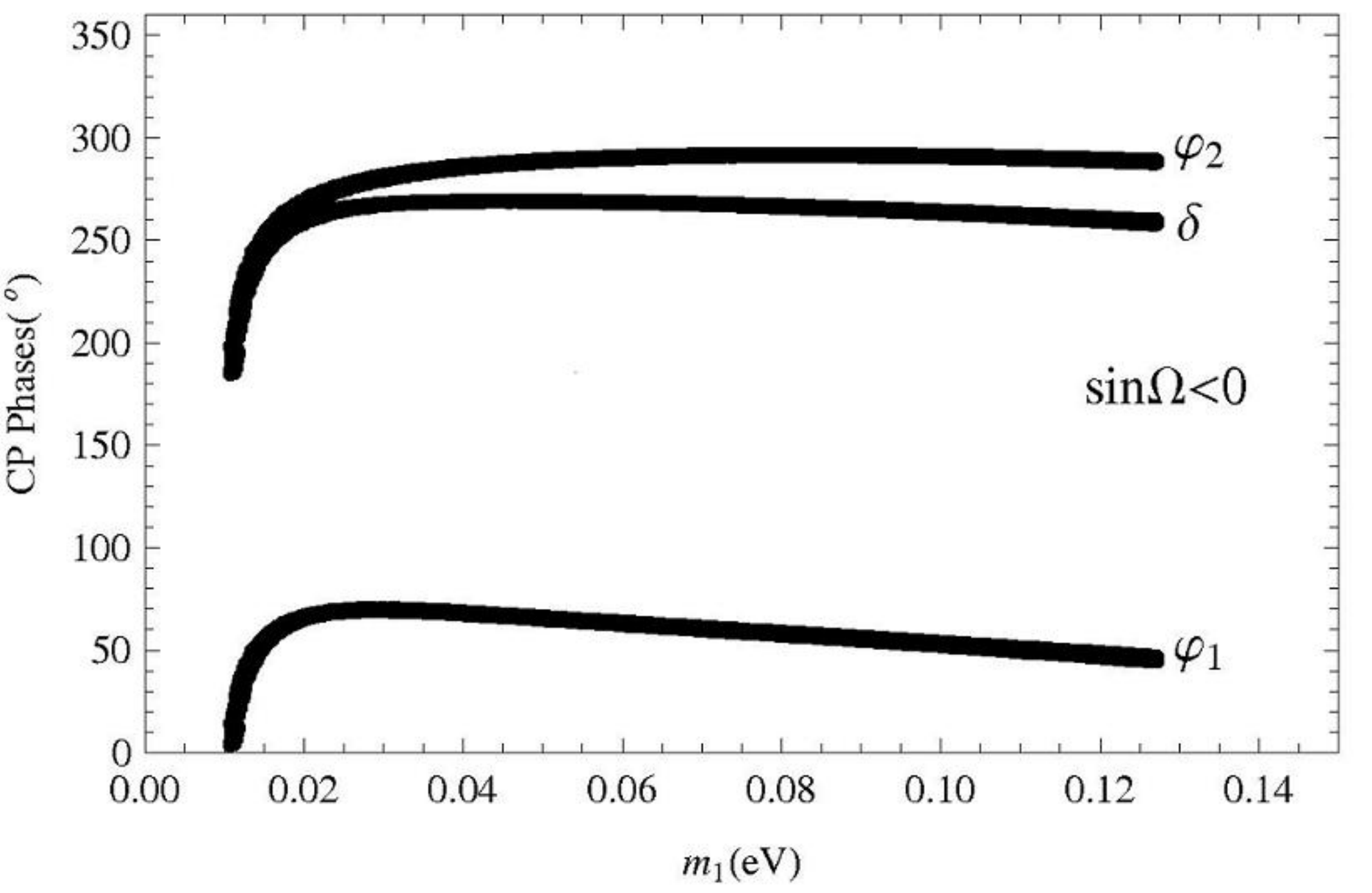}&\hspace{0.5cm}\includegraphics[scale=.33]{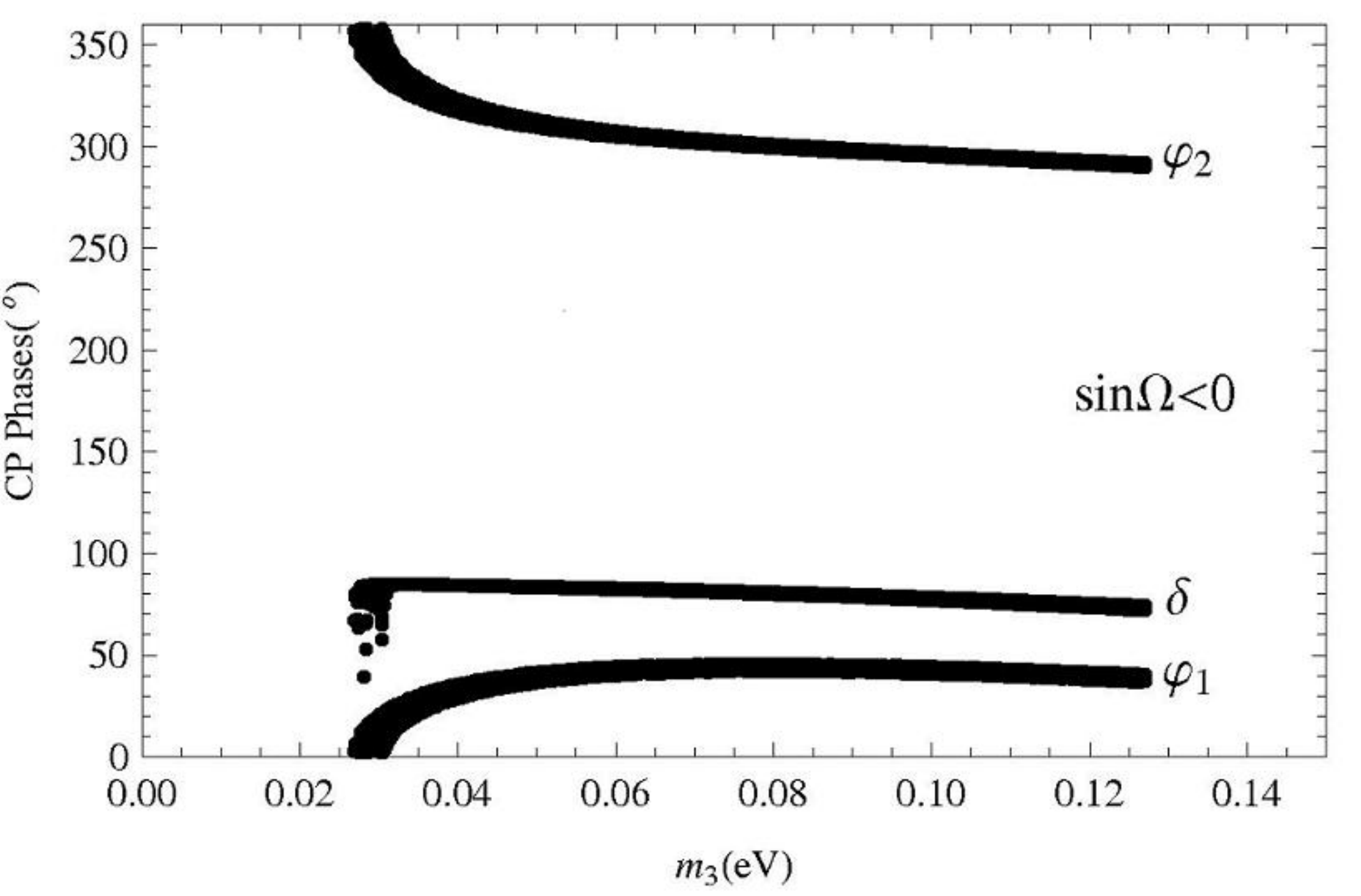}\\
\includegraphics[scale=.34]{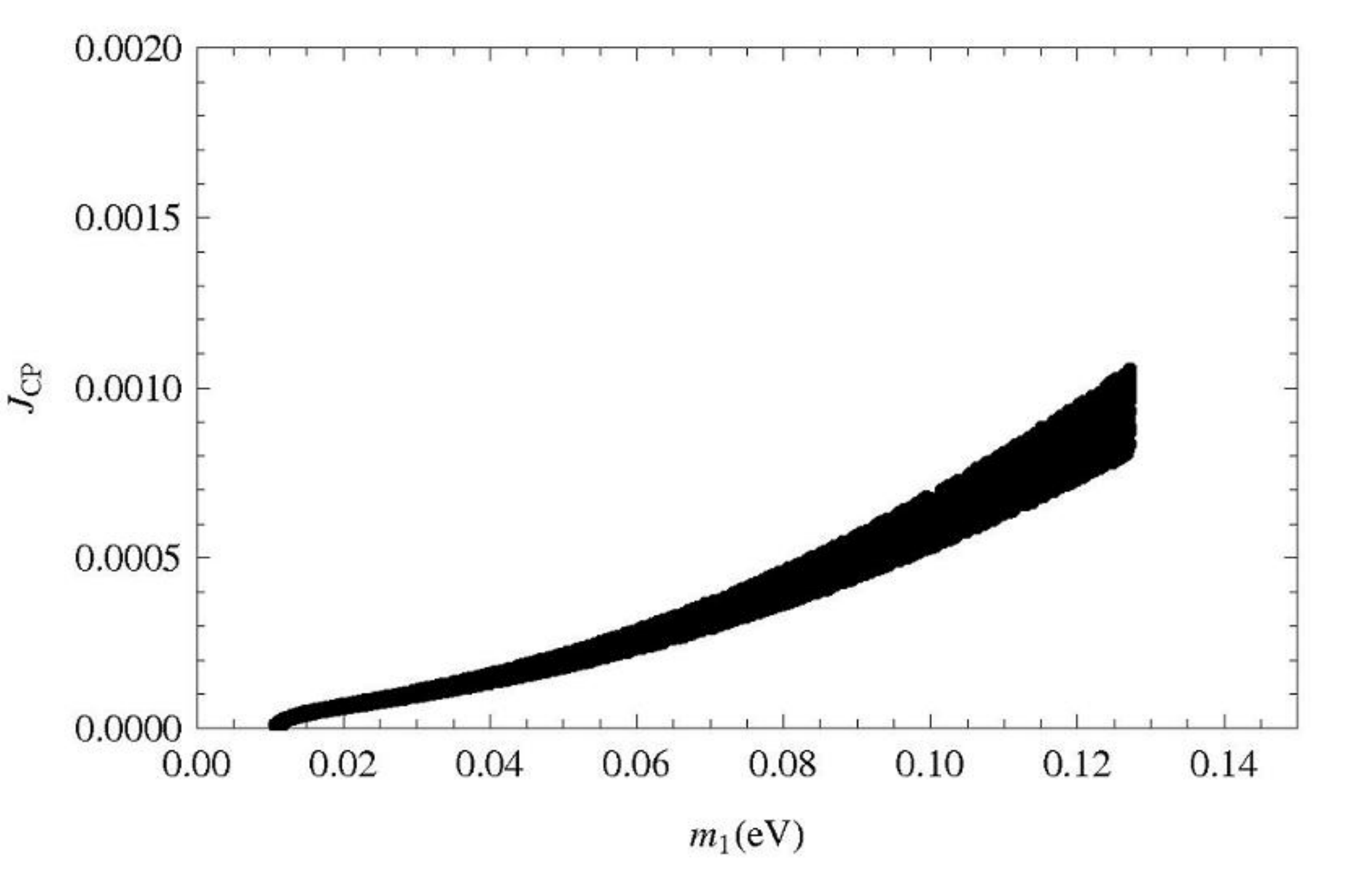}&\hspace{0.5cm}\includegraphics[scale=.34]{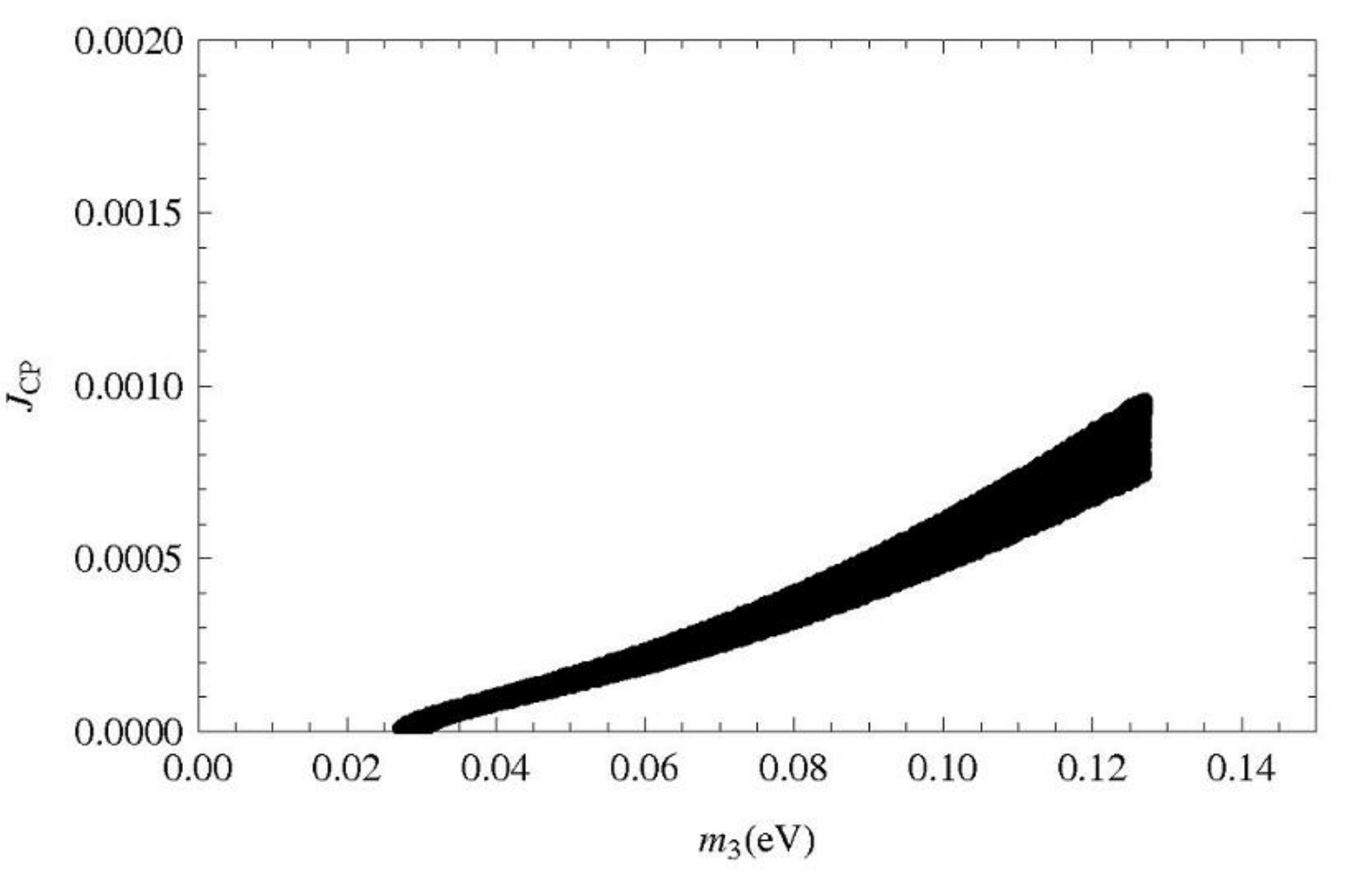}
\end{tabular}
\caption{\label{fig:RGE_BMM_mass2}RGE corrections to the CP phases
and the Jarlskog invariant in the BMM $S_4$ model with
$\tan\beta=10$. The left column of the plots are the results for NH
spectrum, and the right column for the IH case.}
\end{center}
\end{figure}

The variations of Majorana phases $\varphi_1$ and $\varphi_2$, Dirac
CP violating phase $\delta$ and the Jarlskog invariant $J_{CP}$ with
respect to the lightest neutrino mass are also plotted in Fig.
\ref{fig:RGE_BMM_mass2}. We note that Dirac phase $\delta$ arises
from the running effect, even though it is undetermined in the
beginning. The initial value of Jarlskog invariant $J_{CP}$ is zero
due to the vanishing of the $\theta_{13}$ in TB mixing scheme, and
it remains small because of the smallness of the $\theta_{13}$,
although the value of $\delta$ is large. It is remarkable that we
can understand the dependence on the sign of $\sin\Omega$ exactly.
At initial scale, the right-handed neutrino mass matrix $M_N$ shown
in Eq.(\ref{15}) is complex with each other for $\sin\Omega>0$ and
$\sin\Omega<0$ apart from the irrelevant overall phase, and the
neutrino Yukawa coupling matrix $Y_{\nu}$ can be chosen to be real.
Therefore, in the case of $\sin\Omega<0$, the complex conjugates of
$Y_{\nu}$, $Y_{e}$, $M_{N}$ and $\kappa$ run in the same way as the
corresponding quantities of $\sin\Omega>0$ with the same initial
conditions. Consequently the resulting low energy effective neutrino
mass matrix for $\sin\Omega<0$ is the complex conjugate of the
corresponding one of $\sin\Omega>0$. As a result, the RG evolution
of mixing angles and $J_{CP}$ is independent of the sign of
$\sin\Omega$, and summation of the each CP phase for $\sin\Omega>0$
and $\sin\Omega<0$ is equal to $2\pi$. These results have been
confirmed in our numerical analysis explicitly.

\begin{figure}[hptb]
\begin{center}
\begin{tabular}{rr}
\includegraphics[scale=.65]{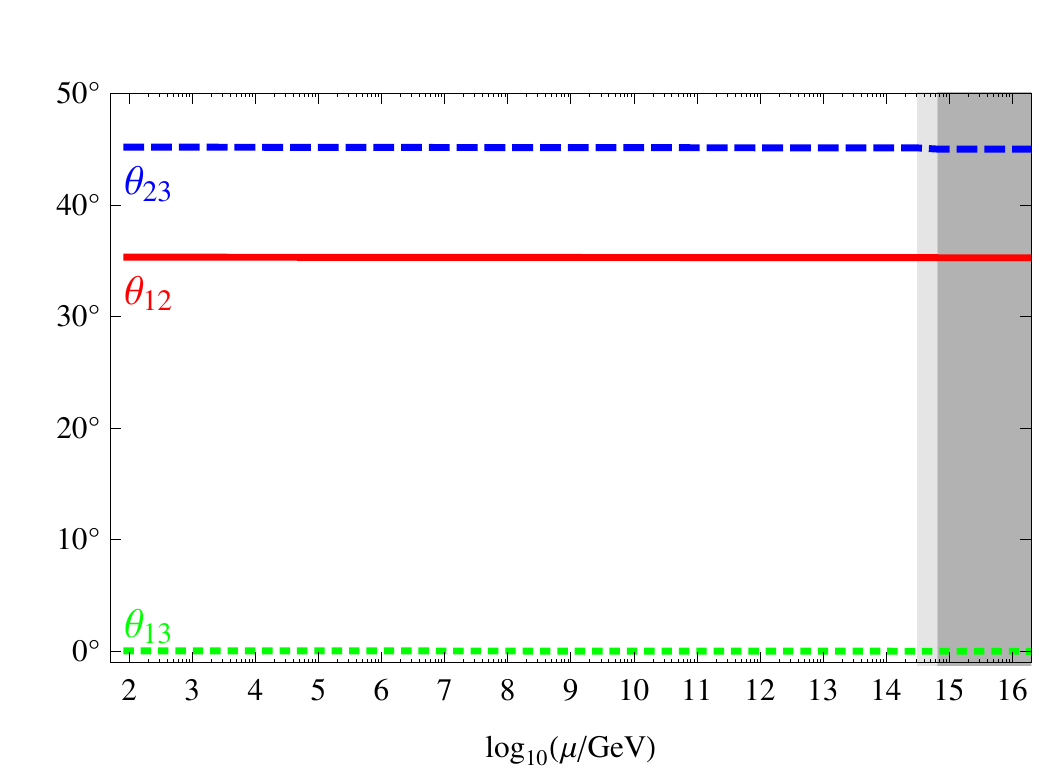}&\hspace{0.5cm}\includegraphics[scale=.65]{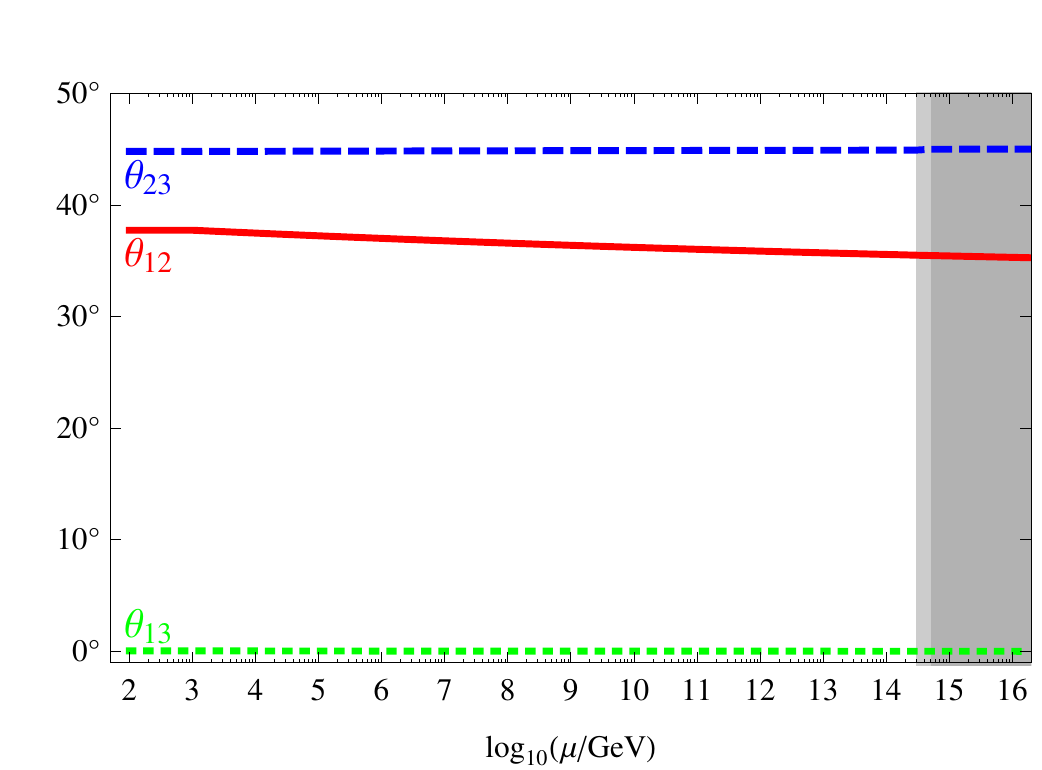}\\
\includegraphics[scale=.65]{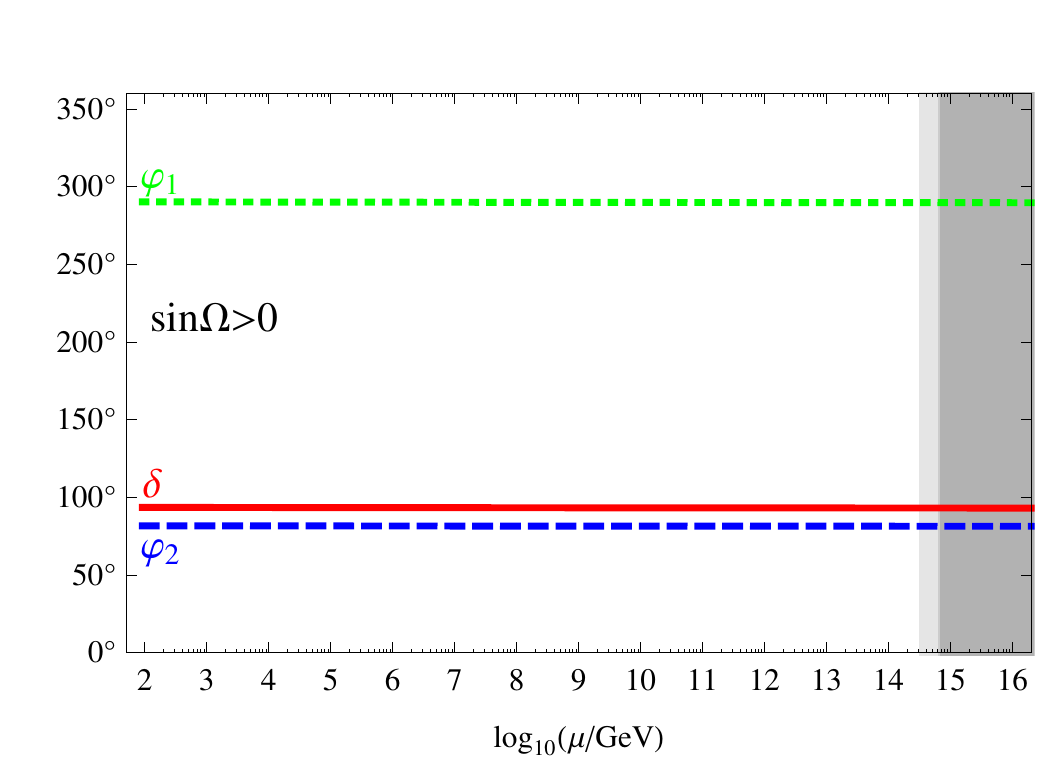}&\hspace{0.5cm}\includegraphics[scale=.65]{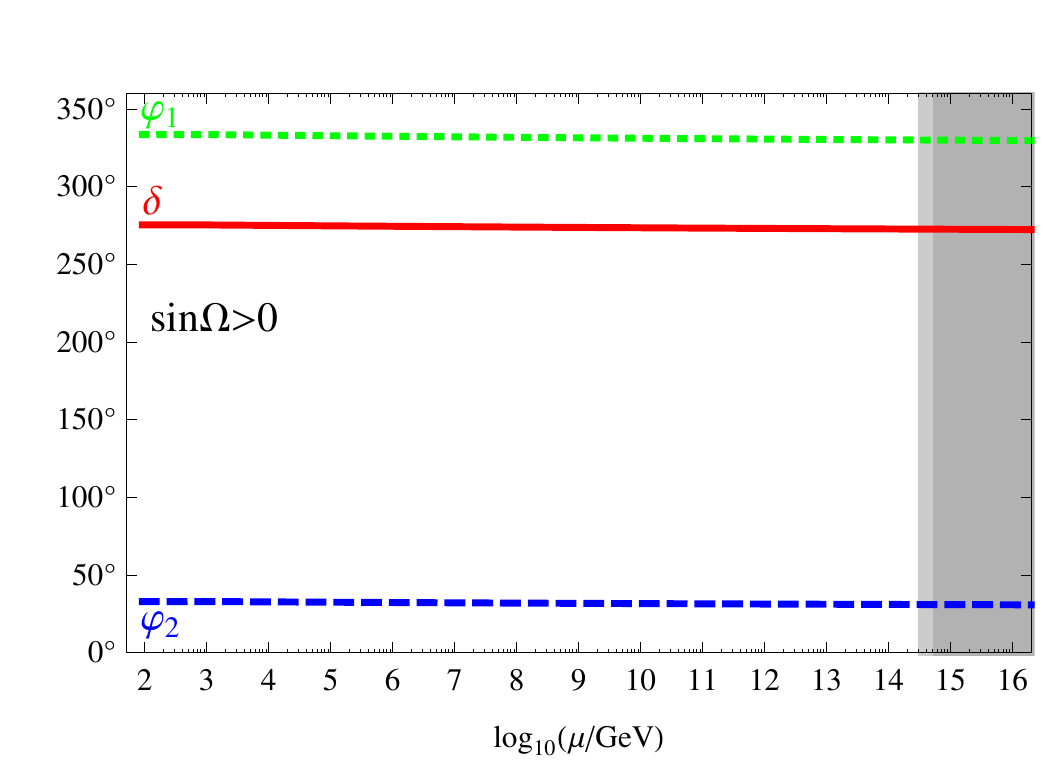}\\
\includegraphics[scale=.65]{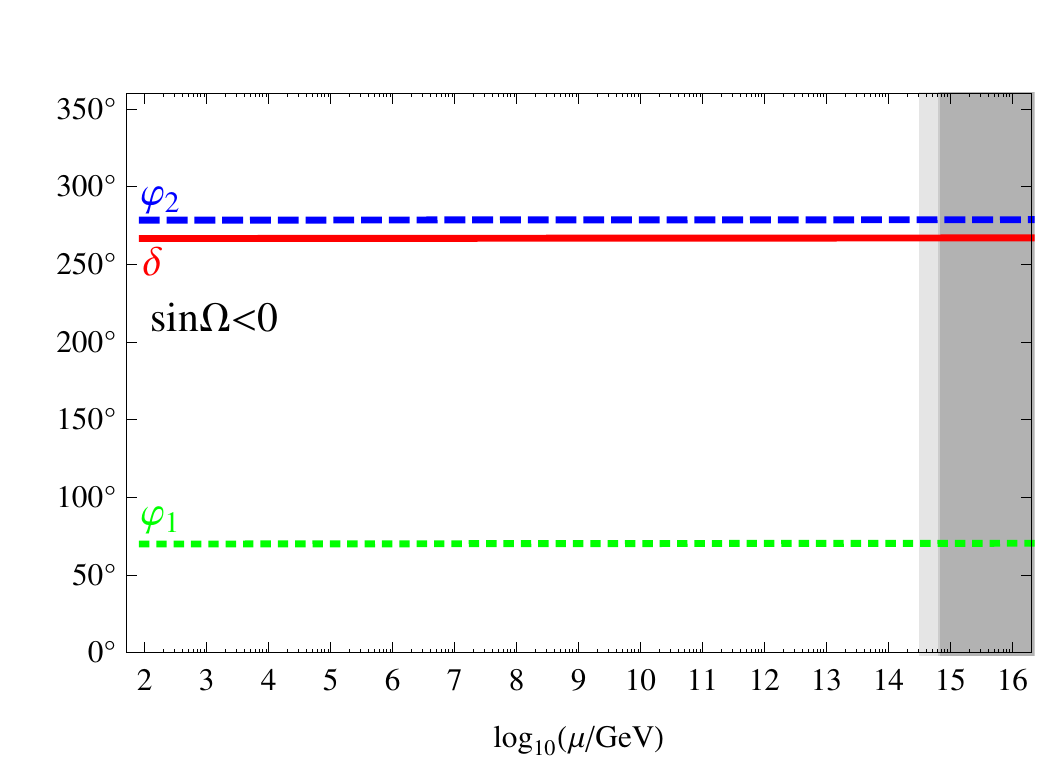}&\hspace{0.5cm}\includegraphics[scale=.65]{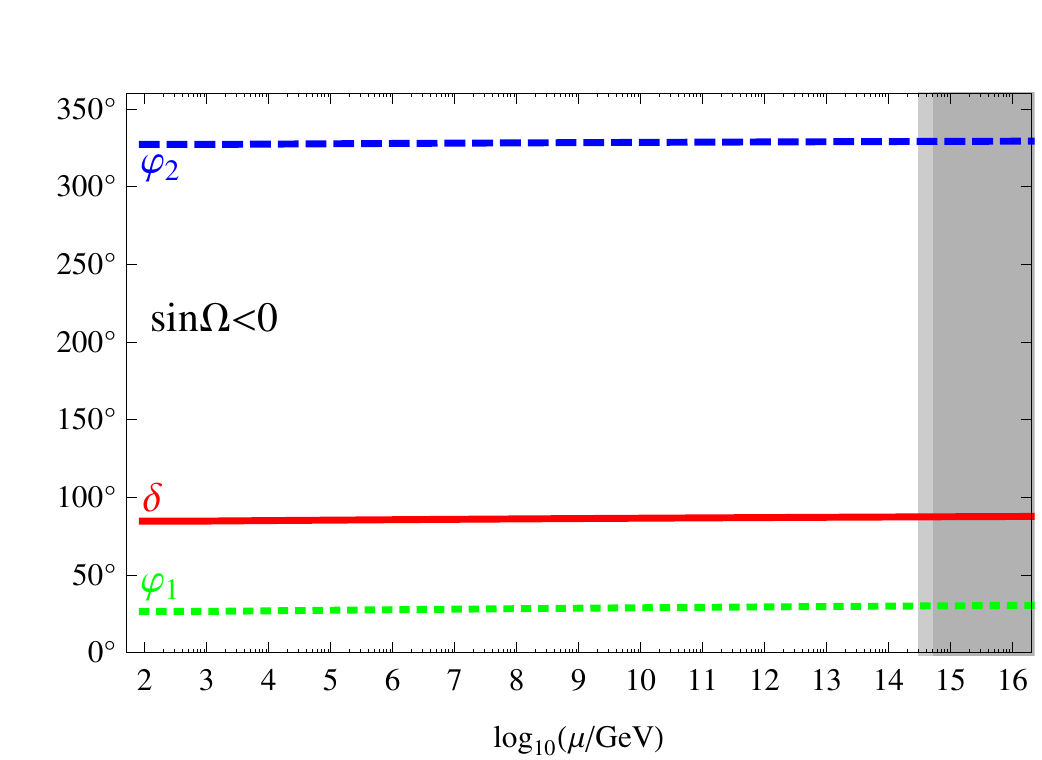}\\
\includegraphics[scale=.65]{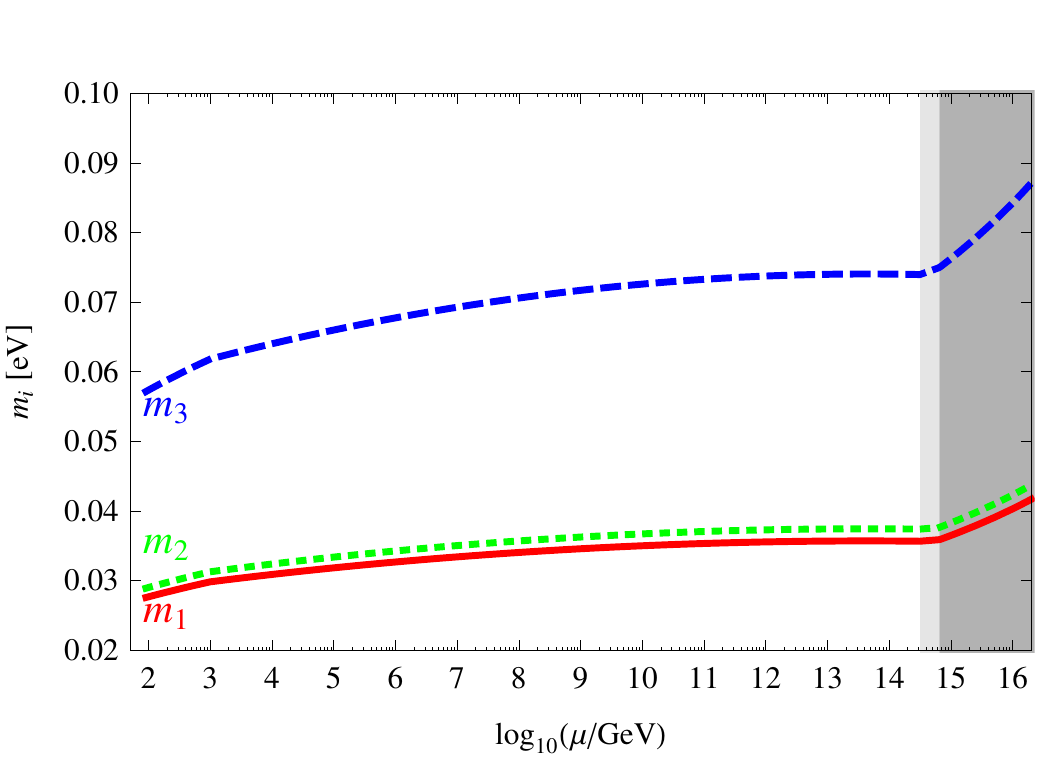}&\hspace{0.5cm}\includegraphics[scale=.65]{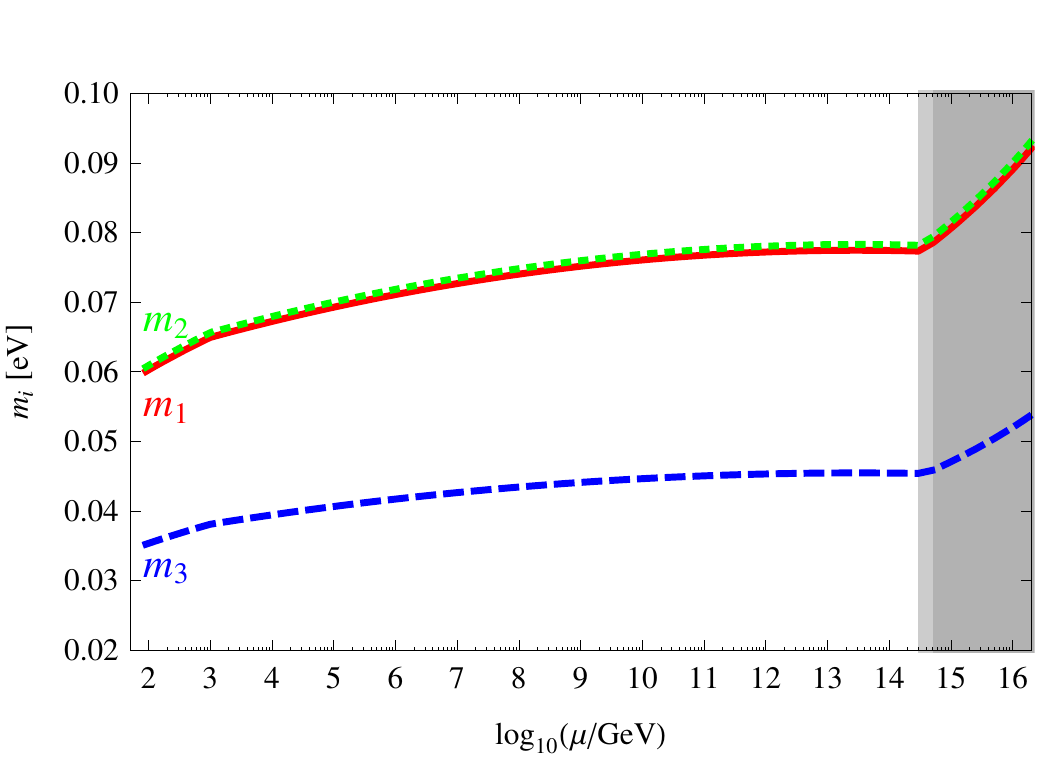}
\end{tabular}
\caption{\label{fig:RGE_BMM_scale} The running of the neutrino
masses and mixing parameters with the energy scale in the BMM $S_4$
model with $\tan\beta=10$ and $M_{SUSY}=1$ TeV. The left column is
the predictions for NH spectrum with $m_1=0.041$ eV, $\Delta
m^2_{\rm sol}=1.76\times10^{-4}{\rm eV^2}$ and $\Delta m^2_{\rm
atm}=5.85\times10^{-3}{\rm eV^2}$. The right column is for the IH
case with $m_3=0.0538$ eV, $\Delta m^2_{\rm
sol}=1.87\times10^{-4}{\rm eV^2}$ and $\Delta m^2_{\rm
atm}=5.58\times10^{-3}{\rm eV^2}$. }
\end{center}
\end{figure}

Concretely the running of neutrino masses and mixing parameters with
the energy scale is displayed in Fig. \ref{fig:RGE_BMM_scale} for
both NH and IH spectrum with $\tan\beta=10$, where the initial
conditions for the NH and IH are chosen to be $m_1=0.041$ eV,
$\Delta m^2_{\rm sol}=1.76\times10^{-4}{\rm eV^2}$, $\Delta m^2_{\rm
atm}=5.85\times10^{-3}{\rm eV^2}$ and $m_3=0.0538$ eV, $\Delta
m^2_{\rm sol}=1.87\times10^{-4}{\rm eV^2}$, $\Delta m^2_{\rm
atm}=5.58\times10^{-3}{\rm eV^2}$ respectively. Reasonable values
for the lower energy oscillation parameters are reached. We see that
the deviation of the solar neutrino mixing angle $\theta_{12}$ from
the TB value can be relative large for the IH spectrum, the mixing
angles $\theta_{23}$ and $\theta_{13}$ and the CP phases $\delta$,
$\varphi_1$ and $\varphi_2$ are stable under the RG evolution, the
corresponding RG corrections are small. Since
$Y^{\dagger}_{\nu}Y_{\nu}=y^2{\bf 1}$, the contribution from the
neutrino Yukawa coupling is universal above the see-saw threshold.
Then, only the charged lepton relevant part $Y^{\dagger}_eY_e$
contributes to the change in mixing angles, and the evolution above
the see-saw scales is essentially the same as below. This is in
contrast with the usual situation where the neutrino Yukawa coupling
plays dominant role in the running of neutrino mass matrix above the
highest see-saw scale. Furthermore, we find that the running of the
neutrino mass $m_{\nu i}$ is approximately given by a common scaling
of the mass eigenvalues, this is the same as the situation below the
see-saw scale \cite{Antusch:2003kp,Chankowski:2001mx}. It is
remarkable that the neutrino mass is reduced by about $2.4$ times at
low energy. We note that the above results about the running
behavior of neutrino masses and mixing parameters are very general,
they almost do not depend on the initial conditions.

%In the MSSM with large $\tan\beta$, while NH still
%gives unobservably small deviations for all the mixing angles, IH is capable of generating
%significant running for $\theta_{12}$. In fact, matching $\theta_{12}$ with the data requires constraining the
%Majorana phases and $\tan\beta$ already at the present stage. For the QD scenario the running
%for all the three cases can be strong.

%In this class of flavour models, the running
%contribution from neutrino Yukawa coupling, which is generally dominant at energies above the seesaw threshold, can be absorbed by a small shift on neutrino mass
%eigenvalues leaving mixing angles unchanged. Consequently, in the whole running
%energy range, the change in mixing angles is due to the contribution coming from
%charged lepton sector

%We find that for normally ordered light neutrinos, the tribimaximal prediction
%is essentially stable under renormalization group evolution. On the other hand, in
%the case of inverted hierarchy, the deviation of the solar angle from its TB value can
%be large depending on mass degeneracy.

\subsection{RG effects in the $S_4$ model of Ding}

\begin{figure}[hptb]
\begin{center}
\begin{tabular}{rr}
\includegraphics[scale=.33]{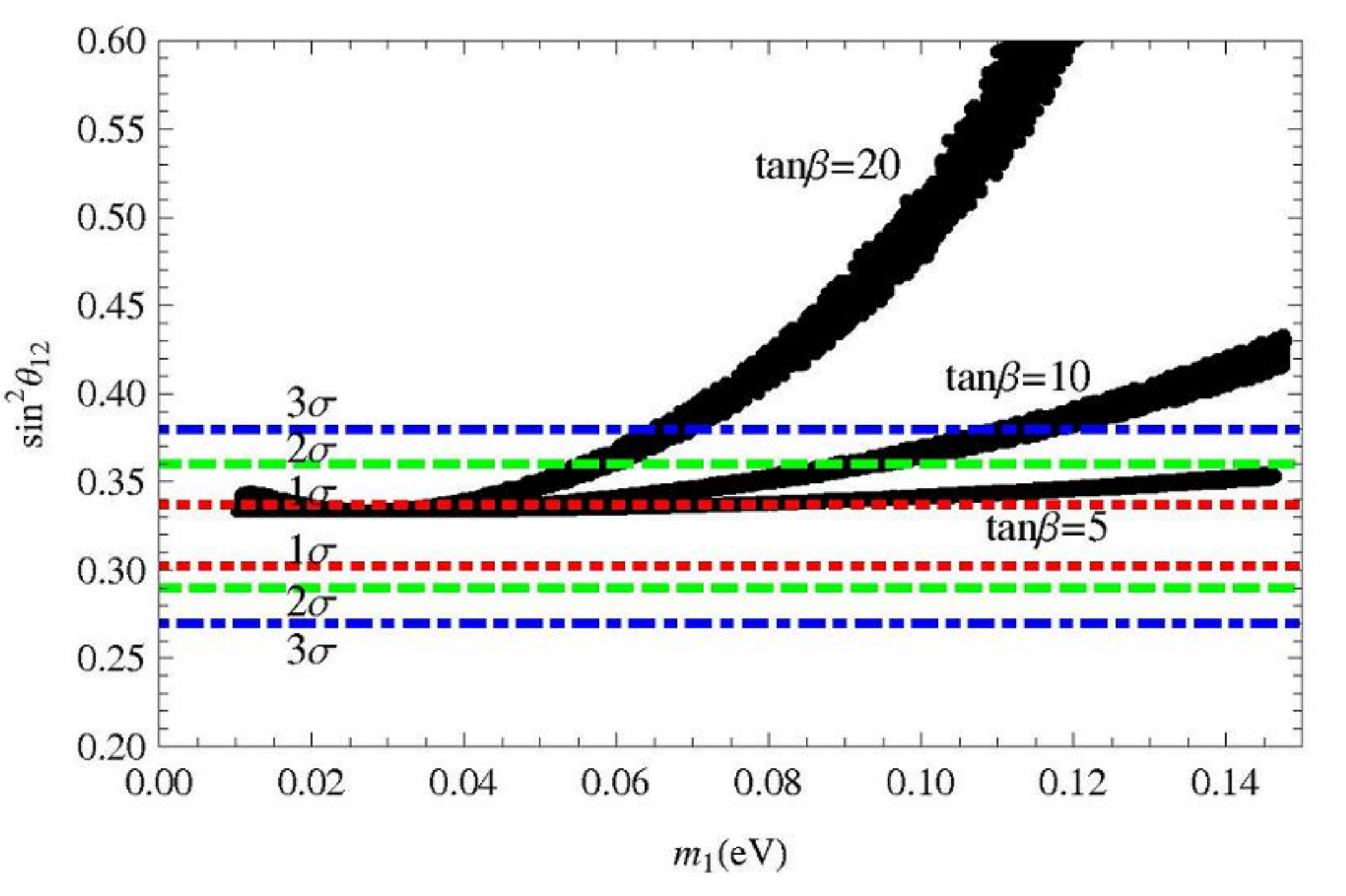}&\hspace{0.8cm}\includegraphics[scale=.33]{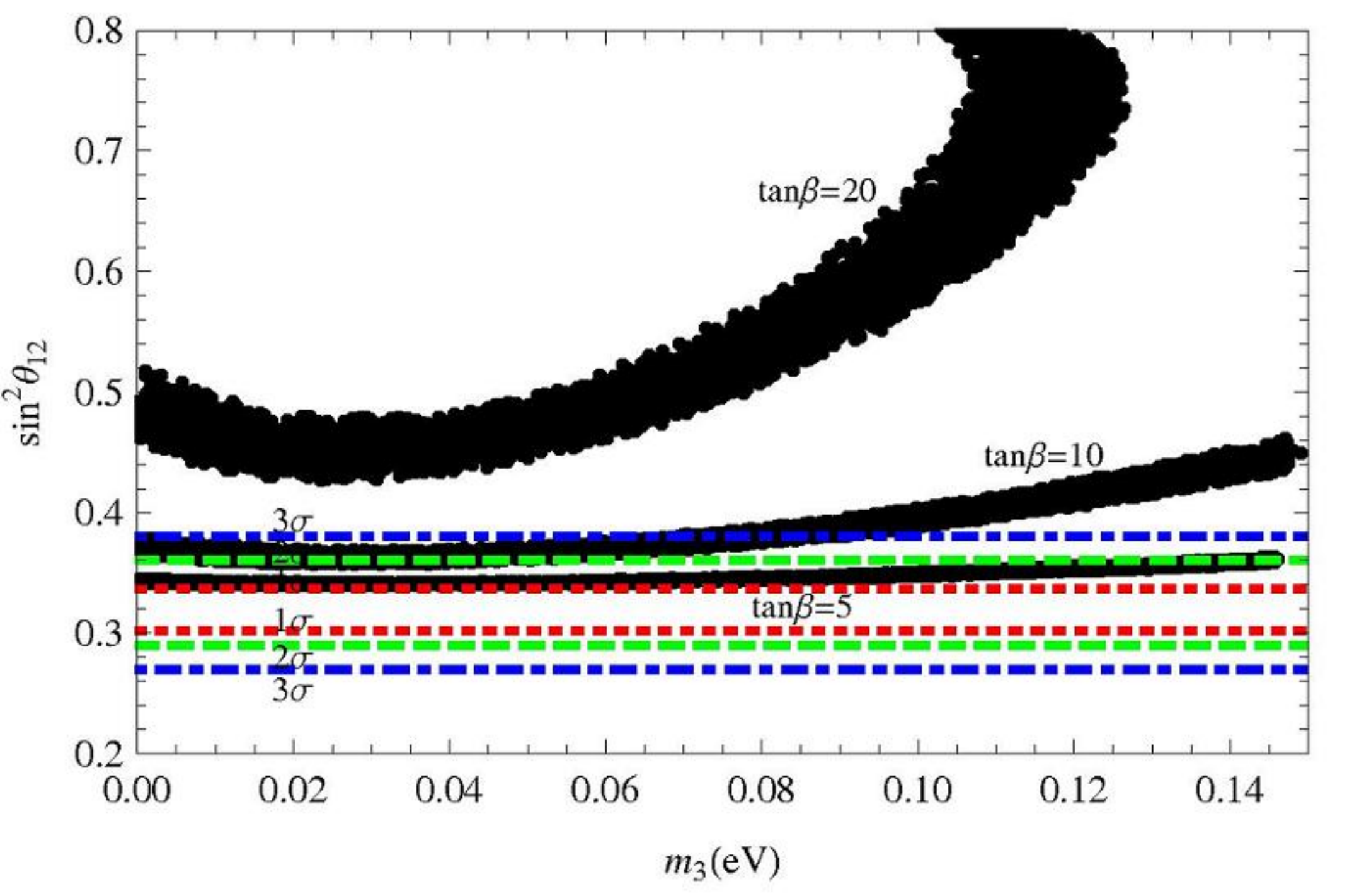}\\
\includegraphics[scale=.33]{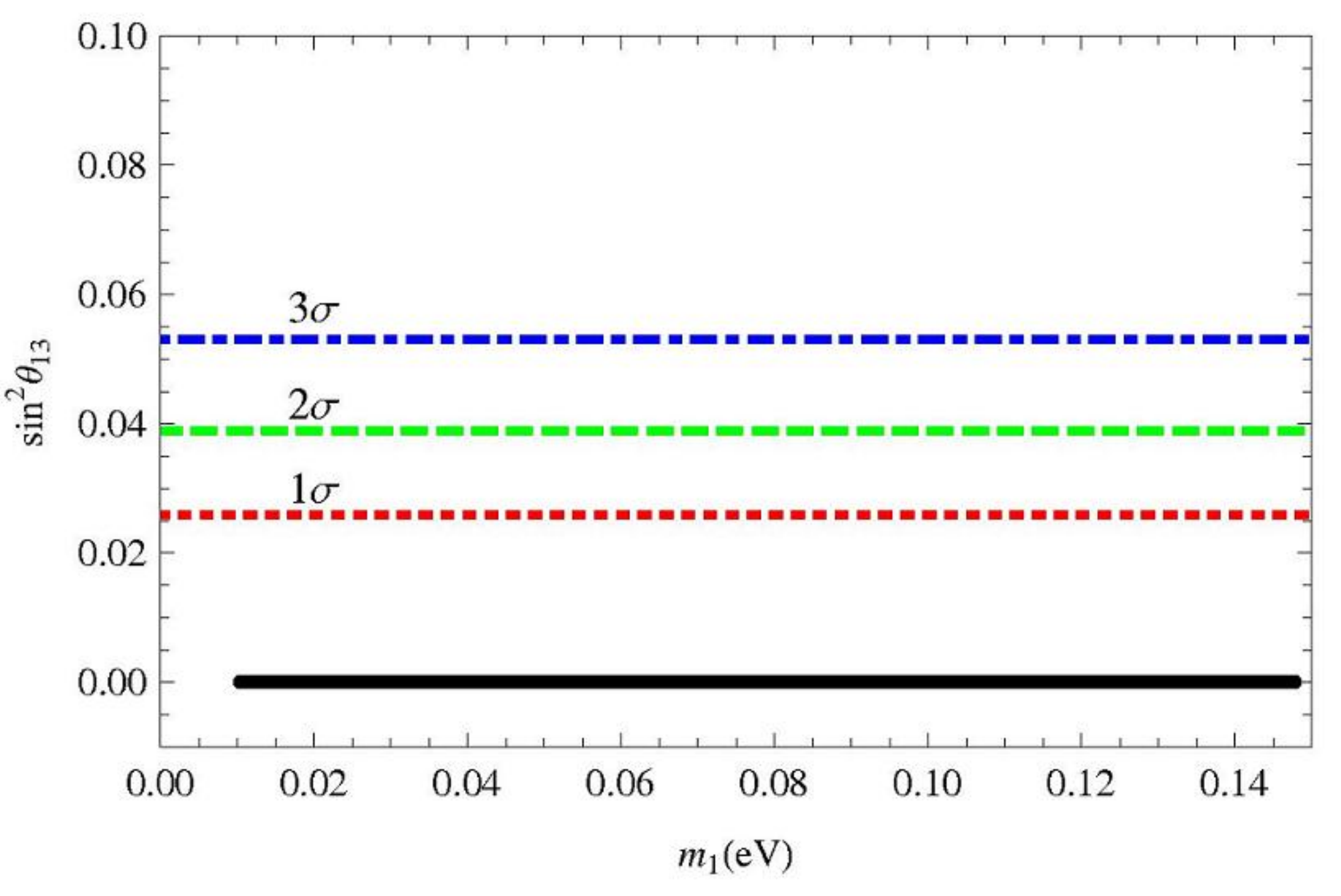}&\hspace{0.8cm}\includegraphics[scale=.33]{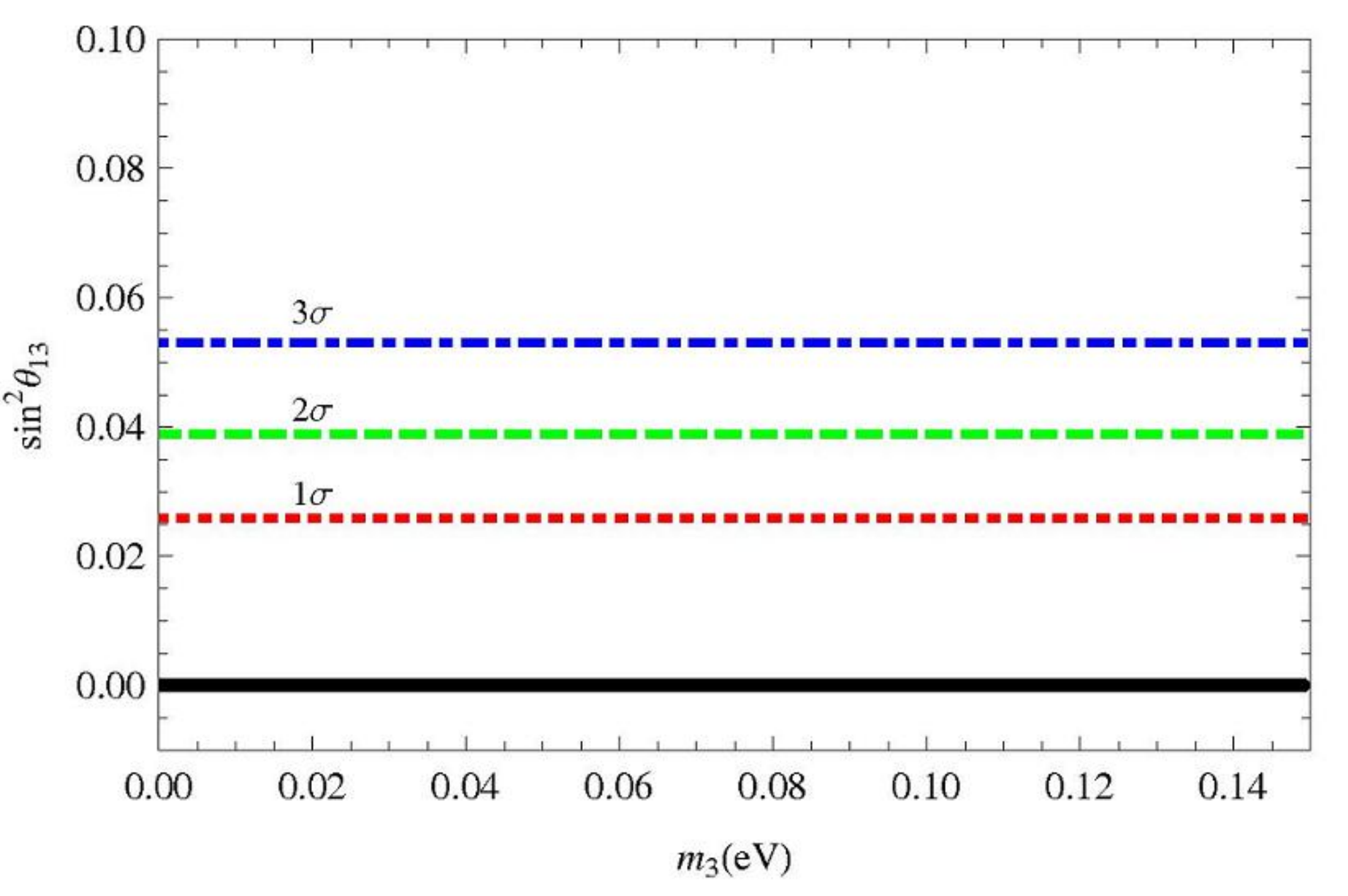}\\
\includegraphics[scale=.33]{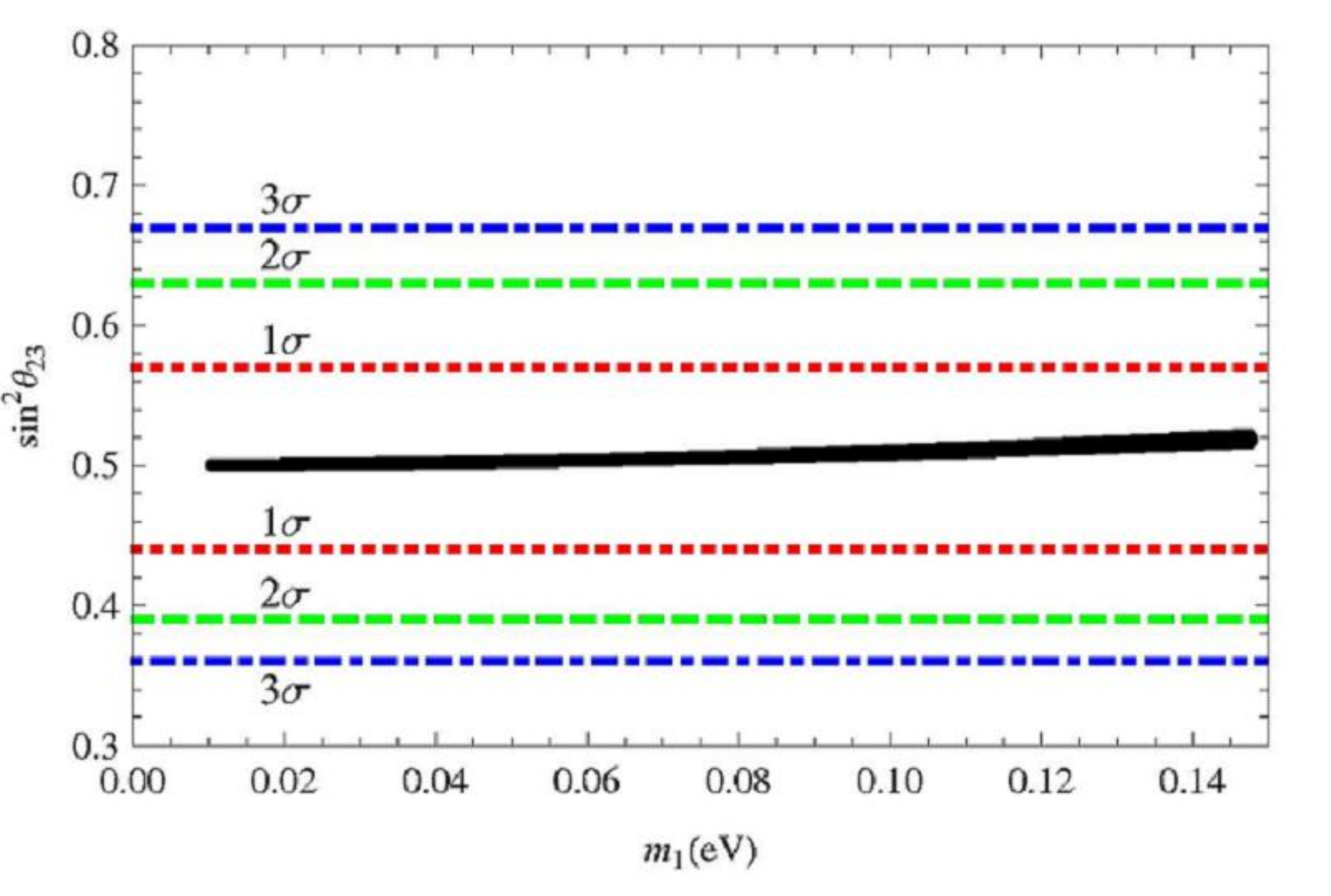}&\hspace{0.8cm}\includegraphics[scale=.33]{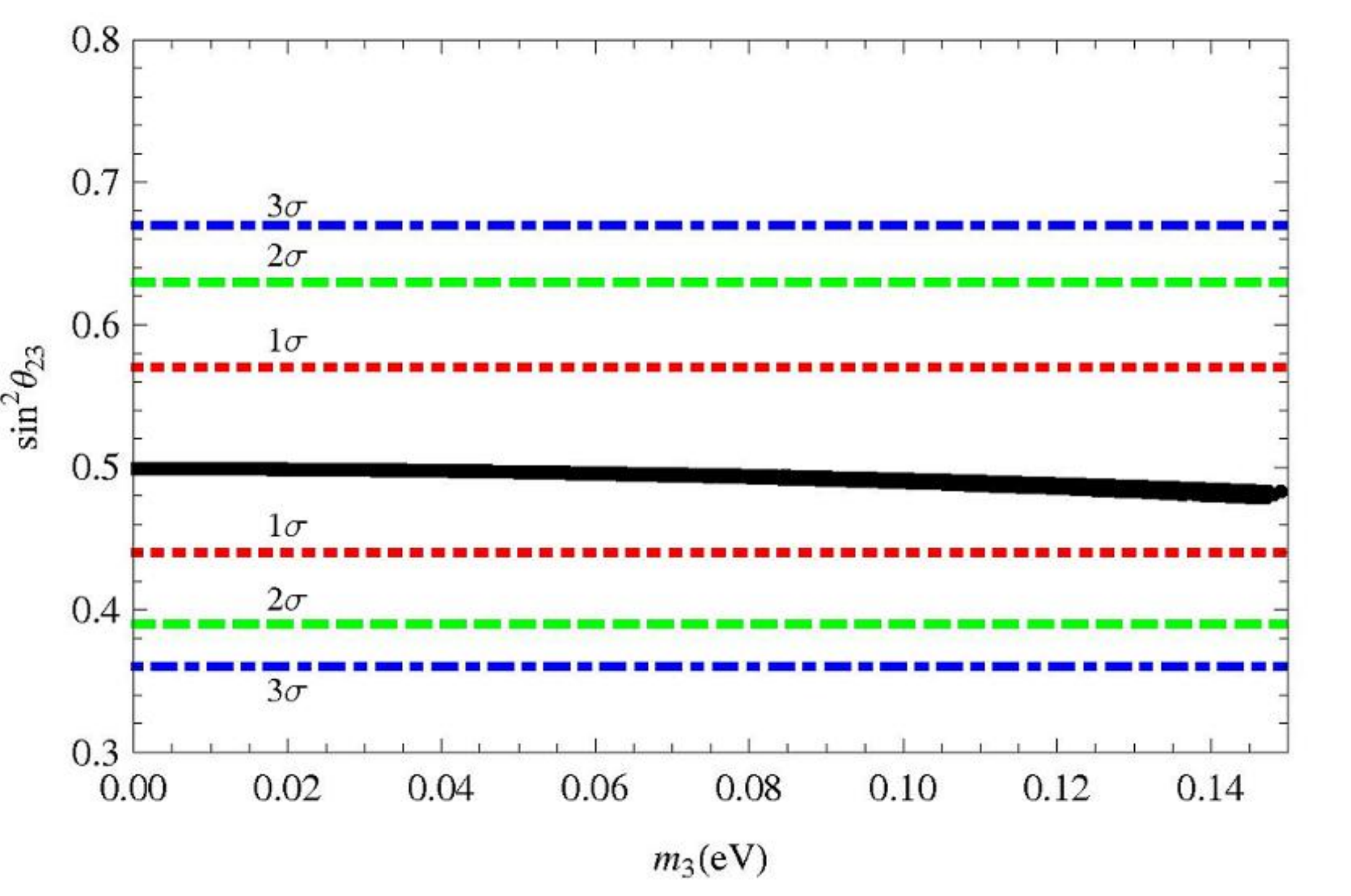}
\end{tabular}
\caption{\label{fig:RGE_Ding_mass1}RGE corrections to the neutrino
mixing angles in the $S_4$ model of Ding with $\tan\beta=10$. The
left column of the plots are the results for NH spectrum, and the
right column for the IH case. In the case of $\sin^2\theta_{12}$,
$\tan\beta=5$ and $\tan\beta=20$ are considered.}
\end{center}
\end{figure}

\begin{figure}[hptb]
\begin{center}
\begin{tabular}{rr}
\includegraphics[scale=.33]{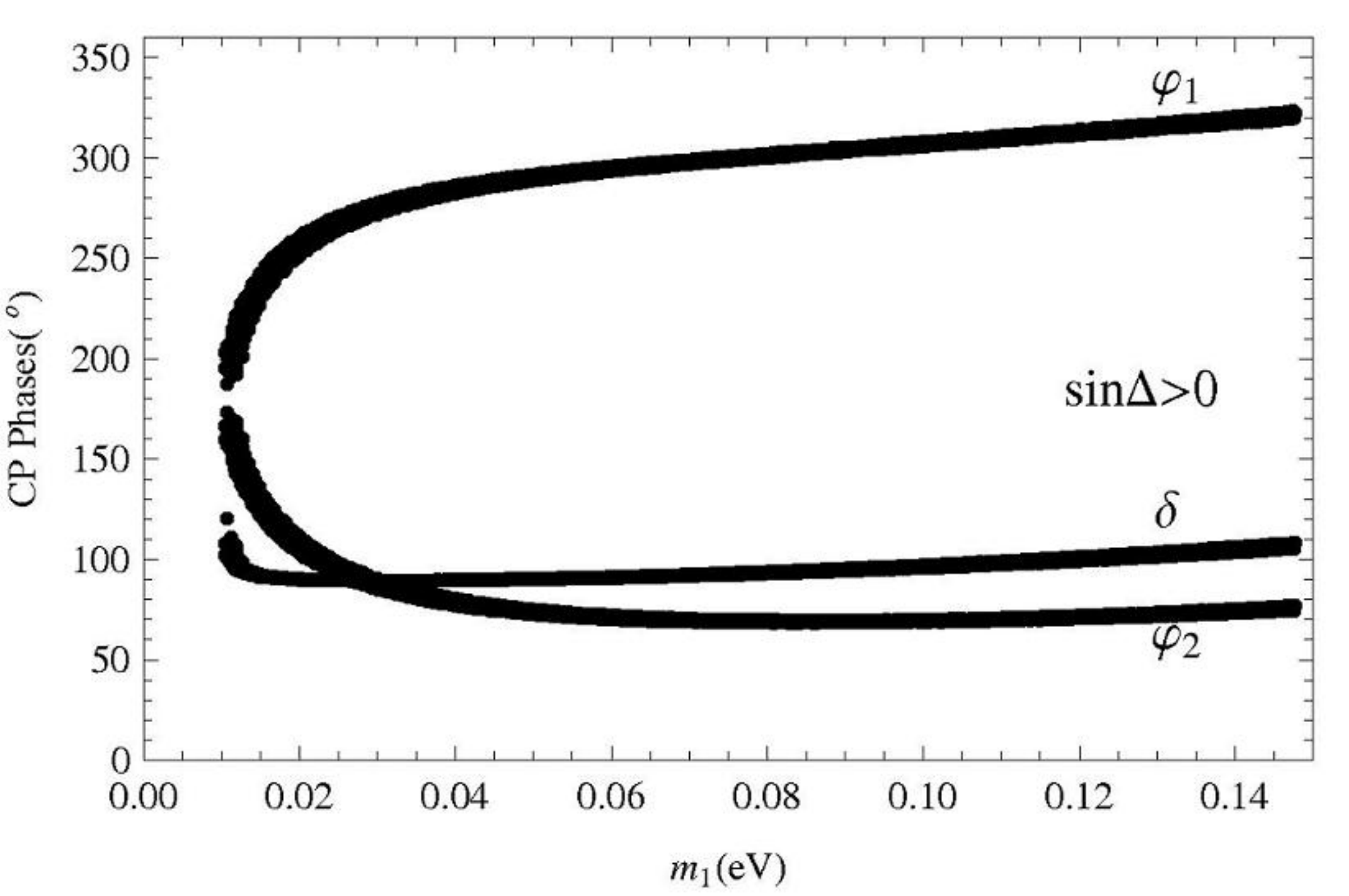}&\hspace{0.8cm}\includegraphics[scale=.33]{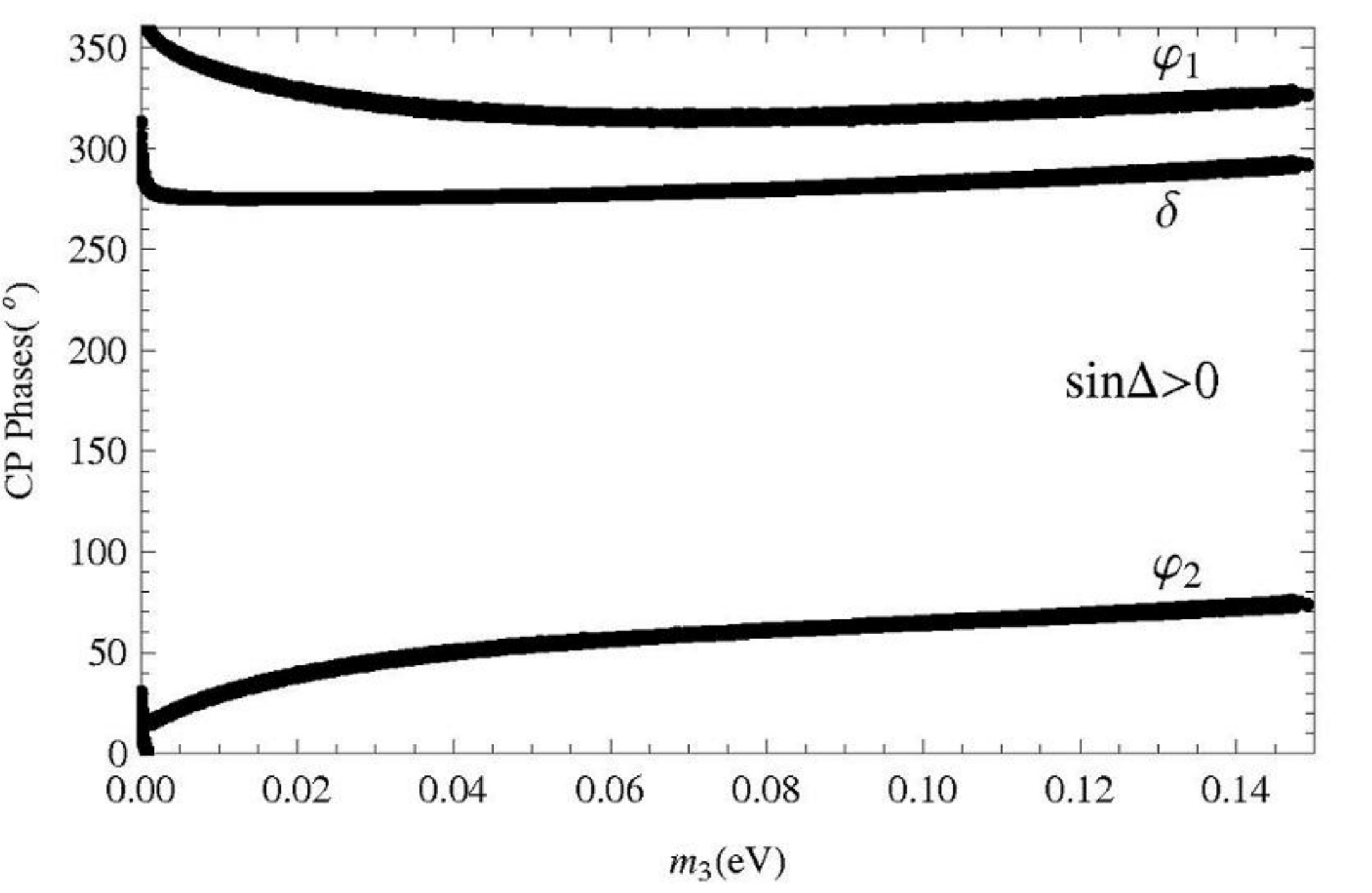}\\
\includegraphics[scale=.33]{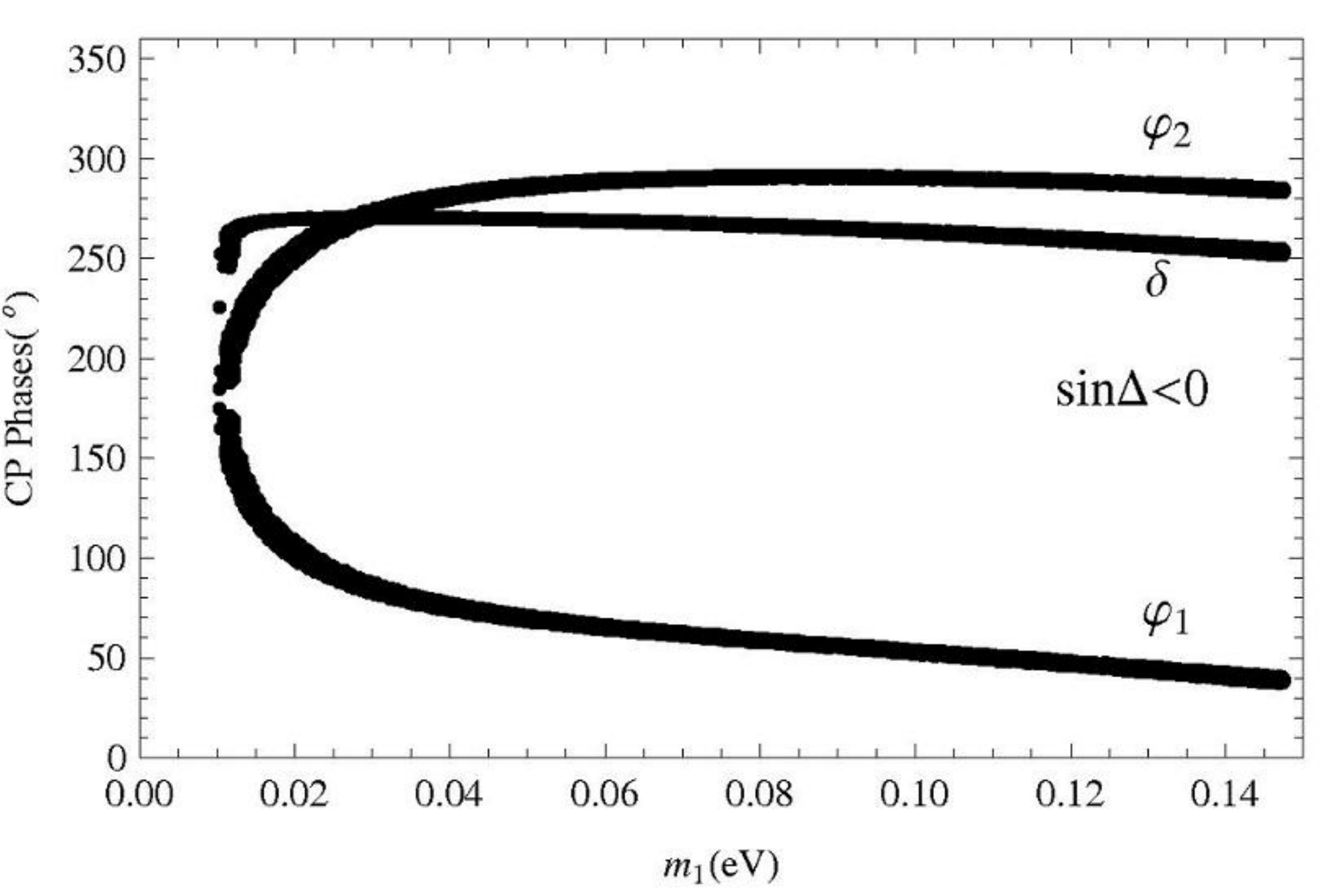}&\hspace{0.8cm}\includegraphics[scale=.33]{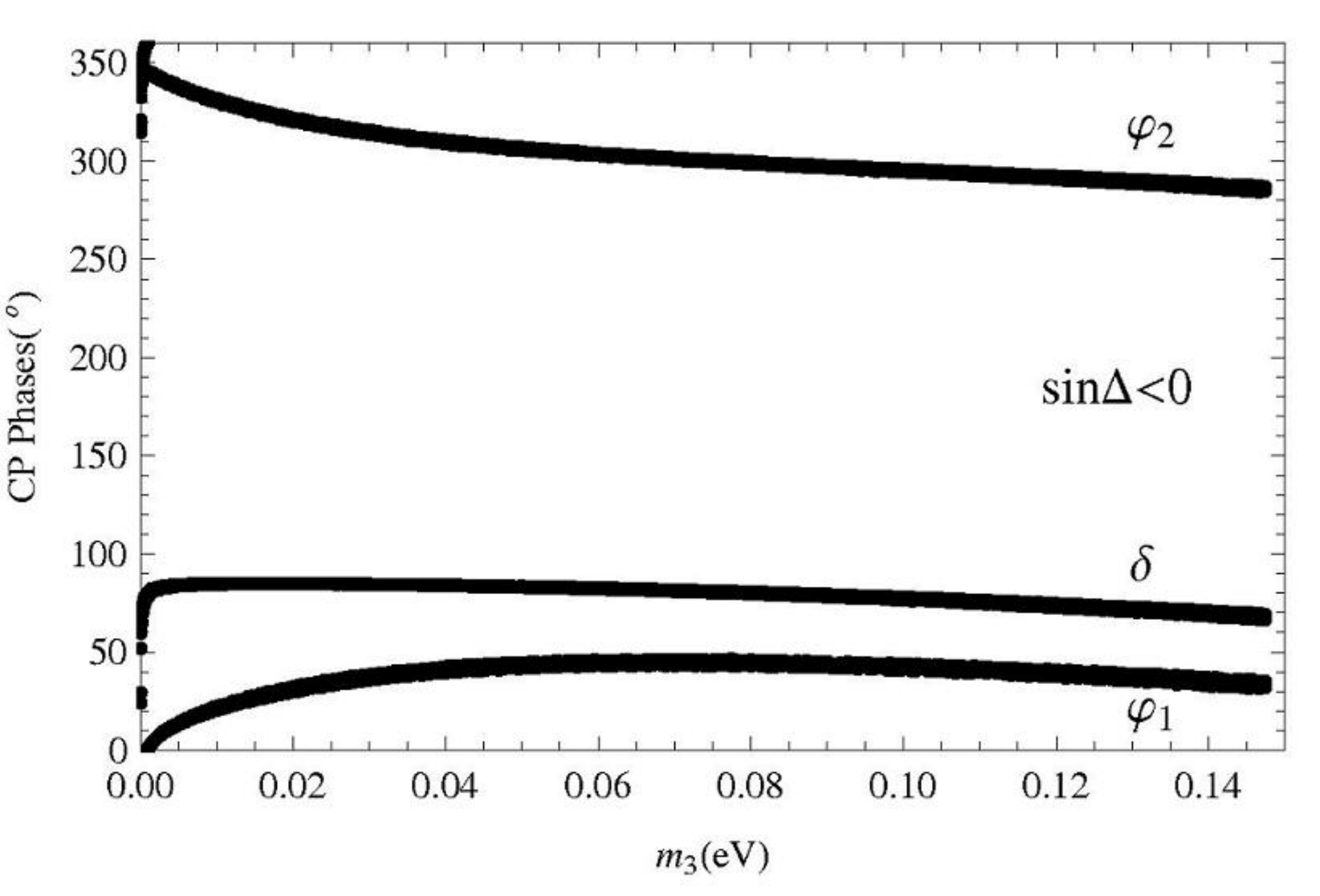}\\
\includegraphics[scale=.34]{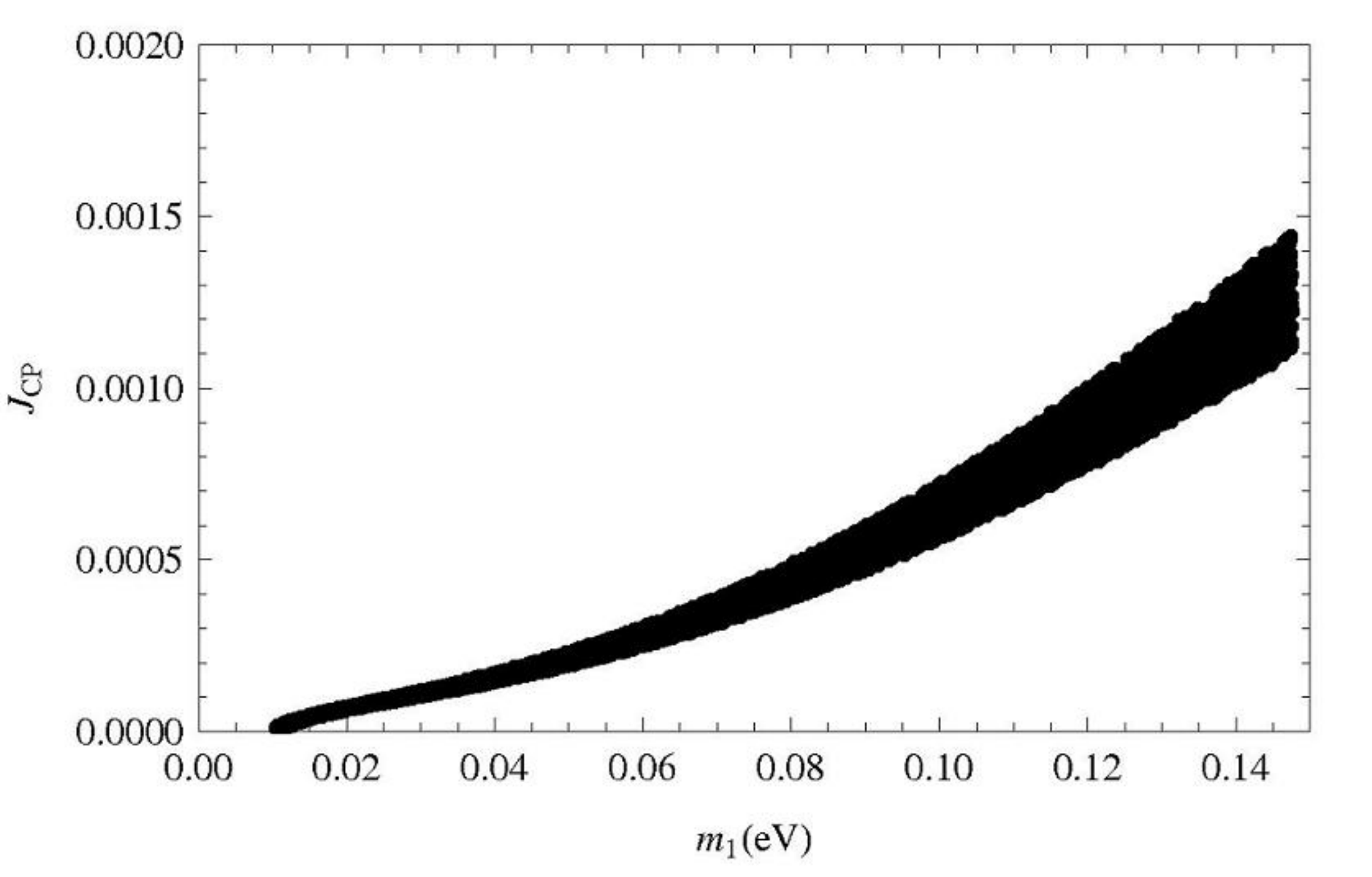}&\hspace{0.8cm}\includegraphics[scale=.34]{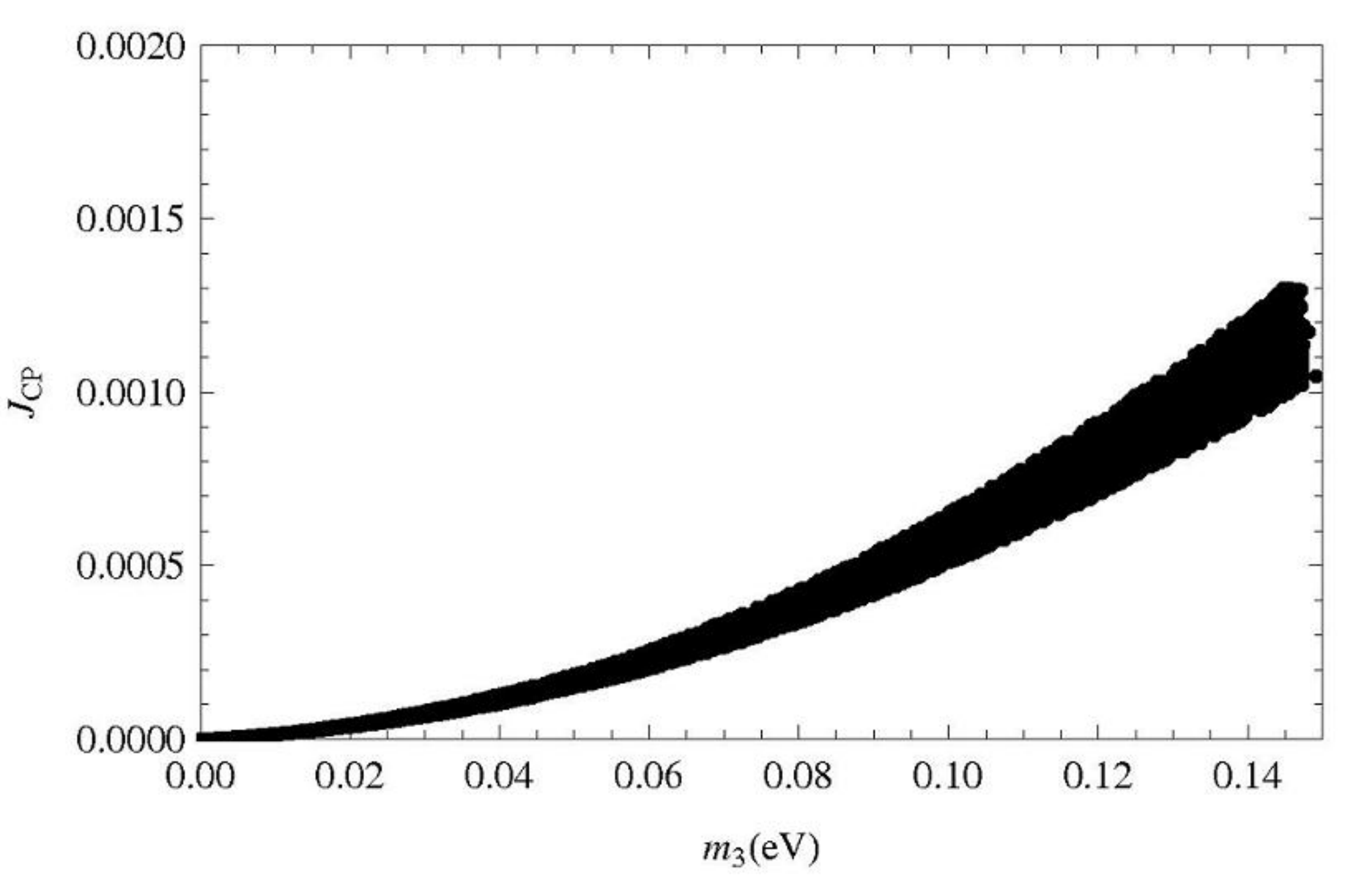}
\end{tabular}
\caption{\label{fig:RGE_Ding_mass2}RGE corrections to the CP phases
and the Jarlskog invariant in the $S_4$ model of Ding with
$\tan\beta=10$. The left column of the plots are the results for NH
spectrum, and the right column for the IH case.}
\end{center}
\end{figure}

It is remarkable that the heavy right-handed neutrinos are
degenerate at LO, and the corrections to the degeneracy arising from
RG running turn out to be so small that could be neglected,
consequently the threshold effects should be very small in this
case. In particular, we note that $Y^{\dagger}_{\nu}Y_{\nu}$ is not
proportional to the unit matrix any more, large RG effects seem
possible. As has been demonstrated in Eq.(\ref{add_EPJC_3}), the
right-handed neutrino mass $M$ is estimated to be of order $
10^{12}\sim10^{13}{\rm GeV}$. Without loss of generality, we shall
choose $M=10^{12}$ GeV in the following numerical analysis, and we
have checked that final results change very slowly with the
parameter $M$. The neutrino mixing angles at electroweak scale as
functions of the lightest neutrino mass are shown in
Fig.\ref{fig:RGE_Ding_mass1}, it is obvious that the lightest
neutrino mass for NH spectrum is bounded from below, and the lower
bound on the lightest neutrino mass in the case of IH spectrum is
still approximately zero. We see that the RG effects on both
atmospheric and reactor angles are rather small, and the running of
$\theta_{12}$ can be large depending on $\tan\beta$ and the mass
degeneracy. Matching $\theta_{12}$ with the data already puts strong
constraints on the lightest neutrino mass spectrum and $\tan\beta$
at the present stage, and a upper bound on the lightest neutrino
mass is usually implied for small value of $\tan\beta$, which means
that the neutrino mass spectrum can not be highly degenerate. In the
case of $\tan\beta=20$, the IH spectrum is ruled out, since the
value of $\sin^2\theta_{12}$ is much larger than its $3\sigma$ upper
bound. While the model is within the $3\sigma$ limit only for small
neutrino mass for NH spectrum, as is displayed in
Fig.\ref{fig:RGE_Ding_mass1}. The predictions for the CP phases and
the Jarlskog invariant are plotted in Fig.\ref{fig:RGE_Ding_mass2}.
In a similar way as section \ref{sec:RG_BMM}, we learn that the
evolutions of mixing angles and $J_{CP}$ do not depend on the
sign of $\sin\Delta$, the summation of each CP phase for
$\sin\Delta>0$ and $\sin\Delta<0$ is equal to $2\pi$. These points
are checked by our detailed numerical analysis.

\begin{figure}[hptb]
\begin{center}
\begin{tabular}{rr}
\includegraphics[scale=.65]{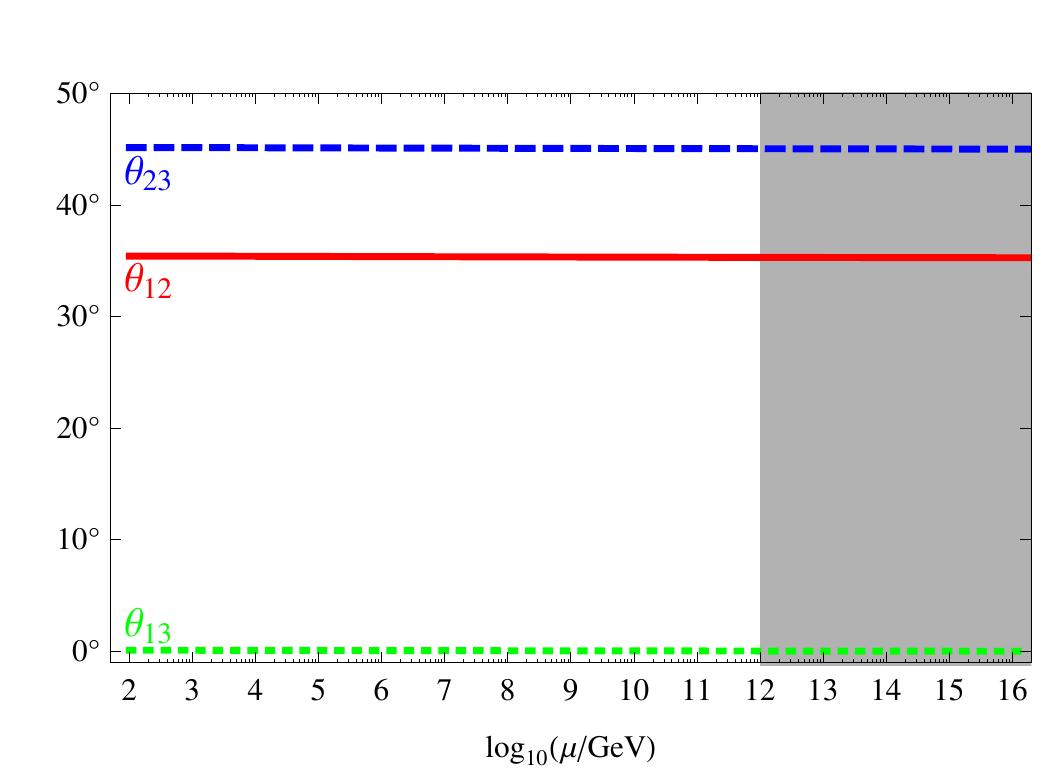}&\hspace{0.5cm}\includegraphics[scale=.65]{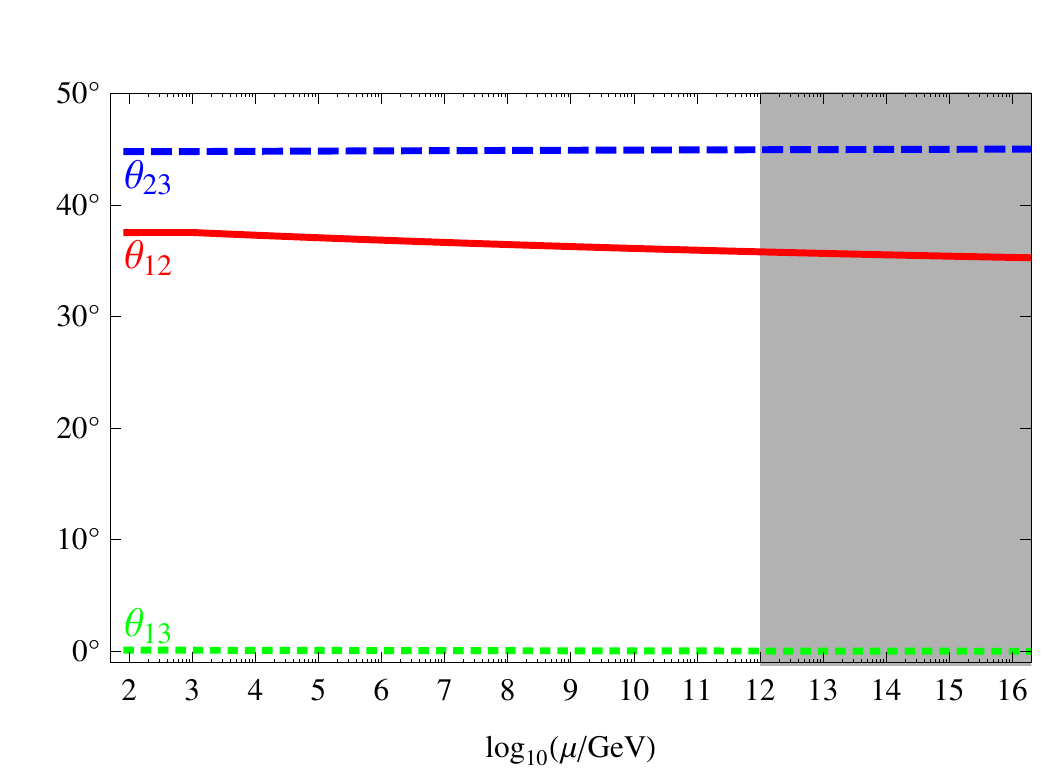}\\
\includegraphics[scale=.65]{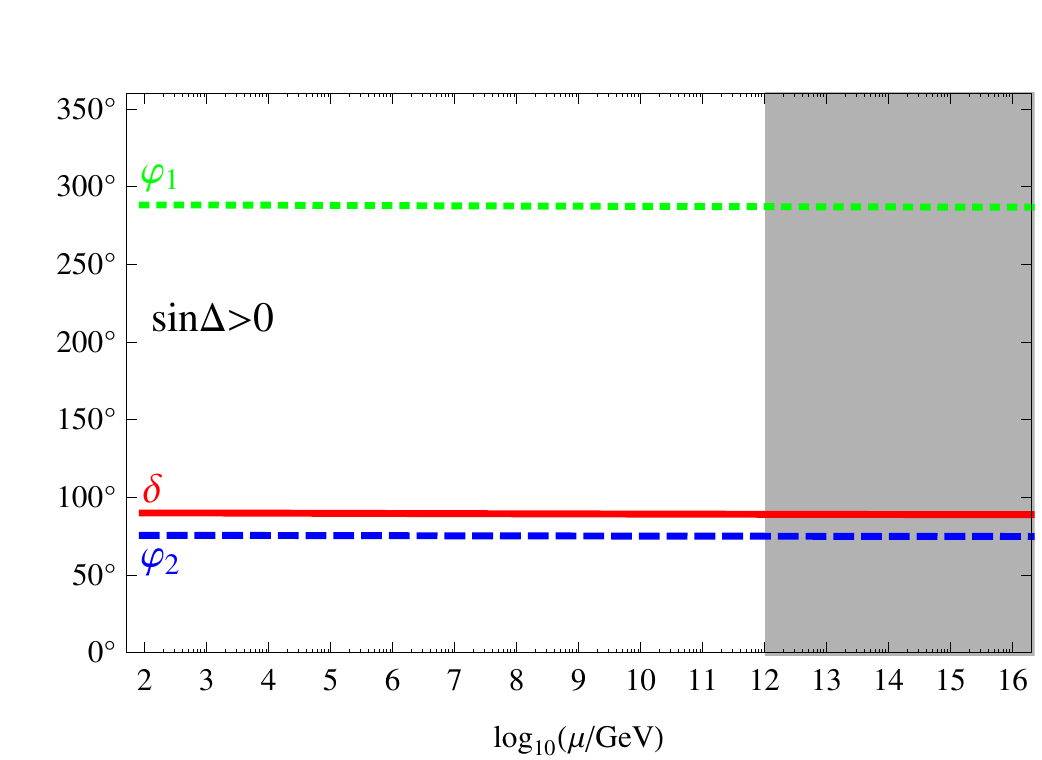}&\hspace{0.5cm}\includegraphics[scale=.65]{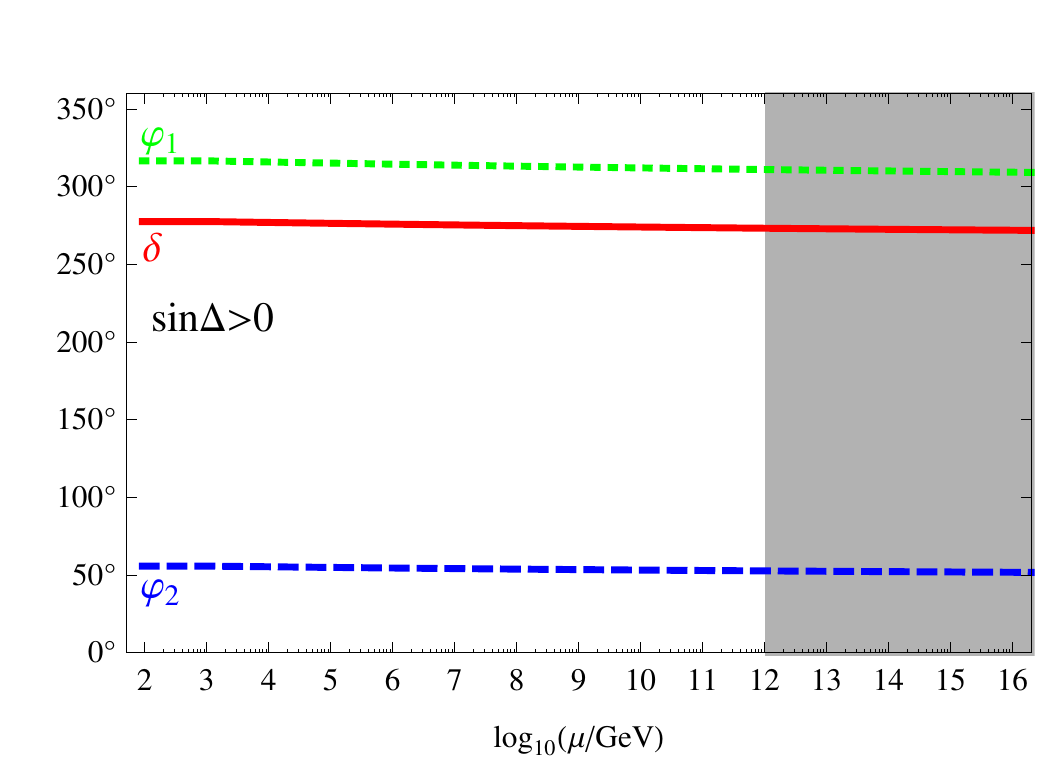}\\
\includegraphics[scale=.65]{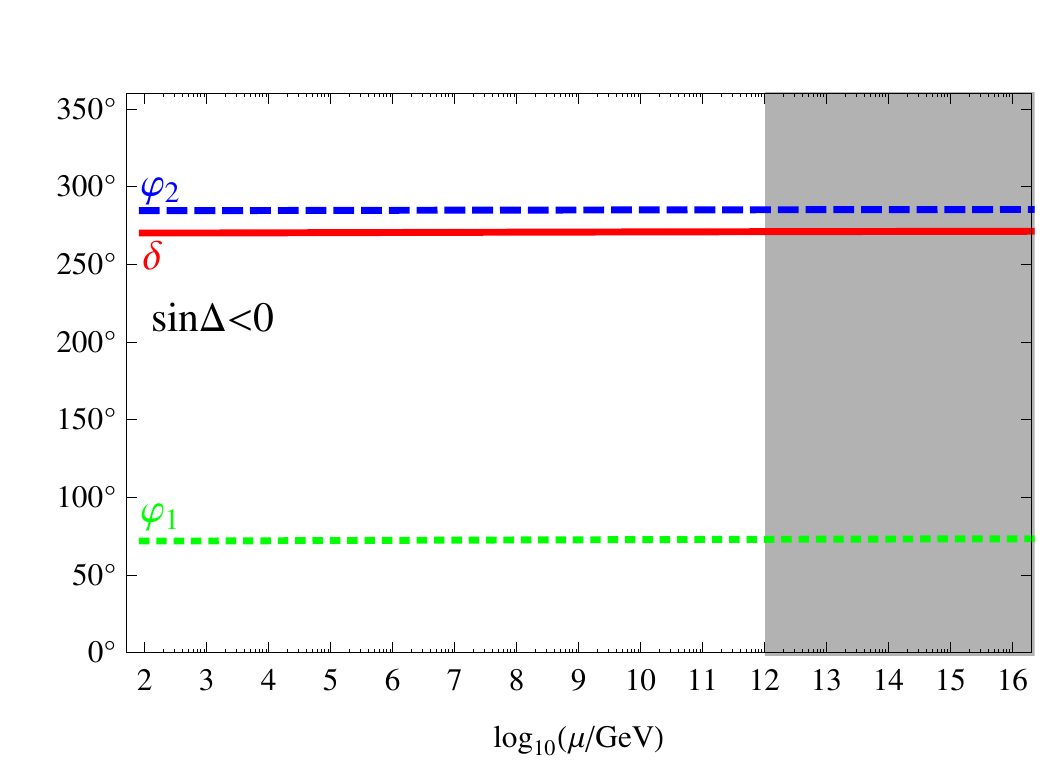}&\hspace{0.5cm}\includegraphics[scale=.65]{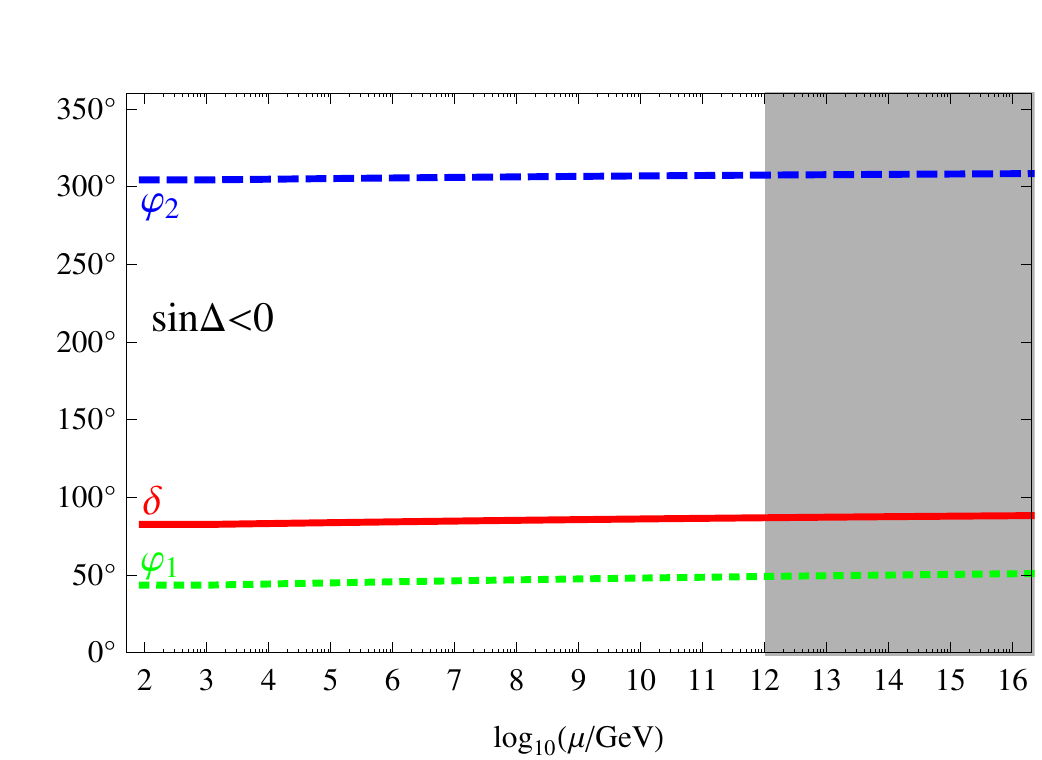}\\
\includegraphics[scale=.65]{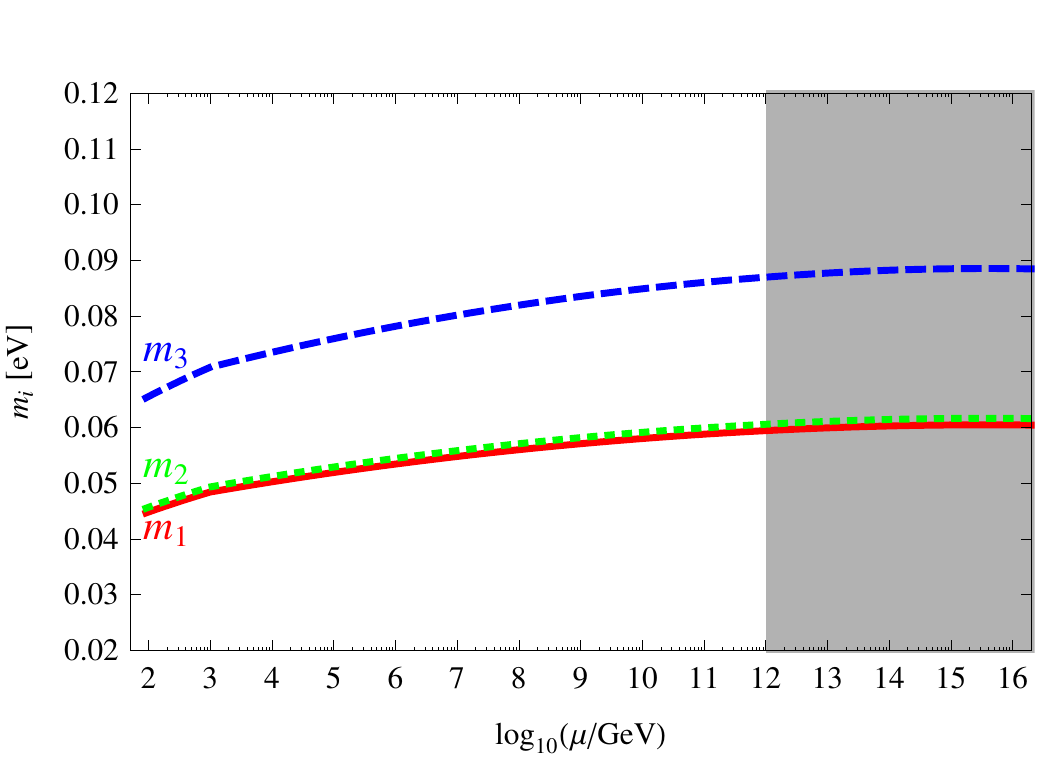}&\hspace{0.5cm}\includegraphics[scale=.65]{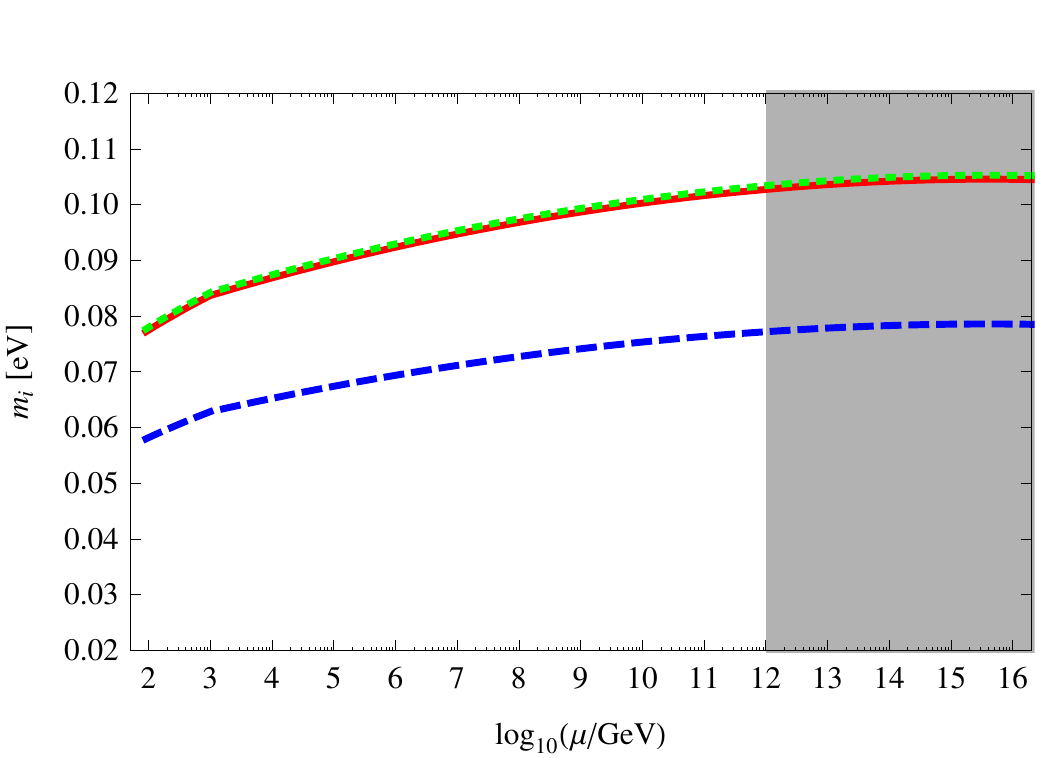}
\end{tabular}
\caption{\label{fig:RGE_Ding_scale} The running of the neutrino
masses and mixing parameters with the energy scale in the $S_4$
model of Ding with $\tan\beta=10$ and $M_{SUSY}=1$ TeV. The left
column is the predictions for NH spectrum with $m_1=0.0604$ eV,
$\Delta m^2_{\rm sol}= 1.43\times10^{-4}{\rm eV^2}$ and $\Delta
m^2_{\rm atm}=4.18\times10^{-3}{\rm eV^2}$. The right column is for
the IH case with $m_3=0.0785$ eV, $\Delta m^2_{\rm
sol}=1.48\times10^{-4}{\rm eV^2}$ and $\Delta m^2_{\rm
atm}=4.75\times10^{-3}{\rm eV^2}$. }
\end{center}
\end{figure}

The running of neutrino masses and mixing parameters with the energy
scale are plotted in Fig.\ref{fig:RGE_Ding_scale}. Being similar to
the situation in the BMM model, the RG corrections to the CP phases
$\delta$, $\varphi_1$ and $\varphi_2$ are typically small, the
corresponding curves are almost straight lines. We see that the
neutrino mixing angles are rather stable under RG evolution except
the solar angle for IH spectrum. The running of neutrino mass can be
approximately described by a common scaling factor, and it reduced by
about 2 times at electroweak scale. In short summary, the evolution
of the neutrino parameters in Ding's $S_4$ model is very similar to
that of BMM model, although the textures of the mass matrices are
totally different.

\section{\label{sec:conclusion}Conclusion}
Flavor models based on discrete flavor symmetry are particularly
interesting, they can produce the tri-bimaximal neutrino mixing (or
some other mass-independent mixing patterns) at LO in an elegant
way. It is a common feature that the LO predictions would be
corrected by the subleading higher dimensional operators, and it
have been shown that the subleading corrections are under control in
some consistent flavor models. Since the tri-bimaximal mixing is
predicted at high energy scale, it is very necessary to investigate
whether the RG effects would push the mixing parameters beyond the
current allowed ranges by experimental data.

In this paper, we have analyzed the RG running of the neutrino mass
and mixing parameters in the BMM model and the $S_4$ model of Ding,
both models predict tri-bimaximal neutrino mixing at LO, but the
textures of the mass matrices are totally different. To study the
running effects, we use the Mathematica package REAP. By detailed
numerical analysis, we find that the evolution of neutrino mixing
parameters displays approximately the same pattern in both $S_4$
models. We see that the atmospheric and reactor neutrino mixing
angles are essentially stable under RG evolution for both NH and IH
spectrum. However, the running of solar neutrino mixing angle
depends on the neutrino mass and the parameter $\tan\beta$, and the
deviation from its TB value could be large. After we take into
account the RG effects, the neutrino mass spectrum is strongly
constrained by the current data on $\theta_{12}$, the lightest
neutrino mass is bounded from both below and up, and the upper bound
decreases with $\tan\beta$. For large $\tan\beta$ ($\tan\beta>10$),
the value of $\sin^2\theta_{12}$ could be larger than its $3\sigma$
upper bound for the whole spectrum in the case of IH spectrum. As a
result, the IH neutrino mass spectrum is disfavored in the case of
large $\tan\beta$. Moreover, we note that the running of light-neutrino masses can be approximately described by a common scaling
factor, and they reduce by about $1.2\sim3$ times at low energy.
This effects is neglected in Ref.\cite{Lin:2009sq}. We note that the
evolutions of mixing angles and $J_{CP}$ don't depend on the sign of
$\sin\Omega$ or $\sin\Delta$, and the sum of each CP phase for both
sign is equal to $2\pi$. These results are confirmed both
analytically and numerically.

Finally we note that running of neutrino parameters in the
Altarelli-Feruglio $A_4$ model, BMM model and Ding's $S_4$ model is
similar to each other, although they produce tri-bimaximal mixing in
different ways. The reason is that the neutrino Yukawa coupling only
contributes to the running of neutrino mass, it doesn't affect the
lepton mixing angles, and the change in mixing angles is due to the contribution from the charged lepton sector. We conclude that
the running of mixing parameters is also severely constrained by the
flavor symmetry in discrete flavor symmetry models.

\section*{Acknowledgements}
We are grateful to Prof. Zhi-Zhong Xing for stimulating discussions
on RGE running.  The author Gui-Jun Ding gratefully acknowledge the
pleasant hospitality of the theory group at the University of
Wisconsin. This work is supported by the National Natural Science
Foundation of China under Grant No.10905053, Chinese Academy
KJCX2-YW-N29 and the 973 project with Grant No. 2009CB825200.
Dong-Mei Pan is supported in part by the National Natural Science
Foundation of China under Grant No.10775124.

\providecommand{\href}[2]{#2}\begingroup\raggedright

\endgroup

%\end{document}
\newpage

\begin{thebibliography}{10}
\bibitem{Strumia:2006db}
  A.~Strumia and F.~Vissani,
  %``Neutrino masses and mixings and..,''
  arXiv:hep-ph/0606054;
  M.~C.~Gonzalez-Garcia and M.~Maltoni,
  %``Phenomenology with Massive Neutrinos,''
  Phys.\ Rept.\  {\bf 460}, 1 (2008)
  [arXiv:0704.1800 [hep-ph]].

\bibitem{Schwetz:2008er}
   T.~Schwetz, M.~Tortola and J.~W.~F.~Valle,
  %``Global neutrino data and recent reactor fluxes: status of three-flavour
  %oscillation parameters,''
  arXiv:1103.0734 [hep-ph];
  T.~Schwetz, M.~A.~Tortola and J.~W.~F.~Valle,
  %``Three-flavour neutrino oscillation update,''
  New J.\ Phys.\  {\bf 10}, 113011 (2008)
  [arXiv:0808.2016 [hep-ph]]; M.~Maltoni and T.~Schwetz,
  %``Three-flavour neutrino oscillation update and comments on possible hints
  %for a non-zero theta_{13},''
  arXiv:0812.3161 [hep-ph].

\bibitem{Fogli:Indication}
G.~L.~Fogli, E.~Lisi, A.~Marrone, A.~Palazzo and A.~M.~Rotunno,
  %``What we (would like to) know about the neutrino mass,''
  arXiv:0809.2936 [hep-ph];
  %%CITATION = ARXIV:0809.2936;%%
G.~L.~Fogli, E.~Lisi, A.~Marrone, A.~Palazzo and A.~M.~Rotunno,
  %``Hints of theta_13>0 from global neutrino data analysis,''
  Phys.\ Rev.\ Lett.\  {\bf 101} (2008) 141801
  [arXiv:0806.2649 [hep-ph]].
  %%CITATION = PRLTA,101,141801;%%

\bibitem{TBmix} P.~F.~Harrison, D.~H.~Perkins and W.~G.~Scott, Phys.\ Lett.\  B {\bf 530}, 167
(2002), hep-ph/0202074; P.~F.~Harrison and W.~G.~Scott, Phys.\
Lett.\  B {\bf 535}, 163 (2002), hep-ph/0203209; Z.~Z.~Xing, Phys.\
Lett.\  B {\bf 533}, 85 (2002), hep-ph/0204049; X.~G.~He and A.~Zee,
Phys.\ Lett.\  B {\bf 560}, 87 (2003), hep-ph/0301092.


\bibitem{Altarelli:2010gt}
  G.~Altarelli and F.~Feruglio,
  %``Discrete Flavor Symmetries and Models of Neutrino Mixing,''
  Rev.\ Mod.\ Phys.\  {\bf 82}, 2701 (2010)
  [arXiv:1002.0211 [hep-ph]].


\bibitem{Ishimori:2010au}
  H.~Ishimori, T.~Kobayashi, H.~Ohki, H.~Okada, Y.~Shimizu and M.~Tanimoto,
  %``Non-Abelian Discrete Symmetries in Particle Physics,''
  Prog.\ Theor.\ Phys.\ Suppl.\  {\bf 183}, 1 (2010)
  [arXiv:1003.3552 [hep-th]].

\bibitem{Balaji:2000au}
  K.~R.~S.~Balaji, A.~S.~Dighe, R.~N.~Mohapatra and M.~K.~Parida,
  %``Radiative magnification of neutrino mixings and a natural explanation  of
  %the neutrino anomalies,''
  Phys.\ Lett.\  B {\bf 481}, 33 (2000)
  [arXiv:hep-ph/0002177];  S.~Antusch and M.~Ratz,
  %``Radiative generation of the LMA solution from small solar neutrino  mixing
  %at the GUT scale,''
  JHEP {\bf 0211}, 010 (2002)
  [arXiv:hep-ph/0208136];  R.~N.~Mohapatra, M.~K.~Parida and G.~Rajasekaran,
  %``High scale mixing unification and large neutrino mixing angles,''
  Phys.\ Rev.\  D {\bf 69}, 053007 (2004)
  [arXiv:hep-ph/0301234].


\bibitem{Antusch:2002hy}
  S.~Antusch, J.~Kersten, M.~Lindner and M.~Ratz,
  %``The LMA solution from bimaximal lepton mixing at the GUT scale by
  %renormalization group running,''
  Phys.\ Lett.\  B {\bf 544}, 1 (2002)
  [arXiv:hep-ph/0206078]; T.~Miura, T.~Shindou and E.~Takasugi,
  %``The renormalization group effect to the bi-maximal mixing,''
  Phys.\ Rev.\  D {\bf 68}, 093009 (2003)
  [arXiv:hep-ph/0308109].


\bibitem{Bazzocchi:2009pv}
  F.~Bazzocchi, L.~Merlo and S.~Morisi,
  %``Fermion Masses and Mixings in a S4-based Model,''
  Nucl.\ Phys.\  B {\bf 816}, 204 (2009)
  [arXiv:0901.2086 [hep-ph]]; F.~Bazzocchi, L.~Merlo and S.~Morisi,
  %``Phenomenological Consequences of See-Saw in S4 Based Models,''
  Phys.\ Rev.\  D {\bf 80}, 053003 (2009)
  [arXiv:0902.2849 [hep-ph]].

\bibitem{Ding:2009iy}
  G.~J.~Ding,
  %``Fermion Masses and Flavor Mixings in a Model with $S_4$ Flavor Symmetry,''
  Nucl.\ Phys.\  B {\bf 827}, 82 (2010) [arXiv:0909.2210 [hep-ph]].

\bibitem{Lam:2008rs}
  C.~S.~Lam,
  %``Determining Horizontal Symmetry from Neutrino Mixing,''
  Phys.\ Rev.\ Lett.\  {\bf 101}, 121602 (2008)
  [arXiv:0804.2622 [hep-ph]];
  C.~S.~Lam,
  %``The Unique Horizontal Symmetry of Leptons,''
  Phys.\ Rev.\  D {\bf 78}, 073015 (2008)
  [arXiv:0809.1185 [hep-ph]];
  C.~S.~Lam,
  %``A bottom-up analysis of horizontal symmetry,''
  arXiv:0907.2206 [hep-ph].


\bibitem{Ma:2005pd}
  E.~Ma,
  %``Neutrino mass matrix from S(4) symmetry,''
  Phys.\ Lett.\  B {\bf 632}, 352 (2006)
  [arXiv:hep-ph/0508231].


\bibitem{Meloni:2009cz}
  D.~Meloni,
  %``A See-Saw $S_4$ model for fermion masses and mixings,''
  J.\ Phys.\ G {\bf 37}, 055201 (2010)
  [arXiv:0911.3591 [hep-ph]].

\bibitem{Altarelli:2009gn}
  G.~Altarelli, F.~Feruglio and L.~Merlo,
  %``Revisiting Bimaximal Neutrino Mixing in a Model with S4 Discrete
  %Symmetry,''
  JHEP {\bf 0905}, 020 (2009)
  [arXiv:0903.1940 [hep-ph]].

\bibitem{Grimus:2009pg}
  W.~Grimus, L.~Lavoura and P.~O.~Ludl,
  %``Is S4 the horizontal symmetry of tri-bimaximal lepton mixing?,''
  J.\ Phys.\ G {\bf 36}, 115007 (2009)
  [arXiv:0906.2689 [hep-ph]].


\bibitem{Dutta:2009bj}
  B.~Dutta, Y.~Mimura and R.~N.~Mohapatra,
  %``An SO(10) Grand Unified Theory of Flavor,''
  JHEP {\bf 1005}, 034 (2010)
  [arXiv:0911.2242 [hep-ph]].

\bibitem{Hagedorn:2010th}
  C.~Hagedorn, S.~F.~King and C.~Luhn,
  %``A SUSY GUT of Flavour with S4 x SU(5) to NLO,''
  JHEP {\bf 1006}, 048 (2010)
  [arXiv:1003.4249 [hep-ph]].


\bibitem{Ishimori:2008fi}
  H.~Ishimori, Y.~Shimizu and M.~Tanimoto,
  %``S4 Flavor Symmetry of Quarks and Leptons in SU(5) GUT,''
  Prog.\ Theor.\ Phys.\  {\bf 121}, 769 (2009)
  [arXiv:0812.5031 [hep-ph]];
  H.~Ishimori, K.~Saga, Y.~Shimizu and M.~Tanimoto,
  %``Tri-bimaximal Mixing and Cabibbo Angle in S4 Flavor Model with SUSY,''
  arXiv:1004.5004 [hep-ph].

\bibitem{Toorop:2010yh}
  R.~de Adelhart Toorop, F.~Bazzocchi and L.~Merlo,
  %``The Interplay Between GUT and Flavour Symmetries in a Pati-Salam x S4
  %Model,''
  JHEP {\bf 1008}, 001 (2010)
  [arXiv:1003.4502 [hep-ph]].


\bibitem{Ding:2010pc}
  G.~J.~Ding,
  %``SUSY adjoint SU(5) grand unified model with S_4 flavor symmetry,''
  Nucl.\ Phys.\  B {\bf 846}, 394 (2011)
  [arXiv:1006.4800 [hep-ph]].



\bibitem{Pakvasa:1978tx}
  S.~Pakvasa and H.~Sugawara,
  %``Mass Of The T Quark In SU(2) X U(1),''
  Phys.\ Lett.\  B {\bf 82}, 105 (1979);
  T.~Brown, N.~Deshpande, S.~Pakvasa and H.~Sugawara,
  %``CP Nonconservation And Rare Processes In S(4) Model Of Permutation
  %Symmetry,''
  Phys.\ Lett.\  B {\bf 141}, 95 (1984);
  Y.~Yamanaka, H.~Sugawara and S.~Pakvasa,
  %``Permutation Symmetries And The Fermion Mass Matrix,''
  Phys.\ Rev.\  D {\bf 25}, 1895 (1982)
  [Erratum-ibid.\  D {\bf 29}, 2135 (1984)];
  T.~Brown, S.~Pakvasa, H.~Sugawara and Y.~Yamanaka,
  %``Neutrino Masses, Mixing And Oscillations In S(4) Model Of Permutation
  %Symmetry,''
  Phys.\ Rev.\  D {\bf 30}, 255 (1984).
\bibitem{Hagedorn:2006ug} D.~G.~Lee and R.~N.~Mohapatra,
  %``An SO(10) x S(4) scenario for naturally degenerate neutrinos,''
  Phys.\ Lett.\  B {\bf 329}, 463 (1994)
  [arXiv:hep-ph/9403201];
  C.~Hagedorn, M.~Lindner and R.~N.~Mohapatra,
  %``S(4) flavor symmetry and fermion masses: Towards a grand unified theory  of
  %flavor,''
  JHEP {\bf 0606}, 042 (2006)
  [arXiv:hep-ph/0602244];
  Y.~Cai and H.~B.~Yu,
  %``An SO(10) GUT Model with $S4$ Flavor Symmetry,''
  Phys.\ Rev.\  D {\bf 74}, 115005 (2006)
  [arXiv:hep-ph/0608022];
  H.~Zhang,
  %``Flavor S(4) x Z(2) symmetry and neutrino mixing,''
  Phys.\ Lett.\  B {\bf 655}, 132 (2007)
  [arXiv:hep-ph/0612214];
  Y.~Koide,
  %``S_4 Flavor Symmetry Embedded into SU(3) and Lepton Masses and Mixing,''
  JHEP {\bf 0708}, 086 (2007)
  [arXiv:0705.2275 [hep-ph]];
  M.~K.~Parida,
  %``Intermediate left-right gauge symmetry, unification of couplings and
  %fermion masses in SUSY $SO(10)\times S_4$,''
  Phys.\ Rev.\  D {\bf 78}, 053004 (2008)
  [arXiv:0804.4571 [hep-ph]].
  %%CITATION = PHRVA,D78,053004;%%

\bibitem{rge1} P.~H.~Chankowski and Z.~Pluciennik, Phys.\ Lett.\  B {\bf 316}, 312 (1993)
  [arXiv:hep-ph/9306333];  K.~S.~Babu, C.~N.~Leung and J.~T.~Pantaleone, Phys.\ Lett.\  B {\bf 319}, 191 (1993)
  [arXiv:hep-ph/9309223].

\bibitem{rge2}
  J.~A.~Casas, J.~R.~Espinosa, A.~Ibarra and I.~Navarro,
  %``Nearly degenerate neutrinos, supersymmetry and radiative corrections,''
  Nucl.\ Phys.\  B {\bf 569}, 82 (2000)
  [arXiv:hep-ph/9905381]; J.~A.~Casas, J.~R.~Espinosa, A.~Ibarra and I.~Navarro,
  %``General RG equations for physical neutrino parameters and their
  %phenomenological implications,''
  Nucl.\ Phys.\  B {\bf 573}, 652 (2000)
  [arXiv:hep-ph/9910420].

\bibitem{rge3} S.~Antusch, M.~Drees, J.~Kersten, M.~Lindner and M.~Ratz,
  %``Neutrino mass operator renormalization revisited,''
  Phys.\ Lett.\  B {\bf 519}, 238 (2001)
  [arXiv:hep-ph/0108005]; S.~Antusch, J.~Kersten, M.~Lindner and M.~Ratz,
  %``Neutrino mass matrix running for non-degenerate see-saw scales,''
  Phys.\ Lett.\  B {\bf 538}, 87 (2002)
  [arXiv:hep-ph/0203233];  M.~Lindner, M.~A.~Schmidt and A.~Y.~Smirnov,
  %``Screening of Dirac flavor structure in the seesaw and neutrino mixing,''
  JHEP {\bf 0507}, 048 (2005)
  [arXiv:hep-ph/0505067].

\bibitem{rge4} S.~F.~King and N.~N.~Singh,
  %``Renormalisation group analysis of single right-handed neutrino
  %dominance,''
  Nucl.\ Phys.\  B {\bf 591}, 3 (2000)
  [arXiv:hep-ph/0006229]; J.~w.~Mei,
  %``Running neutrino masses, leptonic mixing angles and CP-violating  phases:
  %From M(Z) to Lambda(GUT),''
  Phys.\ Rev.\  D {\bf 71}, 073012 (2005)
  [arXiv:hep-ph/0502015]; J.~R.~Ellis, A.~Hektor, M.~Kadastik, K.~Kannike and M.~Raidal,
  %``Running of low-energy neutrino masses, mixing angles and CP violation,''
  Phys.\ Lett.\  B {\bf 631}, 32 (2005)
  [arXiv:hep-ph/0506122].

\bibitem{rge5} W.~Chao and H.~Zhang,
  %``One-loop renormalization group equations of the neutrino mass matrix in the
  %triplet seesaw model,''
  Phys.\ Rev.\  D {\bf 75}, 033003 (2007) [arXiv:hep-ph/0611323];
 M.~A.~Schmidt,  %``Renormalization Group Evolution in the type I + II seesaw model,''
  Phys.\ Rev.\  D {\bf 76}, 073010 (2007) [arXiv:0705.3841 [hep-ph]].

\bibitem{Chakrabortty:2008zh}
  J.~Chakrabortty, A.~Dighe, S.~Goswami and S.~Ray,
  %``Renormalization group evolution of neutrino masses and mixing in the
  %Type-III seesaw mechanism,''
  Nucl.\ Phys.\  B {\bf 820}, 116 (2009)
  [arXiv:0812.2776 [hep-ph]].
  %%CITATION = NUPHA,B820,116;%%


\bibitem{Antusch:2003kp}
  S.~Antusch, J.~Kersten, M.~Lindner and M.~Ratz,
  %``Running neutrino masses, mixings and CP phases: Analytical results and
  %phenomenological consequences,''
  Nucl.\ Phys.\  B {\bf 674}, 401 (2003)
  [arXiv:hep-ph/0305273].

\bibitem{Antusch:2005gp}
  S.~Antusch, J.~Kersten, M.~Lindner, M.~Ratz and M.~A.~Schmidt,
  %``Running neutrino mass parameters in see-saw scenarios,''
  JHEP {\bf 0503}, 024 (2005)
  [arXiv:hep-ph/0501272].


\bibitem{Xing:2007fb}
  Z.~z.~Xing, H.~Zhang and S.~Zhou,
  %``Updated Values of Running Quark and Lepton Masses,''
  Phys.\ Rev.\  D {\bf 77}, 113016 (2008)
  [arXiv:0712.1419 [hep-ph]].
\bibitem{Bergstrom:2010qb}
  J.~Bergstrom, M.~Malinsky, T.~Ohlsson and H.~Zhang,
  %``Renormalization group running of neutrino parameters in the inverse seesaw
  %model,''
  Phys.\ Rev.\  D {\bf 81}, 116006 (2010)
  [arXiv:1004.4628 [hep-ph]].
\bibitem{Blennow:2011mp}
  M.~Blennow, H.~Melbeus, T.~Ohlsson and H.~Zhang,
  %``Renormalization Group Running of the Neutrino Mass Operator in Extra
  %Dimensions,''
  arXiv:1101.2585 [hep-ph].


\bibitem{Barry:2010yk}
  J.~Barry and W.~Rodejohann,
  %``Neutrino Mass Sum-rules in Flavor Symmetry Models,''
  Nucl.\ Phys.\  B {\bf 842}, 33 (2011)
  [arXiv:1007.5217 [hep-ph]].

\bibitem{pdg} K.~Nakamura {\it et al.}  [Particle Data Group],
  %``Review of particle physics,''
  J.\ Phys.\ G {\bf 37}, 075021 (2010).

\bibitem{Boudjemaa:2008jf}
  S.~Antusch and M.~Spinrath,
  %``Quark and lepton masses at the GUT scale including SUSY threshold
  %corrections,''
  Phys.\ Rev.\  D {\bf 78}, 075020 (2008)
  [arXiv:0804.0717 [hep-ph]]; S.~Boudjemaa and S.~F.~King,
  %``Deviations from Tri-bimaximal Mixing: Charged Lepton Corrections and
  %Renormalization Group Running,''
  Phys.\ Rev.\  D {\bf 79}, 033001 (2009)
  [arXiv:0808.2782 [hep-ph]]; M.~Bustamante, A.~M.~Gago and J.~Jones-Perez,
  %``SUSY Renormalization Group Effects in Ultra High Energy Neutrinos,''
  arXiv:1012.2728 [hep-ph].

\bibitem{katrin}
A.~Osipowicz {\it et al.}  [KATRIN Collaboration],
  %``KATRIN: A next generation tritium beta decay experiment with sub-eV
  %sensitivity for the electron neutrino mass,''
  arXiv:hep-ex/0109033;
  %%CITATION = HEP-EX/0109033;%%
  see also: http://www-ik.fzk.de/~katrin/index.html.


\bibitem{Lin:2009sq}
  Y.~Lin, L.~Merlo and A.~Paris,
  %``Running Effects on Lepton Mixing Angles in Flavour Models with Type I
  %Seesaw,''
  Nucl.\ Phys.\  B {\bf 835}, 238 (2010)
  [arXiv:0911.3037 [hep-ph]].

\bibitem{Chankowski:2001mx}
  P.~H.~Chankowski and S.~Pokorski,
  %``Quantum corrections to neutrino masses and mixing angles,''\
Int.\ J.\ Mod.\ Phys.\  A {\bf 17}, 575
(2002)[arXiv:hep-ph/0110249].



\end{thebibliography}
\end{document}